\documentclass{aa}

\usepackage{graphicx}

\usepackage{txfonts}
\usepackage[retainorgcmds]{IEEEtrantools}

\usepackage{hyperref}
\hypersetup{colorlinks=true, linkcolor=blue, citecolor=blue}

\numberwithin{equation}{section}
\usepackage{color}
\usepackage{longtable}
\usepackage{appendix}
\usepackage{supertabular}
\usepackage{ragged2e}
\begin{document}

   \title{HR-Cosmos\thanks{Based on observations made with ESO Telescopes at the La Silla Paranal Observatory under programme ID 083.A-0935.}: Kinematics of Star-Forming Galaxies at z $\sim$ 0.9}

   \author{D. Pelliccia\inst{1},  L. Tresse\inst{2}, B. Epinat\inst{1}, O. Ilbert\inst{1}, N. Scoville\inst{3}, P. Amram\inst{1},  B. C. Lemaux\inst{4}, G. Zamorani\inst{5}
          }

   \institute{Aix Marseille Université, CNRS, LAM (Laboratoire d’Astrophysique de Marseille) UMR 7326, 13388, Marseille, France\\
              \email{debora.pelliccia@lam.fr}
         \and
			Univ Lyon, Ens de Lyon, Univ Lyon1, CNRS, Centre de Recherche Astrophysique de Lyon UMR5574, F-69007, Lyon, France
			\and
		California Institute of Technology, MC 249-17, 1200 East California Boulevard, Pasadena, CA 91125, USA
		\and
		 University of California Davis, 1 Shields Avenue, Davis, CA 95616
		\and
		INAF - Osservatorio Astronomico di Bologna, via Ranzani 1, 40127 Bologna, Italy}

   \date{Received Month day, year; accepted Month day, year}

  \abstract{We present the kinematic analysis of a sub-sample of 82 galaxies at $\mathrm{0.75<z<1.2}$ from our new survey HR-COSMOS aimed to obtain the first statistical sample to study the kinematics of star-forming galaxies in the treasury COSMOS field at $\mathrm{0<z<1.2}$. 
We observed 766 emission line galaxies using the multi-slit spectrograph ESO-VLT/VIMOS in high-resolution mode (R=2500). To better extract galaxy kinematics, VIMOS spectral slits have been carefully tilted along the major axis orientation of the galaxies, making use of the position angle measurements from the  high spatial resolution \textit{HST}/ACS COSMOS images.
We constrained the kinematics of the sub-sample at $0.75<z<1.2$ by creating high-resolution semi-analytical models.
We established the stellar-mass Tully-Fisher relation at $z\simeq 0.9$ with high-quality stellar mass measurements derived using the latest COSMOS photometric catalog, which includes the latest data releases of UltraVISTA and \textit{Spitzer}. 
 In doubling the sample at these redshifts compared with the literature, we estimated the relation without setting its slope, and found it consistent with previous studies in other deep extragalactic fields assuming no significant evolution of the relation with redshift at $z\lesssim1$. We computed dynamical masses within the radius R$_{2.2}$ and found a median stellar-to-dynamical mass fraction equal to 0.2 (assuming Chabrier IMF), which implies a contribution of gas and dark matter masses of 80\% of the total mass within  R$_{2.2}$, in agreement with recent integral field spectroscopy surveys. We find no dependence of the stellar-mass Tully-Fisher relation with environment probing up to group scale masses. This study shows that multi-slit galaxy surveys remain a powerful tool to derive kinematics for large numbers of galaxies at both high and low redshift.
  }

   \authorrunning{D. Pelliccia et al.}

   \maketitle
   
%

\section{Introduction}
In the local Universe, we distinguish a clear  bimodality distribution of galaxies in two types with red and blue rest-frame colors, commonly referred to as the red sequence and blue cloud, with a lack of galaxies at intermediate colors \citep{Blanton2003, Baldry2004}. These color classes are strongly correlated with morphology, so that the blue cloud region is mostly populated by rotating star-forming galaxies (usually spiral) and on the red sequence we find generally quiescent ellipticals, though there is a rare population of red spirals which overlaps the red sequence and may arise from truncated star formation \citep{Masters2010, Cortese2012}. Star-forming galaxies at high redshift ($\mathrm{z>0.1}$) exhibit a larger diversity of kinematics than in the local Universe ($\mathrm{z\ll 0.1}$), and higher fractions of such galaxies are not dominated by rotation or have complex kinematic signatures owing to merging processes \citep{Conselice2003, Hammer2009, Kassin2012, Puech2012}. 

Kinematics enables us to trace galaxy properties from high to low redshift and to constrain the galaxy formation scenario using, for example, the well established scaling relation for rotating galaxies observed for the first time by \citet{Tully1977} between the optical luminosity and HI line width.
The Tully-Fisher (TF) relation is a key to understand the structure and evolution of disc galaxies, since it probes the  dark matter halos with their luminous baryonic components, and thus constrains the galaxy formation and evolution models \citep[e.g.,][]{Dalcanton1997, Sommer-Larsen2003}.
Later works have shown that the TF relation becomes tighter when defined using microwave to infrared luminosity \citep{Verheijen1997}, and even tighter when luminosity is replaced by the galaxy's stellar mass \citep{Pizagno2005} or total baryonic mass \citep{McGaugh2005}.

The stellar mass Tully-Fisher (smTF) relation has been studied in the local Universe \citep{BelldeJong2001, Pizagno2005, Courteau2007, Reyes2011} and it appears to be remarkably tight. However, the relation is not yet well defined at higher redshift. Past studies have investigated possible evolution of the smTF relation with conflicting results.
\citet{Conselice2005} was the first to look at the smTF relationship at intermediate redshift using the longslit spectroscopy technique with a sample of 101 disc gala\-xies with near-IR photometry, fin\-ding no evidence for an evolution of the relation from $\mathrm{z = 0}$ to $\mathrm{z=1.2}$. 
 \citet{Puech2008, Puech2010} constrained the smTF relation at   
$\mathrm{z\sim0.6}$ for a small sample of 16 rotating disc galaxies (out of 64 analyzed), part of the IMAGES survey, and they found an evolution of the relation towards lower stellar masses. A slit-based survey with DEIMOS multi-object spectrograph on Keck \citep{Faber2003} was presented by \citet{Miller2011} who have sampled a larger number of galaxies (N=129) at  
$\mathrm{0.2<z<1.3}$, showing no statistically significant evolution of the smTF relation, though an evident evolution of the B-band TF with a decrease in luminosity of $\Delta M_B = 0.85\pm 0.28 \:\mathrm{mag}$ from $\mathrm{\langle z \rangle \simeq 1.0}$ to 0.3.

Furthermore, it is unclear whether an evolution with redshift of the intrinsic scatter around the TF relation exists. \mbox{\citet{Kannappan2002}} have shown how the scatter around the TF relation in the local Universe increases by broadening their spiral sample, for which the relation is computed, to include all morphologies. Since galaxies at higher redshift show more irregular morphology \citep{Glazebrook1995, ForsterSchreiber2009}, we expect to observe an increase in the scatter with redshift.  \citet{Puech2008, Puech2010} argue that the increase in intrinsic scatter around the TF relation was due to the galaxies with complex kinematics. \citet{Kassin2007}, as well,  looking at the smTF relationship of 544 galaxies ($\mathrm{0.1 < z < 1.2}$)  from the DEEP2 redshift survey \citep{Newman2013_deep2},  have found a large scatter ($\sim$ 1.5 dex in velocity) dominated by the more disturbed morphological classes and the lower stellar masses. Conversely \citet{Miller2011} reported a very small intrinsic scatter around the smTF relation throughout their whole sample at $\mathrm{0.2<z<1.3}$. 

The large scatter around the TF relation for broadly selected samples, may also depend  on properties external to the \mbox{galaxies}. Environment, for example,  is known to have a strong impact on galaxy parameters \citep{Baldry2006, Bamford2009, Capak2007,  Sobral2011, Scoville2013} and we could expect to see  environmental effects also on the TF relation. An alteration in the TF relation in a dense environment, for example, could be due to a temporary increase of the star formation activity or to kinematic asymmetry owing to dynamical interaction of galaxies.
Nevertheless, studies on the effects of the environment on the TF relation are still few. 
\citet{Mocz2012} constrained the TF relation in the \textit{u}, \textit{g}, \textit{r}, \textit{i}, and \textit{z} bands and smTF for a local Universe sample of $25\,698$ late spiral galaxies from the Sloan Digital Sky Survey (SDSS) and investigated any dependence of the relation from the environment defined using the local number density of galaxies as a proxy to their host region. 
They found no strong or statistically significant changes in the TF relation slope or intercept. At intermediate redshifts, some works constrained the TF relation for cluster and field galaxies, but the results are highly discordant. \citet{Ziegler2003} compared the B-band TF relation for a small sample of 13 galaxies from three clusters at $\mathrm{z = 0.3}$, $\mathrm{z = 0.42}$ and  $\mathrm{z = 0.51}$ and 77 field galaxies at  $\mathrm{0<z<1}$ observed with the same set up of the clusters, and they have found no significant differences. The same result of no environmental effect  was presented by \citet{Nakamura2006} for a still small sample of 13 cluster galaxies and 20 field galaxies between $\mathrm{z = 0.23}$ and $\mathrm{z = 0.58}$.
Conversely, \citet{Bamford2005} with a larger sample of 80 galaxies at $\mathrm{0<z<1}$ (22 in clusters and 58 in the field) have found a systematic offset of $\Delta M_B = 0.7\pm 0.2 \:\mathrm{mag}$  in the B-band TF relation for the cluster galaxies with respect to the field sample.  \citet{Moran2007}, as well, computed the scatter around the Ks-band (proxy of the stellar mass) and \mbox{V-band} TF relations at $\mathrm{0.3 \lesssim  z \lesssim 0.65}$ for 40 cluster galaxies and 37 field galaxies, showing that cluster galaxies are more scattered than the field ones. More recent work of \citet{Jaffe2011} included in their sample group galaxies, for a total of $\sim$150 \mbox{galaxies} at  
$\mathrm{0.3<z<0.9}$,  and found no correlation between the scatter around the B-band TF and the environment. Whilst \citet{Bosch2013}, for a sample of 55 cluster galaxies at $\mathrm{z\sim 0.17}$ and 57 field galaxies at 
$\mathrm{\langle z \rangle = 0.25}$,  found a slight shift in the intercept of the B-band TF relation to brighter values for field galaxies and no evolution on intercept of the smTF relation. 

We emphasize that defining the environment as two extreme regions in the density distribution of galaxies, i.e., galaxy cluster and general field, is an overly coarse binning of the full dynamical range of the density field at these redshifts. There are, indeed, intermediate environments, like galaxy groups, outskirts of clusters, and filaments which are also important \citep{Fadda2008, Porter2008, Tran2009, Coppin2012,  Darvish2014, Darvish2015}. A larger and more homogeneous study to investigate the environmental effect on the kinematics is still needed.

The recent advent of Integral Field Spectroscopy has provided the opportunity to map the kinematics of intermediate/high redshift galaxies. Various surveys have been obtained so far, by making use of the new generation of Integral Field Unit (IFU), such as the SINS \citep{ForsterSchreiber2009} and MASSIV \citep{Contini2012} surveys with  the near-infrared spectrograph SINFONI  \citep{Eisenhauer2003}, KMOS$\mathrm{^{3D}}$ \citep{Wisnioski2015} and KROSS \citep{Stott2016} surveys with the Multi-Object Spectrograph KMOS \citep{Sharples2013}. Kinematic analysis has been also presented with the first set of observations taken with the Multi Unit
Spectroscopic Explorer (MUSE) on VLT \citep{Bacon2015, Contini2015} which was able to resolve galaxies in the lower stellar mass range down to $10^8\, M_\odot$. One of the principal limiting factors of the IFU surveys to date has been the necessity to observe one object at a time, because of the IFU small field of view (FOV). This limitation has been improved a lot with the new multi-IFU spectrograph KMOS that operates using 24 IFUs, each one with $\mathrm{2.8\arcsec \times 2.8\arcsec}$ FOV, and the $\mathrm{1\arcmin \times 1\arcmin}$ wide field mode provided by MUSE. However the FOV of those spectrographs remains much smaller than the one that multi-slit spectrographs can provide, such as VLT-VIMOS, which is made of four identical arms, each one with a FOV of  $\mathrm{7\arcmin \times 8\arcmin}$, enabling us to observe $\sim\,$120 galaxies per exposure.

Our new survey HR-COSMOS aims to obtain the first statistical sample to study the kinematics of star-forming galaxies in the COSMOS\- field \citep{Scoville2007}  at $\mathrm{0<z<1.2}$. We seek to obtain a sample that is as representative as possible. Thus we apply no morphological pre-selection other than a cut on the axial ratio $b/a<0.84$ and the position angle $\mid PA\mid \,\leq 60\,^{\circ}$ for an optimal extraction of the kinematic parameters. We make use of the VIsible Multi-Object Spectrograph at ESO-VLT (VIMOS) in high-resolution (HR) mode, which provides the advantageous facility to perfectly align the spectral slits on our masks along the galaxy major axis measured from the high spatial resolution \textit{HST}/ACS images. 
In this paper, we present the kinematic analysis of a sub-sample at $\mathrm{0.75<z<1.2}$. We constrain the smTF relation at z$\,\sim\,$0.9, using rotation velocities extracted with high-resolution semi-analytical models and stellar masses measured  employing the latest COSMOS photometric catalog, which includes UltraVISTA and \textit{Spitzer} latest data releases. We investigate possible dependence of the smTF on the environment defined using the local surface density measurements by \citet{Scoville2013}, who make use of the 2D Voronoi tessellation techni\-que. 
We extend the same analysis to the whole sample to investi\-gate any  kinematic evolution with redshift in Pelliccia et al. (\textit{in~prep.}), whilst our data set will be described in Tresse et al. (\textit{in prep.}).

The paper is organized as follows. In $\S$\ref{sec:Data} we present our HR-COSMOS data set, describing the selection criteria, the data reduction process, and the measurements of the stellar masses. In  $\S$\ref{sec:kinematic_model} we describe the method used to create the kinematic models and to fit them with the observed data. The results are presented in  $\S$\ref{sec:results}, showing the constrained smTF relation at z$\sim$ 0.9, the dynamical mass measurements and the smTF relation as a function of the environment.
We summarize our results in $\S$\ref{sec:conclusion}.

Throughout this paper, we adopt a \citet{Chabrier2003} initial mass function and a standard $\Lambda$CDM cosmology with
H$_0$ = 70 km s$^{-1}$, $\Omega_\Lambda$ = 0.73, and $\mathrm{\Omega_M}$ = 0.27. Magnitudes are given in the AB system. 
\nocite{*}
%
\section{Data} \label{sec:Data}
\subsection{The HR-COSMOS survey} \label{subsec:the_survey}
Our data set at $\mathrm{z\sim 0.9}$ is part of an observational  campaign \mbox{(PI: L. Tresse)} aimed to measure the kinematics of $\sim$800 emission line galaxies at redshift $\mathrm{0<z<1.2}$ observed with VIMOS in HR mode ($\S$ \ref{subsec:Observation}). The description of the whole survey will be presented in a following paper.

We selected our targets within the Cosmic Evolution Survey (COSMOS) field \citep{Scoville2007}, a $\sim$2 deg$^2$ equatorial field imaged with the Advanced Camera for Survey (ACS) of the \textit{Hubble Space Telescope} (\textit{HST}). This \textit{HST}/ACS treasury field has been observed by most of the major space-based telescope (\textit{Spitzer}, \textit{Herschel}, \textit{Galex}, \textit{XMM-Newton}, \textit{Chandra}) and by many large ground based telescopes (Subaru, VLA, ESO-VLT, ESO-VISTA, UKIRT, CFHT, Keck, and others). Thus we have access to a wide range of physical parameters of the COSMOS galaxies. 
Emission-line targets have been selected using the spectroscopic information from the $\mathrm{zCOSMOS}$ \mbox{\textit{10k}-bright} sample \citep{Lilly2007} acquired with VIMOS at medium resolution ($R=580$), and selected with  $I_{AB}<22.5$ and $\mathrm{z<1.2}$.
We chose spectra with reliable spectroscopic redshifts (flags=2,3,4) and emission line flux $f_{line}>10^{-17}erg/s/cm^2$ for at least one emission line amongst [O{\small II}]$\lambda$3727, H$\beta \lambda$4861, [O{\small III}]$\lambda$5007, H$\alpha \lambda$6563, lying away from sky lines and visible in VIMOS HR ($\mathrm{R=2500}$) grism spectral range.
To fill the VIMOS masks we added a smaller sample of galaxies that have photometric redshift information \citep{Ilbert2009}. For this sample, to increase the probability to target rotating galaxies, we made use of the GINI parameter as morphology measure \citep{Abraham2003, Lotz2004, Capak2007} and selected galaxies with GINI parameter in a range between 0.200 and 0.475.

\begin{figure}[!tb]
   \centering
   \includegraphics[width=0.99\hsize]{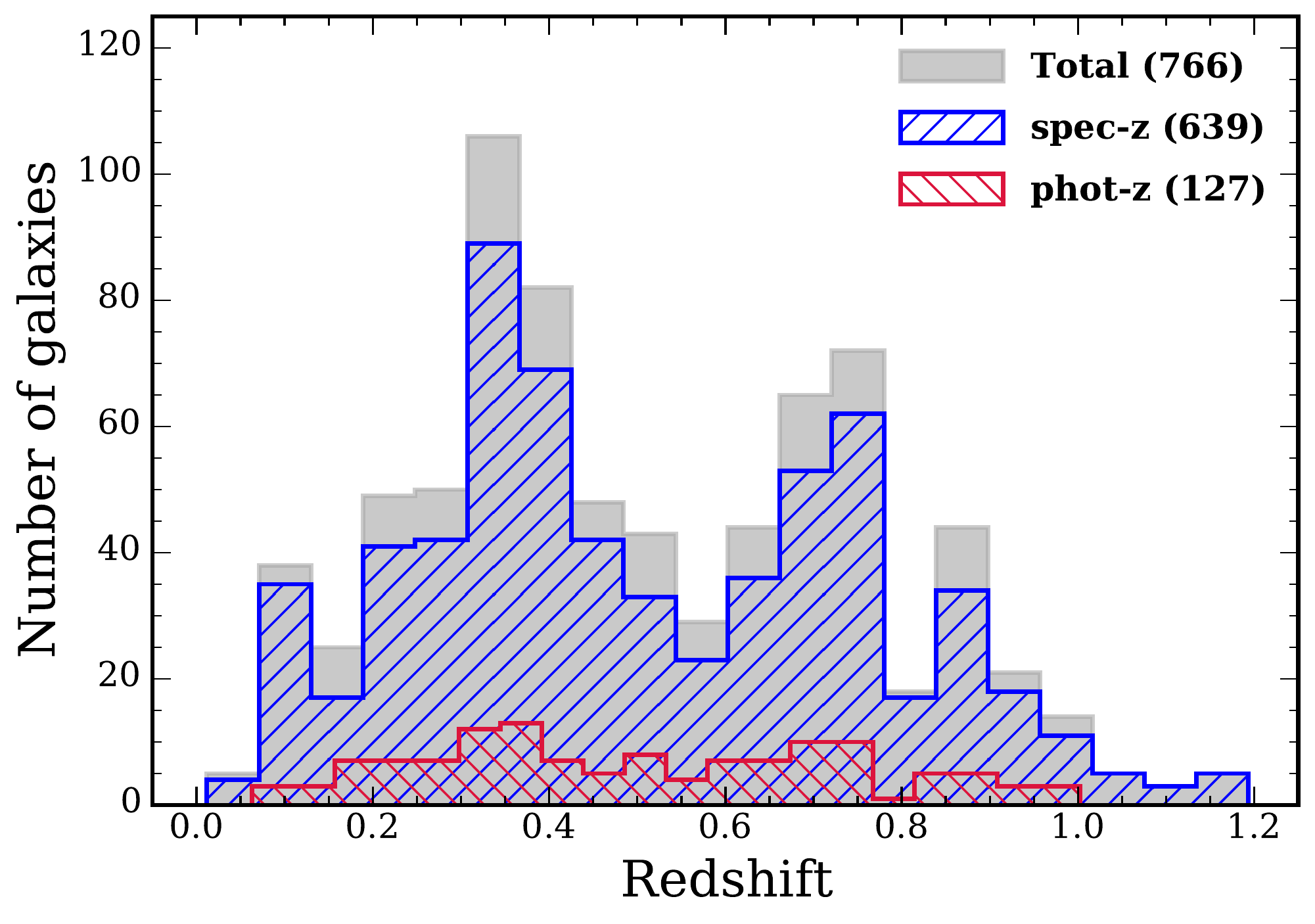}
      \caption{Redshift distribution of the 766 star-forming galaxies part of our HR-COSMOS survey. The gray filled histogram shows the overall sample and the hashed blue and red histograms represent the sub-samples selected with known spectroscopic redshifts and photometric redshifts, respectively.
              }
         \label{fig:z_histo}
   \end{figure}

In Figure~\ref{fig:z_histo} we show the redshift distribution of the whole sample, of which 83$\%$ was selected  with known spectroscopic redshift.
\begin{table*}[!htb]
\caption{Table of observations. Run ID: 083.A-0935(B)}   
\label{tab:observ_log}     
\centering         
\begin{tabular}{c c c c c c c}
\hline\hline 
\noalign{\smallskip}       
Pointing  & $\alpha$ & $\delta$ & t$_{exp}$ & Seeing & Date & Period \\[5pt]
 & J2000 & J2000& \textit{h:m:s} & \textit{arcsec} & & \\  [5pt]
 (1) & (2)  & (3)  & (4) & (5) &(6)  & (7)\\
\hline
\noalign{\smallskip} 
P01 &	10:00:14.400	& 02:15:00.000 & 1:40:45 &  0.796 & Dec. 2009, Jan. 2010 & P84   \\  [2pt]
P02 &	10:01:26.400	& 02:15:00.000 & 1:40:45   &  0.893 & Mar. 2010 & P84 \\ [2pt]
P03 &	09:59:02.400	& 02:15:00.000 & 1:40:45   &  0.886 & Apr. 2009, Dec. 2009 & P83, P84 \\ [2pt]
P04 & 	10:01:00.000	& 02:30:00.000 & 1:55:20  &  0.732 &	Mar. 2010, Apr. 2010 & P84, P85 \\ [2pt]
P05 &	09:59:45.600	& 02:30:00.000 &	1:40:45  &  0.786  & Feb. 2010, Mar. 2010 &P84      \\ [2pt]
P06 &	10:01:00.000	& 02:00:00.000 &	1:40:45 &   0.829 & Jan. 2010 & P84      \\ [2pt]
P07 &	09:59:45.600	& 02:00:00.000 &	 1:40:45  &  0.739 & Jan. 2010, Feb. 2010 & P84     \\ [2pt]
\hline
\end{tabular}
\flushleft
{\footnotesize (1) Observation pointings, (2) and (3) central coordinates, (4) exposure time, (5) mean seeing over the four quadrants, computed as the FWHM from the spatial profile of the collapsed star spectra present in each quadrant (see text $\S$ \ref{subsec:DataReduction}). (6) Observation dates, (7) ESO Period.}
\end{table*}
To keep our kinematic sample as representative as possible, we have applied a little pre-selection of the targets:  
\begin{center}
$z = 0. - 1.2$ \\[6pt]
$I_{AB} \leq 22.5$\\[6pt]
$f_{line} >10^{-17} erg/s/cm^2$\\[6pt]
$-60\,^{\circ} \leq PA \leq 60\,^{\circ}$ \\[6pt]
$0 \leq b/a \leq 0.84$.
\end{center}
The first two criteria, the redshift and $I_{AB}$ magnitude, are in common with the zCOSMOS parent sample \citep{Lilly2007}. The condition on the line flux is to ensure the detection of the line visible in VIMOS HR grism spectral range, and the last two selection criteria are chosen for an optimal extraction of the kinematics along the slit. To derive high-quality kinematics, VIMOS spectral slits have been carefully tilted along the major axis orien\-tation of the galaxies. We retained star-forming galaxies  with position angle, defined as the angle (East of North) between the North direction in the sky and the galaxy major axis, $\mid PA\mid \,\leq 60\,^{\circ}$ since this is the limit to tilt the individual slits of VIMOS from the N-S spatial axis. The condition on axial ratio $b/a$ (ratio between galaxy minor $b$ and  major $a$ axis) reflects a condition on the inclination, defined as the angle between the line of sight and the normal to the plane of the galaxy ($i$ = 0 for face-on galaxies), and it was applied to analyze the kinematic of galaxies with inclination greater than $30\,^{\circ}$, since for smaller inclinations the rotation velocity is highly uncertain (small changes in the inclination will result in large changes in the rotation velocity). 
Both parameters, $PA$ and $b/a$, have been measured from the high spatial resolution \textit{HST}/ACS F814W COSMOS images by  \citet{Sargent2007} using the GIM2D (Galaxy IMage 2D) IRAF software package, which was designed for the quantitative structural analysis of distant galaxies \citep{Simard2002}. These measurements are included in the Zurich Structure and Morphology Catalog\footnote{\href{http://irsa.ipac.caltech.edu/data/COSMOS/tables/morphology/}{http://irsa.ipac.caltech.edu/data/COSMOS/tables/morphology/}}.

In this paper we analyse our data in our highest redshift bin ($\mathrm{0.75<z<1.2}$), where the doublet [O{\small II}]$\lambda \lambda$3726,3729 is expected. 
\subsection{Observations} \label{subsec:Observation}
The observations for our kinematic survey were collected over a series of runs between April 2009 and April 2010 with the $ $VIsible Multi-Object Spectrograph (VIMOS, \citealt{leFevre2003}) mounted on third Unit (Melipal) of the ESO Very Large Telescope (VLT) at the Paranal Observatory in Chile. The targeted area within boundaries of zCOSMOS survey was covered by seven VIMOS pointings, each of them  composed of four quadrants, with $\sim$30-35 slits per quadrant. Three exposures were taken for each  pointing,  for a total exposure time of $\sim$1.5 hours per galaxy. Observations details are listed in Table \ref{tab:observ_log}.

The 2D spectra have been obtained using the high-resolution grism \textit{HR-red} (R=2500) with  the 1$\arcsec$ width slits tilted to follow each galaxy major axis.
The manual slit placement was performed using the ESO software VMMPS  \citep[VIMOS Mask Preparation Software,][]{Bottini2005}, which enabled us to visually check the correctness of the adopted $PA$, and in some cases to decide to adjust the slit tilt to get better kinematic extraction. The wavelength coverage of each observed spectrum is 2400 $\AA$ wide, but its range changes according to the position of the slit on the mask. This allowed us to observe emission lines in a wide wavelength range from 5700$\AA$ to 9300$\AA$. 
When slits were tilted, the slit width was automatically adjusted to keep constant the width along the dispersion direction, and thus a spectral resolution constant regardless of slit $PA$. The instrument provides an image scale of $0. 205\arcsec /pixel$ and a spectral scale of $0.6\,\AA/pixel$. For the sample analyzed in this paper, at $\mathrm{z\sim 0.9}$ the spectral pixel scale is equal, on average, to 25~km/s.
Star spectra have also been obtained, three or four per quadrant, to measure the seeing during each exposure.
In total, 766 spectra of emission line galaxies (and $\sim$90 spectra of stars) have been obtained. In the redshift range ($\mathrm{0.75<z<1.2}$) analyzed in this paper we have 119 galaxies, of which 17$\%$ are part of the targets selected with known  photometric redshift and within the specific range of the GINI parameter (see $\S$ \ref{subsec:the_survey}). Although we did not target active galactic nucleus (AGN) galaxies,  we found two galaxies of our sample as being classified as non-broad-line (NL)  AGNs in the catalog of optical and infrared identifications for X-ray sources detected in the XMM-COSMOS survey by \citet{Brusa2010}.

\subsection{Data reduction} \label{subsec:DataReduction}
We performed spectroscopic reduction using the  VIMOS Interactive Pipeline and Graphical Interface (VIPGI, \citealt{Scodeggio2005}). The software provides powerful data organizing capabilities, a set of data reduction recipes, and dedicated data browsing and plotting tools to check the results of all critical steps and to plot and analyze final extracted 1D and 2D spectra. 
The global data reduction process was rather traditional: (1) average bias frame subtraction, (2) location of the spectral traces on the raw frames and computation of the inverse dispersion solution, (3) extraction of rectified 2D spectra and application of the wavelength calibration using the computed inverse dispersion solution, (4) sky subtraction, (5) combination of the three different exposures taken for each target, (6) extraction of 1D spectra. 
We note that the wavelength calibration task for VIMOS spectra observed in HR mode is somewhat tedious, since the wavelength coverage changes from slit to slit according to where the slit is located on the mask. We have also used the VIMOS ESO pipeline (version 2.9.9), but it does not allow us to combine exposures from different observing nights, therefore we did not proceed further with this tool. Although standard stars have been observed during each observing night, we were not able to perform the spectrum absolute flux calibration. Standard stars spectra were obtained positioning the stars, for each quadrant, at the center of the mask, which did not allow us to produce the calibration for the wide wavelength range of our observations that were obtained thanks to the different positions of the slits on the mask ($\S$ \ref{subsec:Observation}).

The presence of three or four reference star spectra in each quadrant of each pointing allowed the determination of the seeing conditions during spectroscopy acquisitions. This measurement is clearly superior to estimates inferred from the Differential Image Motion Monitor (DIMM) \citep{Bosch2013}, since it not only measures this quantity integrated over the entire exposure time of the spectra, but also accounts for systematic effects resulting from observing and co-adding processes. 
After the reduction (including the combining process of the three different exposures for each target) we collapsed each star spectrum along the spectral direction,  and determined the full width at half maxi\-mum (FWHM), by fitting a Gaussian function to its spatial profile. We therefore computed seeing conditions in each quadrant as the mean of the FWHMs measured, and we used the determined values in our dynamical analysis. The measured values of the seeing range between 0.57$\arcsec$ and 1.24$\arcsec$, with a median value of 0.8$\arcsec$ during all the observations.
 \begin{figure*}[!htb]
   \centering
   \includegraphics[width= 0.45 \hsize]{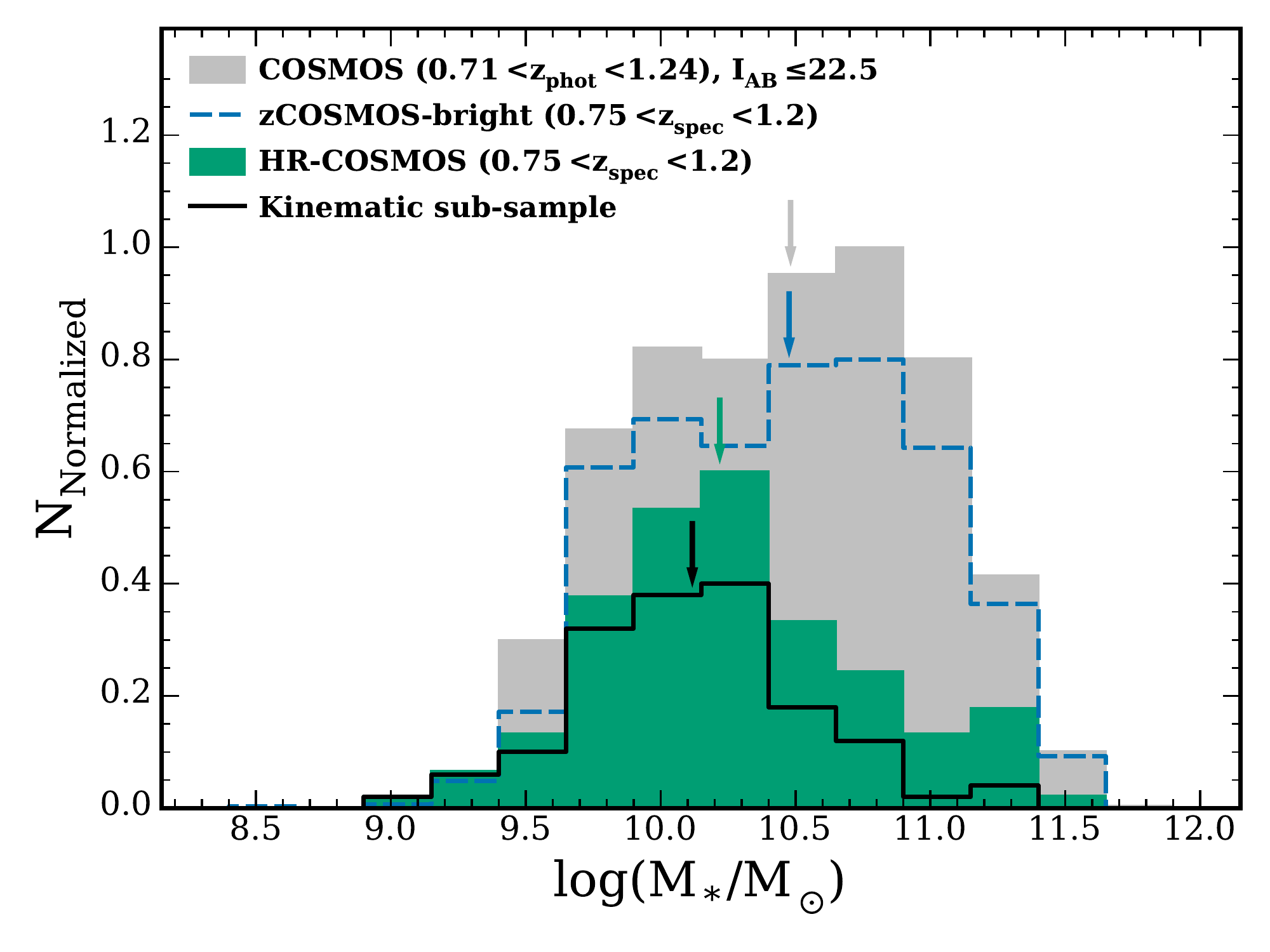}
    \includegraphics[width=0.45 \hsize]{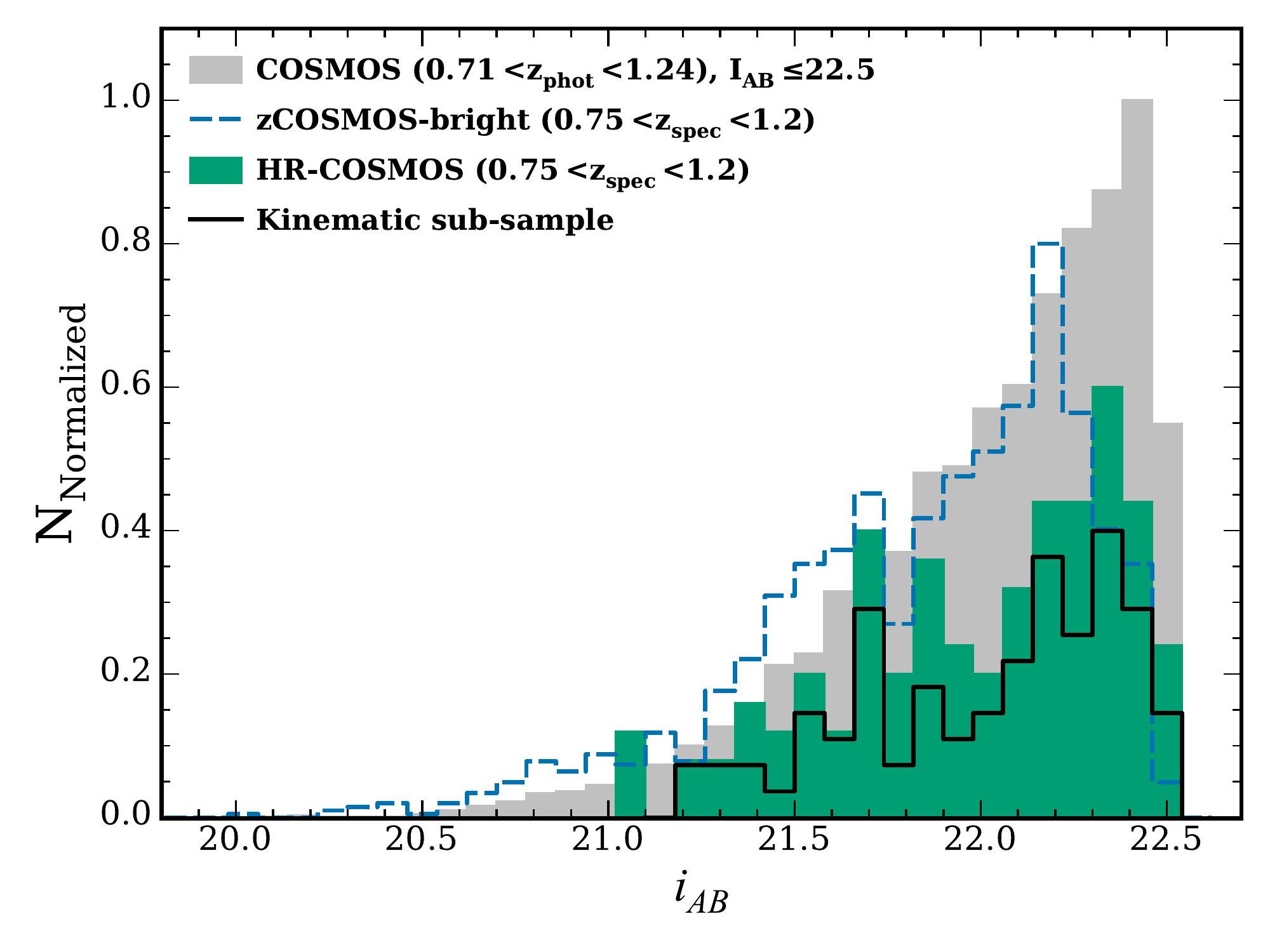}
     \includegraphics[width=0.45 \hsize]{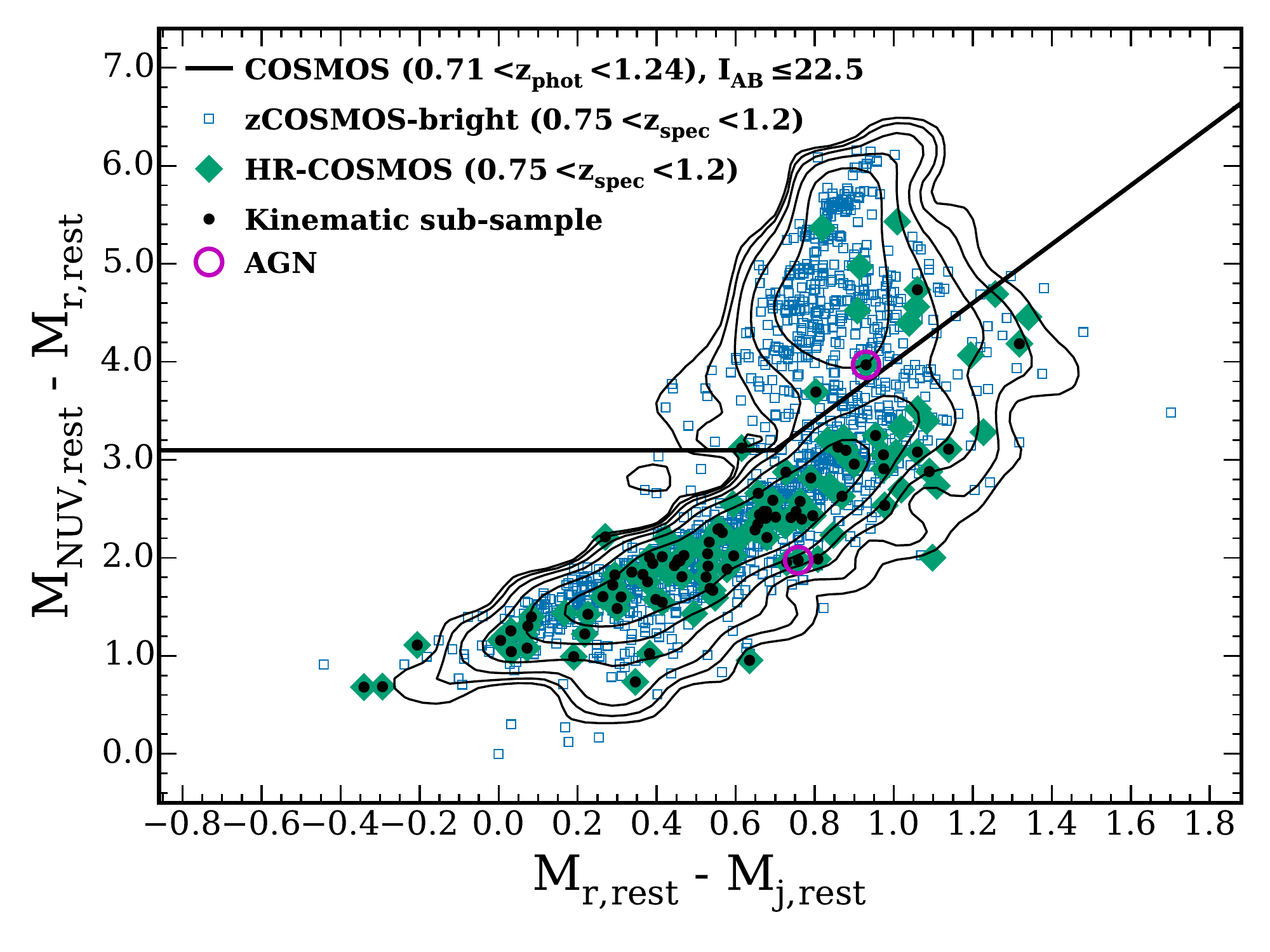}
      \includegraphics[width=0.45 \hsize]{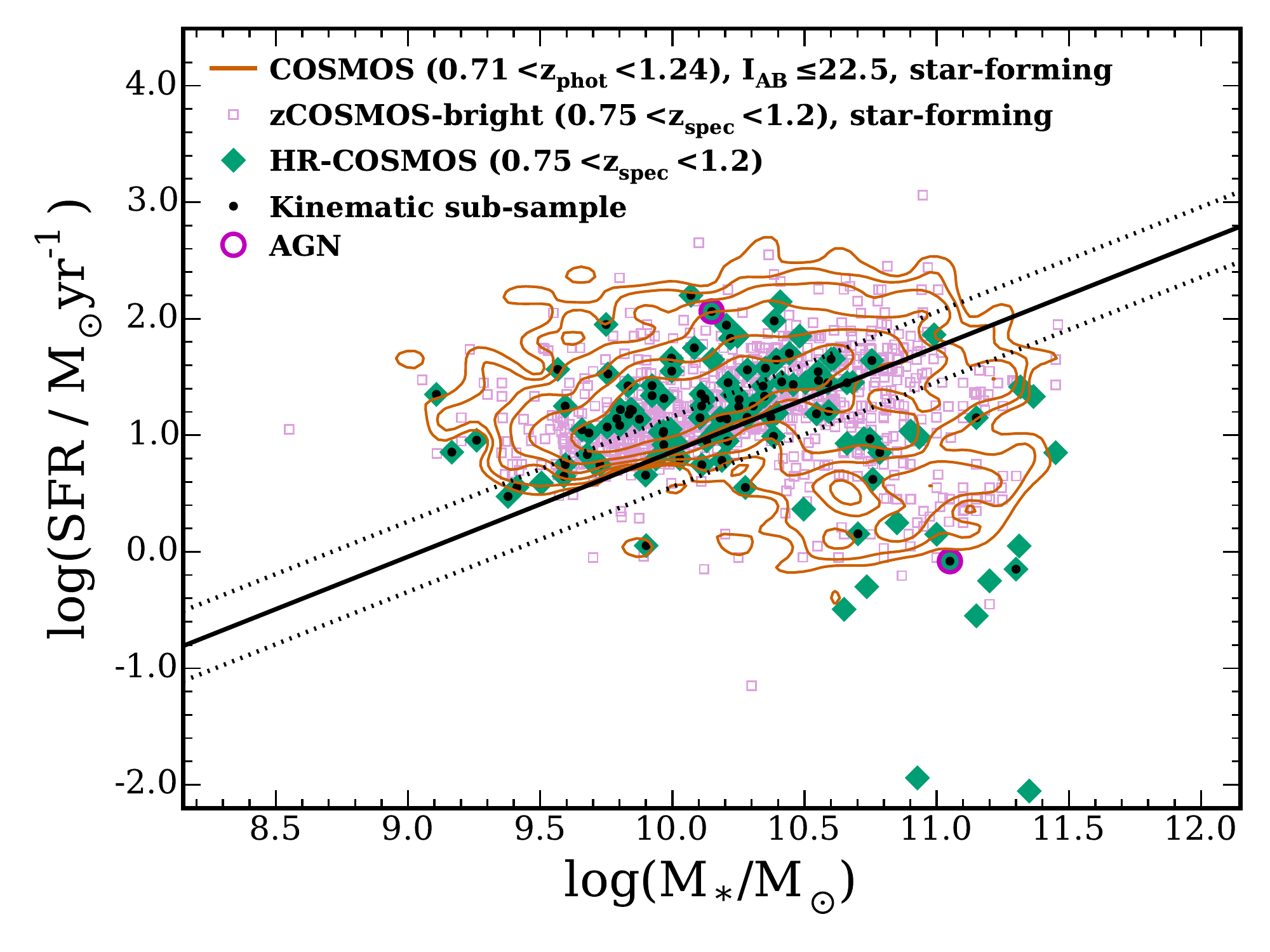}
      \caption{\textit{Top panels:} Stellar mass (\textit{left}) and $I_{AB}$ selection magnitude (\textit{right}) distributions of the 119 galaxies in our sample at $\mathrm{0.75<z<1.2}$ (green histograms). We show for comparison the distributions of the parent samples in the same redshift range (COSMOS in gray and zCOSMOS-bright in blue) and in black the distribution of 82 galaxies in our kinematic sub-sample (see $\S$ \ref{subsec:kin_params}). The photometric redshift range for the COSMOS sample takes into account the precision of the measurements (see text $\S$ \ref{subsec:stellarmass}). The arrows in the left panel show the median of each distribution. The histograms are re-normalized for a better visual comparison of their spreads and peaks, therefore their normalization does not reflect the actual scale.  \textit{Bottom left panel:} Rest-frame $\mathrm{M_{NUV}- M_r}$ versus $\mathrm{M_r - M_J}$ color-color diagram. Colors are the same as in the top panels. The contours show the distribution of the COSMOS parent sample with relative number of galaxies equal to 50\%, 25\%, 13\%, 6\%, 3\%, 1.5\% of the maximum of the distribution. The back solid line divides quiescent galaxies (above the line) from star-forming galaxies (below the line) and is determined following the technique adopted by \citet{Ilbert2013}. \textit{Bottom right panel:} Relationship between SED-derived star formation rate (SFR) and stellar mass. Green diamonds and black points represent the galaxies at $\mathrm{0.75<z<1.2}$  in our full sample and in our kinematic sub-sample, respectively. The dark-orange contours and the plum open squares show the distribution for the star-forming galaxies \citep[as defined by][]{Laigle2016} in COSMOS and zCOSMOS-bright samples, respectively. The contours represent the relative number of galaxies at the 50\%, 25\%, 13\%, 6\%, 3\%, 1.5\% of the maximum of the distribution. The solid and dotted lines represent the correlation, with the uncertainties, between the SFR and stellar mass of blue star-forming galaxies at $\mathrm{0.8<z<1.2}$ by \citet{Elbaz2007}. The magenta circles show the AGNs.  }
         \label{fig:Mstar}
   \end{figure*}
The nominal spectral resolution (3$\AA$, for the grism central wavelength $\lambda_c$=7400$\AA$) was checked along the observed spectral range. To that end, we measured the FWHM of four skylines (7275.7$\AA$, 7750.6$\AA$, 7841.0$\AA$, 7913.2$\AA$) in each slit by fitting a Gaussian function to every single skyline spectral profile. The distribution of the spectral FWHMs ranges between 2.6$\AA$ to 3.2$\AA$. Since we used  the nominal spectral resolution for our models,  the measured variation has been taken into account in the error computation of the modeled kinematic parameters (see $\S$ \ref{subsec:uncetainties}). Although we were provided with the redshift measurement from zCOSMOS, our kinematic analysis requires high precision in  measuring the centroid of the emission line, and given the higher resolution of our spectra compared to the ones from zCOSMOS,  we re-measured the galaxy redshifts from our 1D reduced spectra.
 
To facilitate the comparison of the models with the observations we chose to extract $\sim40\AA$ wavelength cutouts around key emission lines of interest (mostly [O{\small II}] doublet and a few H$\beta$ in our sub-sample at $\mathrm{z\sim 0.75-1.2}$)  using the HR redshift measurement, and we subtracted the stellar continuum spectrum, since we did not include it in the kinematic models (a cutout with continuum information was also saved for each observation). To perform the continuum subtraction we selected two background regions at the left and right of the emission line (avoiding reduction artifacts and other spectral features), we stacked them, and we computed the median along the wavelength direction, obtaining a spatial profile of the continuum that was subtracted for the entire cutout.   During this process we visually inspected all the emission lines of the 119 galaxies in our sub-sample and we decided to discard 28 for which it would have been impossible to retrieve the kinematic information, either because the emission line was too faint, or not detected, or it was shifted at the same wavelength of skylines (even if we selected galaxies with emission lines away from skylines, this could be possible for the targets selected with known photometric redshift for which the measurements uncertainties are larger than for the spectroscopic redshift). 

Although the slits were always carefully  placed along the major axis of the galaxies ($\S$ \ref{subsec:Observation}), we visually double-checked their correctness and we found that for one galaxy the slit was not well placed.
The inclination, adopted for both the sample selection and the kinematic analysis, was also visually inspected by overlapping ellipses with the axial ratio $b/a$ from the Zurich Structure and Morphology Catalog on the \textit{HST}/ACS F814W images of all the galaxies in our sub-sample (see $\S$\ref{appendix_incl}). During this visual inspection, we also found that one galaxy was obviously face-on, despite the fact that we selected galaxies with inclination $\gtrsim$ 30$^{\circ}$ (see $\S$ \ref{subsec:the_survey}). Thus, these two galaxies (one with the slit misplaced and one face-one) were not included in our kinematic analysis.  This reduced our sample to 89 galaxies, which have been modeled following the method described in $\S$~\ref{sec:kinematic_model}. 

\subsection{Stellar mass measurements} \label{subsec:stellarmass}
The stellar masses were estimated using the latest COSMOS photometric catalog \citep{Laigle2016}, which makes use of deep \textit{Spitzer} SPASH IRAC imaging \citep{Steinhardt2014} and the latest release of COSMOS UltraVISTA survey \citep{McCracken2012}. These measurements are extremely accurate since they are based on deep thirty-band UV–IR photometry that covers all galaxy spectral types, and they were obtained using our high-resolution spectroscopic redshift measurements.
The measurement technique follows the same recipes presented in \citet{Ilbert2015}, based on the Le Phare software \citep{Arnouts2002, Ilbert2006}.
Briefly, the galaxy stellar masses have been derived  using a library of synthetic spectra generated with the stellar population synthesis (SPS) package developed by \citet{Bruzual2003}. We assume a universal IMF from \citet{Chabrier2003},  as well as exponentially declining and delayed star formation histories. For all these templates, two metallicities (solar and half-solar) are considered. Emission lines are added following \citet{Ilbert2009}, and  two attenuation curves are included: the starburst curve of \citet{Calzetti2000} and a curve with a slope 0.9 \citep[Appendix A of][] {Arnouts2013}. The stellar continuum extinction $\mathrm{E_s(B-V)}$ is allowed to take values in the range [0-0.7]. The values of the stellar masses were assigned using the median of the stellar mass probability distribution marginalized over all other parameters.

In Figure \ref{fig:Mstar}, we show the distribution of various properties of the HR-COSMOS sample of 119 galaxies in the redshift range $\mathrm{0.75<z<1.2}$. These are contrasted with the parent zCOSMOS-bright and COSMOS samples. The values for both parent samples are drawn from the latest COSMOS photometric and SED-fitting catalogs \citep{Laigle2016}. For the former parent sample, we compare only those galaxies in the same spectroscopic redshift range as the HR-COSMOS sample presented here. For the latter, the redshift range is extended slightly to $\mathrm{0.75 - 3\sigma_{\Delta z/(1+z_s)} <z< 1.2 + 3\sigma_{\Delta z/(1+z_s)}}$ to account for the uncertainties in the photometric redshifts, where $\mathrm{\sigma_{\Delta z/ (1+z_s)}} = 0.007$.

In the top panels, we present the stellar mass (left) and $I_{AB}$ selection magnitude (right) distributions for our sample of 119 galaxies at $\mathrm{0.75<z<1.2}$ with green histograms and the distributions of the parent samples in gray (COSMOS) and blue (zCOSMOS-bright). Galaxies in our sample have stellar masses spanning from $1.3\times 10^9 M_\odot$ to $2.0\times 10^{11} M_\odot$ and values of $I_{AB}$  spanning from 21.0 mag to 22.5 mag.
The black histograms show the distributions of the galaxies used in the kinematic analysis (see $\S$ \ref{sec:results}). Since we are not interested in comparing the normalization of the distributions, we re-normalized the histograms  to better visually compare their spreads and peaks, therefore their normalization does not reflect the actual scale. We measure that the median of the parent samples distributions is equal to $3.0\times 10^{10} M_\odot$, while the median for the distribution of our samples shifts to lower values ($1.7\times 10^{10} M_\odot$ for the whole sample of 119 galaxies and $1.3\times 10^{10} M_\odot$ for the kinematic sub-sample). This may be a result of the fact that our sample is composed primarily of star-forming galaxies and, therefore, the high mass end of the parent samples distributions (probably populated by massive quiescent galaxies) is suppressed in our sample.  

In the bottom panels of Figure, \ref{fig:Mstar} we plot the rest-frame $\mathrm{M_{NUV}- M_r}$ versus $\mathrm{M_r - M_J}$ color-color diagram (left) and the relation between star formation rate (SFR) and stellar mass (right) for our sample compared to the parent samples. The color-color diagram is a diagnostic plot that enables us to separate star-forming from quiescent galaxies. We divide quiescent galaxies from star-forming galaxies following the technique adopted by \citet{Ilbert2013}. In the bottom left panel, we show how our sample follows the underling distribution of star-forming galaxies closely. 
This is supported by the SFR-M$_{\ast}$ relation displayed in the bottom right panel for the star-forming galaxies \citep[as defined by][]{Laigle2016} in COSMOS and zCOSMOS-bright samples. The SFR computed from SED fitting is known to be not as precise as that computed using other SFR estimators \citep{Ilbert2015, Lee2015}. Following \citet{Ilbert2015} we compared the 
SED-based SFR to the 24$\mu m$ IR SFR for the parent COSMOS sample in the redshift range $\mathrm{0.75<z<1.2}$, and found an offset of 0.15~dex towards larger SED-based SFR. To take into account this discrepancy, we applied a shift of $-0.15$~dex to the SFR measurements.
Our sample follows the star-forming galaxy distribution of the parent samples.

\section{Kinematic modeling} \label{sec:kinematic_model}
To study the galaxy kinematics, we created  high-resolution semi-analytic models.  The advantage of using a model to constrain the kinematics is that we can compare it  directly to the observations, after taking into account all the degradation effects owing to the instrumental resolution, and avoid the intermediate steps of data analysis that can introduce additional noise \citep{Fraternali2006}.  
\subsection{The model} \label{subsec:model}
Our first step for modeling the galaxy kinematics was to describe an astrophysically-motivated picture of a galaxy at high-resolution. Ionized gas was described as rotating in a thin disc with a certain inclination \textit{i} and position angle \textit{PA} of the major axis.
A pseudo-observation was created combining high-resolution  models of intrinsic flux distribution of the emission line, rotation velocity and dispersion velocity using the following relation:
\begin{equation} 
\\ \\   I(r,V)=\dfrac{\Sigma(r)}{\sqrt{2\pi}\sigma(r)} \exp \left\{- \dfrac{[V - V(r)]^2}{2\sigma^2(r)}\right\} .
\end{equation}
This models the galaxy emission at each radius r and  reconstructs  a 2D emission line to be compared to the observation. If the observed emission was a doublet (as in the case of [O{\small II}], the main investigated feature in this study) another contribution was added to the relation, having the same functional form and velocity dispersion $\sigma$ as the previous one, but shifted in velocity by a fixed value, that was equal to the separation between the two lines of the doublet, and scaled in intensity by a ratio  $\mathrm{R_{[O{\small II}]}}$, an additional parameter of the model.
 The intrinsic line-flux distribution in the plane of the disc was described by a truncated exponential disc function:  
\begin{equation} \label{eq:surface_bright}
\\ \\  \qquad \Sigma(r)=\left\{\begin{array}{ll}
\Sigma_{0}\: e^{-r/r_0} & \text{if }  r <  r_{trunc} \\[6pt]
0 & \text{if }  r > r_{trunc} \end{array} , \right.
\end{equation}
where $\Sigma_{0}$ is the line flux at the center of the galaxy, $r_0$ is the scale radius and $r_{trunc}$ is the maximum radius, after which the emission is no longer detected. The truncation was necessary since we allowed $r_0$ to assume negative values to better match the observed line flux distribution, in particular when the distribution is characterized by bright clumps in the peripheries of the galaxies.

The velocity along the line of sight, assuming that for spiral galaxies the expansion and the vertical motions are negligible with respect to the rotation, was described  as
\begin{equation} \label{eq:velocity_short}
\\ \\ \\ V(r)= V_{sys}+V_{rot}(r) \sin i \cos \theta ,
\end{equation}
where $V_{sys} $ is the systemic velocity, that is the velocity (corresponding to the systemic redshift) of the entire galaxy with respect to our reference system. For very distant galaxies this quantity is usually dominated by the Hubble flow. Since we are interested only in the galaxy internal kinematics we set it to zero. The  galaxy inclination $i$ was measured as the angle between the line of sight and the normal to the plane of the galaxy ($i=0\,^{\circ}$ for face-on galaxy) and $\theta $ is the angle in the plane of the galaxy between the galaxy PA and the position where the velocity is measured. Since our model reproduces the emission coming from the slit aligned to the galaxy PA, we assume $\theta$ equal zero.

To model the rotation velocity $V_{rot}$, which results from both baryonic (stars and gas) and dark matter potentials, three different functions have been used. All of them are described by two parameters: the maximum velocity $V_t$ and the transition radius $r_t$. 
A velocity profile used is the \textit{Freeman disc} \citep{Freeman1970}, which fits a galaxy with a gravitational potential generated by an exponential disc mass distribution.
It is expressed as 
\begin{equation}
\qquad \qquad \quad \quad V_{rot}(r)=\frac{r}{h} \sqrt{\pi G \mu_0 h\, (I_0 K_0 - I_1 K_1)} \, , \label{eq:exp}
\end{equation}
where $\mu_{0}$ is the central surface density and $h$ is the disc scale-length of the surface density distribution. We note that $\mu_{0}$ and $h$ are not constrained by $\Sigma_0$ and $r_0$ of Equation \ref{eq:surface_bright} since the rotation velocity model does not only describe baryons. $I_i$ and $K_i$ are the i-order modified Bessel function evaluated at $0.5\,r/h$. 
The maximum velocity 
\begin{equation}  \label{eq:vmax_exp}
\\ \qquad \qquad \qquad V_t\sim 0.88\sqrt{\pi G \mu_0 h} \, ,
\end{equation}
is reached at the transition radius $r_t\sim2.15h$. Substituting the Equation \ref{eq:vmax_exp} and the transition radius $r_t$ in the Equation \ref{eq:exp}, we obtain an expression of the rotation velocity $V_{rot}$ described by the only two parameters  $V_t$ and $r_t$.
Unlike the Freeman disc, the two other rotation curve models are analytic functions that have no physical derivation nor assumed mass distribution, but they appear to describe well the observed rotation curves of local galaxies.
 The \textit{flat model} consists of a two slopes model, which describes a rotation velocity distribution that has a sharp transition at $r_t$ and flats at $V_t$. It takes the form:
 \begin{equation} \label{eq:flat}
\\ \\ \quad	V_{rot}(r)=    \left\{\begin{array}{ll}
	V_t \times \displaystyle{ \frac{r}{r_t}} & \text{if } r < r_t\\[6pt]
V_t  & \text{if } r \geq r_t \end{array} \right. .
\end{equation}
Last velocity profile adopted is  an \textit{arctangent} function \citep{Courteau1997}, expressed as
 \begin{equation} \label{eq:arctan}
 \\ \\ \quad V_{rot}(r)= V_t \frac{2}{\pi} \arctan \frac{2r}{r_t} \, ,
 \end{equation}
which smoothly rises and reaches a maximum $V_t$ asymptotically at an infinite radius. The transition radius $r_t$ is defined as the radius for which the velocity is 70$\%$ of  $V_t$.
The galaxy velocity dispersion $\sigma$ has been modeled to be constant at each radius.
This assumption is considered reasonable since it is based on what has been observed in the local Universe in the galaxies from the GHASP sample \citep{Epinat2008}. It has also been shown by some authors \citep{Weiner2006_1, Epinat2008} that the peak observed in the velocity dispersion profile for galaxies at high redshift is due to the blurring effect of the resolution. 
The high-resolution model sampling (both spectral and spatial) was chosen to be a sub-multiple of the VIMOS data sampling. We  therefore created high-resolution models with spectral and spatial sampling equal to 0.15$\AA$ and  0.05125$\arcsec$, respectively.
To make the high-resolution models comparable with the observations, we performed a spatial and spectral smoothing by convolving the models with the spatial resolution given by the seeing measured for each quadrant, and the spectral resolution of the instrument (see $\S$ \ref{subsec:DataReduction}). The effect due to the slit width was also taken into account by convolving the models with a window function.  Finally, the models were re-binned to match the VIMOS sampling ($1 pixel = 0.205\arcsec= 0.6 \AA$).

\subsection{The fitting method} \label{subsec:fitting}
The comparison with the observation has been done through the $\chi^2$-minimization fitting using the Python routine MPFIT  \citep{Markwardt2009}, based on the Levenberg–Marquardt least-squares technique.
Preliminary steps were necessary to define a set of initial parameters.

The rotation center, $r_{cen}$ was measured as the center of the continuum spectrum's spatial profile, or, when the continuum emission was too faint, as the center of the emission line's spatial distribution around its observed central wavelength. This last quantity, $\lambda_{line}(\mathrm{z})$, was a parameter of the model that was kept fixed during the fitting process. Using our measured redshift ($\S$ \ref{subsec:DataReduction}) we derived the systemic velocity $V_{sys}$, which quantifies the motion of the entire galaxy with respect to our reference system. However, for galaxies at high redshift $V_{sys}$ is dominated by the Hubble flow, therefore it is normal practice to set it to zero, to focus only on the internal kinematics. The systemic velocity is allowed to freely vary in the fitting to take into account the effect of the redshift measurement uncertainties.
The inclination \textit{i} is derived from the axial ratio, computed from \textit{HST} images, found in the literature (Zurich Structure and Morphology catalog), and corrected for projection effects using Holmberg’s oblate spheroid description \citep{Holmberg}: 
\begin{equation}
\\ \\  \quad i = \arccos \sqrt{\dfrac{(b/a)^2 - q_0^2}{1-q_0^2}} \, ,
\end{equation}
where $b/a$ is the observed axis ratio, $q_0 = c/a$ is the axial ratio of a galaxy viewed edge-on, $a$ and $b$ are the disc major and minor axes, and $c$ is the polar axis. The value of $q_0$ is known to vary from 0.1 to 0.2 depending
on galaxy type \citep{Haynes1984}. In this study we assumed $q_0 = 0.11$, driven by the necessity to take into account the smallest value of $b/a$ in our sample, that is $0.12$, knowing form \citet{Pizagno2005} that the $q_0$  range $0.1 - 0.2$ corresponds to a small variation (typically $\sim1\, km\, s^{-1}$ ) in the velocity. Owing to the degeneracy between velocity and inclination \citep{Begeman1987}  we decide to keep $i$ fixed during the fit. The position angle $PA$ is  taken from the literature and it is the same used to align the slit along the major axis of the galaxies during the observations (see $\S$\ref{subsec:Observation}). The $PA$ is  fixed in the $\chi^2$ minimization fitting process. 
Although photometric-kinematic $PA$ misalignment may exist, it has been demonstrated that $PA$ measurements from high-resolution F814W images are generally in reasonable  agreement with the kinematic $PA$. \citet{Epinat2009}, for a sample of nine galaxies between $\mathrm{1.2 < z < 1.6}$ observed with SINFONI, found that the agreement between morphological and kinematic $PA$ is better than 25$\degr$, except for galaxies that have a morphology compatible with a galaxy seen face-on. A similar result was found by \citet{Wisnioski2015}, who studied the galaxy kinematics of a sample of galaxies at $\mathrm{0.7\leq z\leq 2.7}$ with KMOS, and have found that the agreement between photometric and kinematic $PA$ is better than 15$\degr$ for 60\% of the galaxies, and for 80\% of the galaxies the agreement is better than 30$\degr$, while misalignments greater than 30$\degr$ typically have axis ratios $b/a$ > 0.6. In our sample only 23 galaxies (out of 82) have $b/a$ > 0.6.
\begin{figure*}[!htb]
   \centering
   \includegraphics[width=0.33\textwidth]{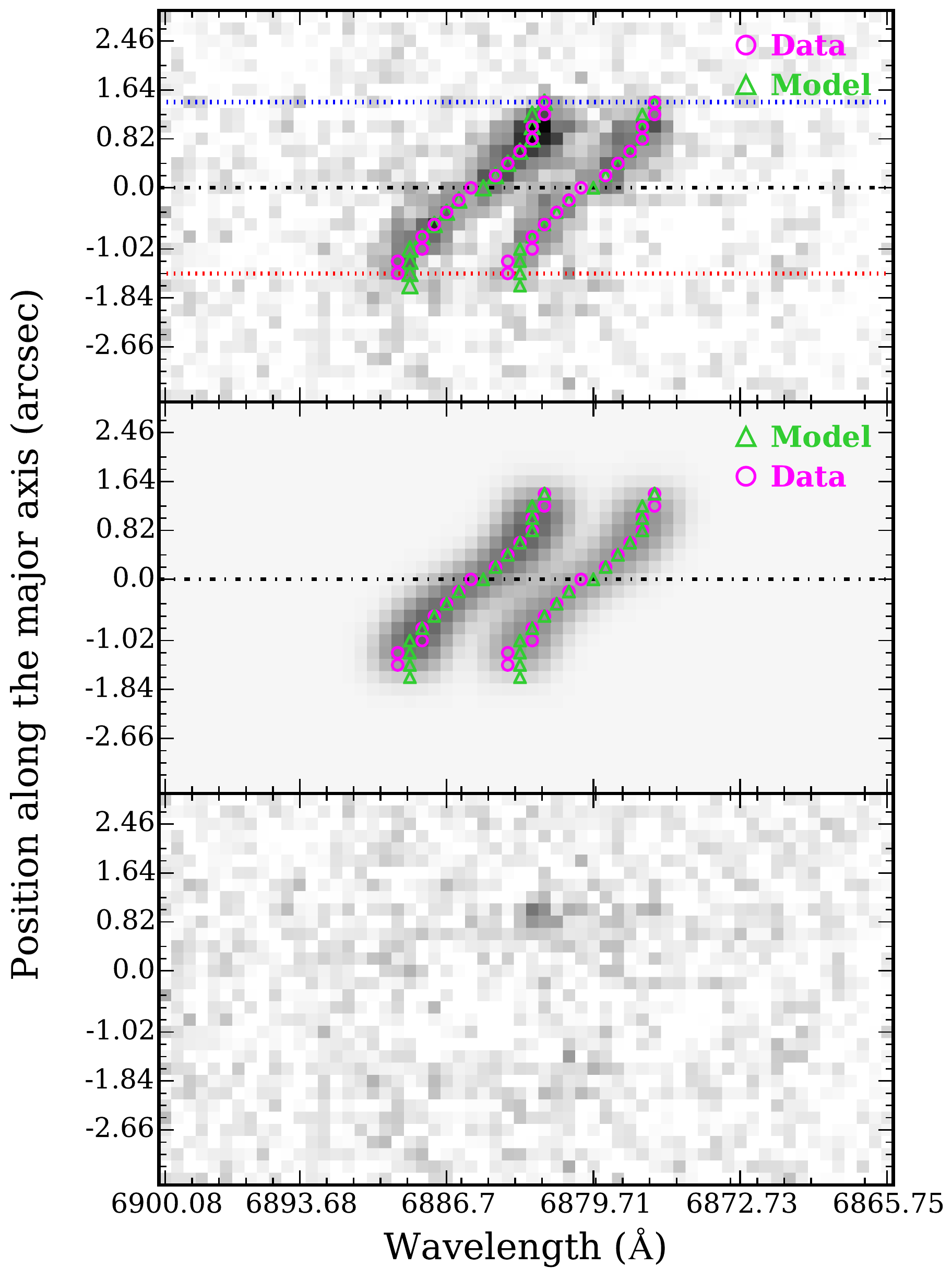}
    \includegraphics[width=0.29\textwidth]{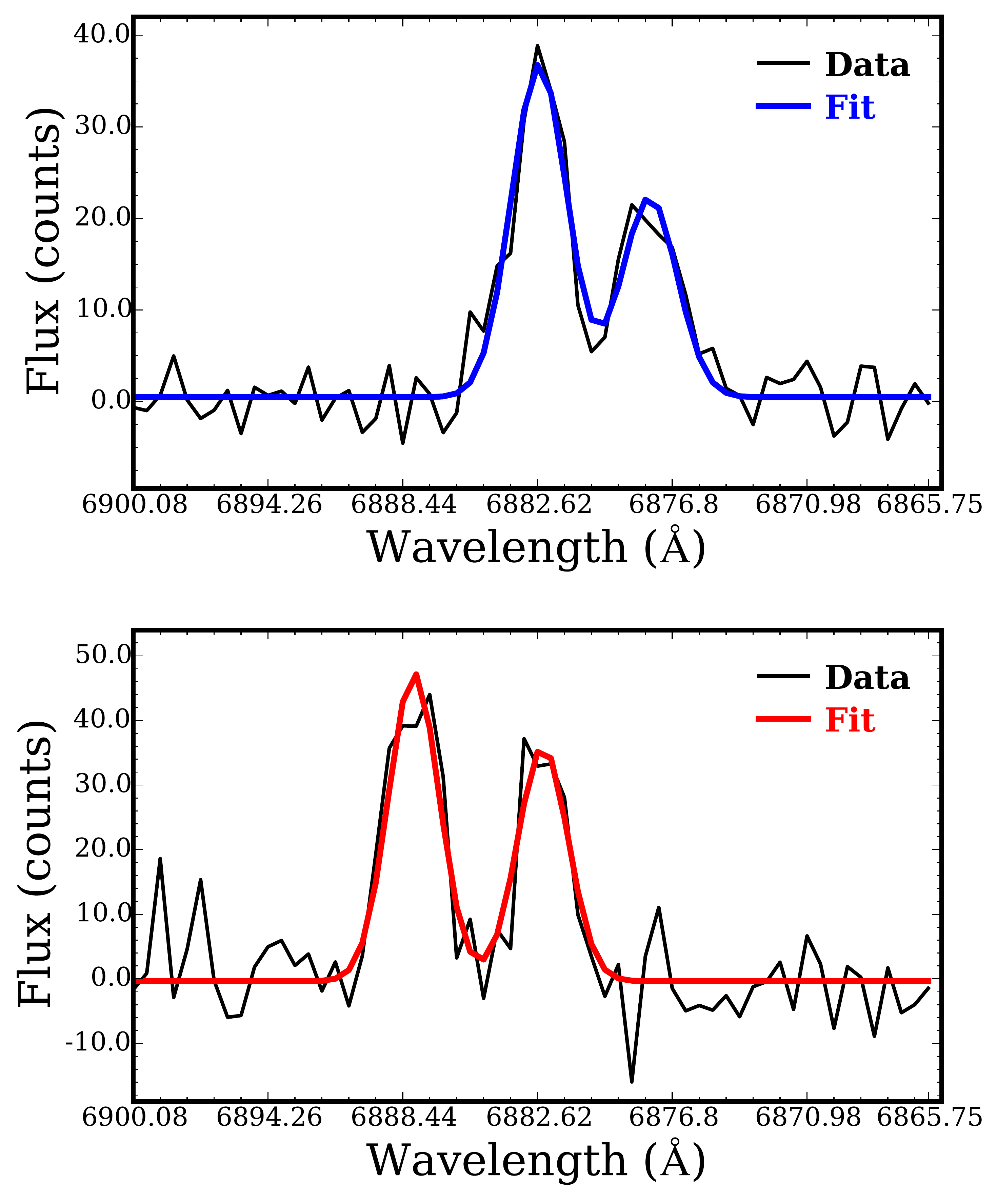}
   \includegraphics[width=0.36\textwidth]{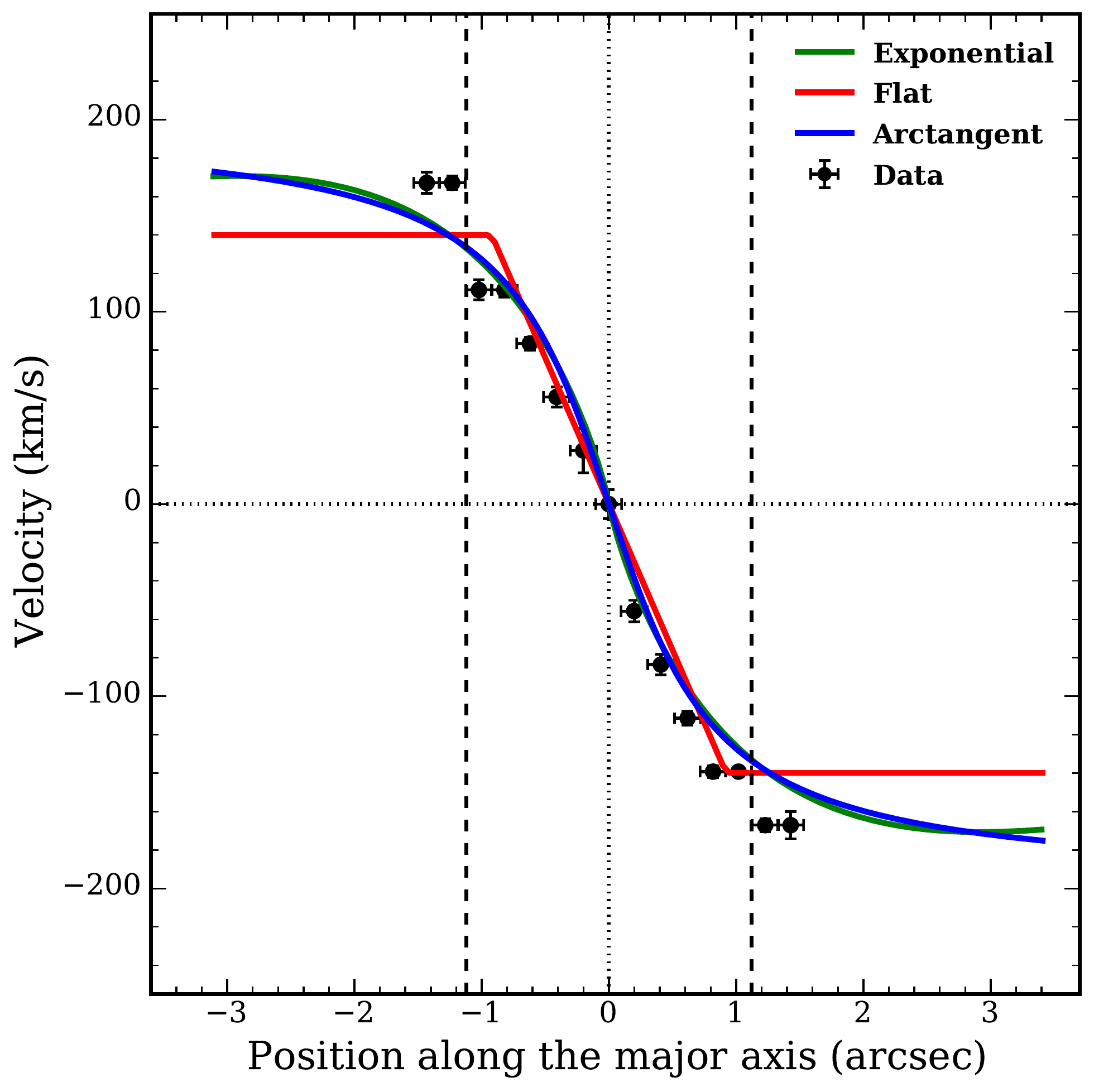} 
      \caption{Example of the output from the model-fitting process and the vision inspections to test the quality of the fit. \textit{Left panels:} From top to bottom: continuum-subtracted 2D spectrum centered at [O{\small II}], best-fit kinematic model to the line emission, and residual image between the 2D spectrum and the best-fit model on the same intensity scale as the 2D spectrum. The vertical and horizontal axes are the spatial position and wavelength, respectively. In the first two panels, the black dot-dashed lines indicate the best-fit parameter~$r_{cen}$. In the top panel, the blue and red dotted lines denote the positions from where the plots in the middle panels are extracted.  The 2D~spectrum from the VIMOS mask has been rotated to have always the North direction in the top. For this reason the wavelength direction is opposite to the conventional representation with values increasing (getting redder) towards left.
The technique of tracing the emission line as a function of the position (see text $\S$ \ref{subsec:fitting}) is also shown in the top and middle pannels. Fuchsia circles show the central wavelength of the emission line at each position, and the green triangles show the centroids of the best-fit model. 
\textit{Middle panels:} Examples of the double Gaussian fit to measure the centroids of the the emission line at two most external galaxy radii.
 \textit{Right panel:} High-resolution rotation curve models (green line: exponential disc; red line: flat model; blue line: arctangent function), not corrected for the inclination, compared to the observed rotation curve (black circles). The errorbars on the y-axis direction represent the uncertainties of the fit to the line at each position to recover the centroid;  the bars on the position direction indicate the bin size. The vertical dotted line indicates the galaxy rotation center, the horizontal dotted line indicates the systemic velocity, and the two vertical dashed lines indicates the position of the radius R$_{2.2}$ at which the velocity is measured to derive the scaling relations ($\S$ \ref{subsec:kin_params}).
     }
              \label{fig:model}
\end{figure*}

The parameters that describe the rotation velocity  $V_t$ and $r_t$, the velocity dispersion  $\sigma$, the doublet line flux ratio $\mathrm{R_{[O{\small II}]}}$, and the exponential disc truncation radius $r_{trunc}$  have been guessed tracing the continuum-subtracted emission line as a function of the spatial position. 
This procedure consists of fitting a Gaussian (double Gaussian with a rest-frame separation equal to 2.8 $\AA$ between the lines of the doublet) to the line profile in the wavelength direction, and it returns the values of the central wavelength, the dispersion (both converted in velocities), and flux ratio (in case of doublet)  in each spatial bin. 
The tracing procedure terminates when the emission is no longer detected above the local noise level (S/N < 3), and this gives us a guess on the parameter $r_{trunc}$. Each trace has been visually inspected to ensure that the fit was not influenced by spurious reduction artifacts. An example of this procedure is shown in Fig. \ref{fig:model} (middle panel).
We defined the initial rotation velocity parameter $V_t$ as the average value between the maximum velocities (corrected for the inclination) reached in both \textit{approaching} and \textit{receding} side of the galaxy and the transition radius $r_t$ as the average between the radii at which the two maximum velocities have been first reached. For the $\sigma$, since we decided to use a constant value, this was set as the mean over the traced dispersion, after correcting it for the instrumental dispersion following  \citet{Weiner2006_1}. The line flux ratio $\mathrm{R_{[O{\small II}]}}$ has been defined as the mean over all the line flux ratio values measured along every spatial bin.

To recover the initial parameters that describe the surface brightness, we collapsed the emission line along the spectral direction (after subtracting the continuum), obtaining a spatial profile of the emission that has been fitted with a truncated exponential disc (a sum of two truncated exponential discs in case of a doublet)  convolved with the spatial resolution. As a result, we get a first guess of the parameters $\Sigma_0$ and $h$. 
Except for the $i$, $PA$ and $\lambda_{line}(\mathrm{z})$, the other parameters are let to vary freely in the fitting process for a total of eight (nine, in the case of doublet) free parameters ($\Sigma_0$, $h$, $r_{trunc}$, $r_{cen}$,  $V_t$, $r_t$, $V_{sys}$, $\sigma$, $\mathrm{R_{[O{\small II}]}}$).
\begin{figure*}[!htb]
   \centering
   \includegraphics[width=0.14\hsize]{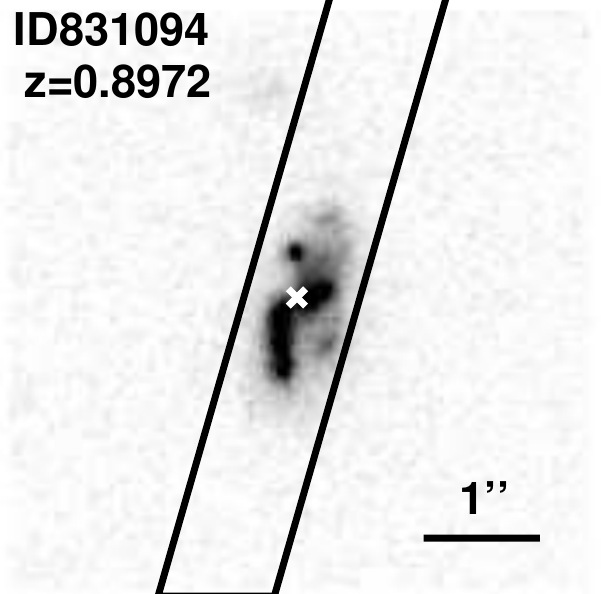}
   \includegraphics[width=0.68\hsize]{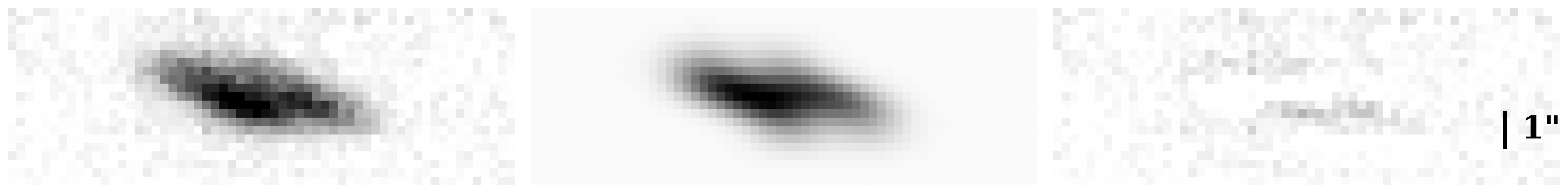}\\
   \medskip
    \includegraphics[width=0.14\hsize]{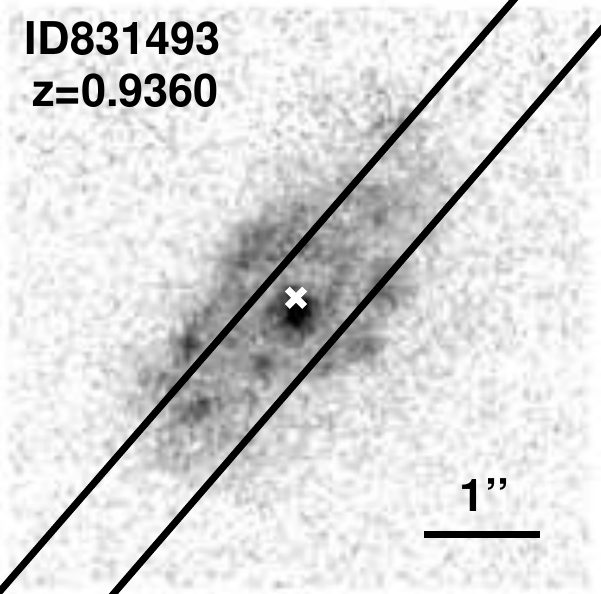}
   \includegraphics[width=0.68\hsize]{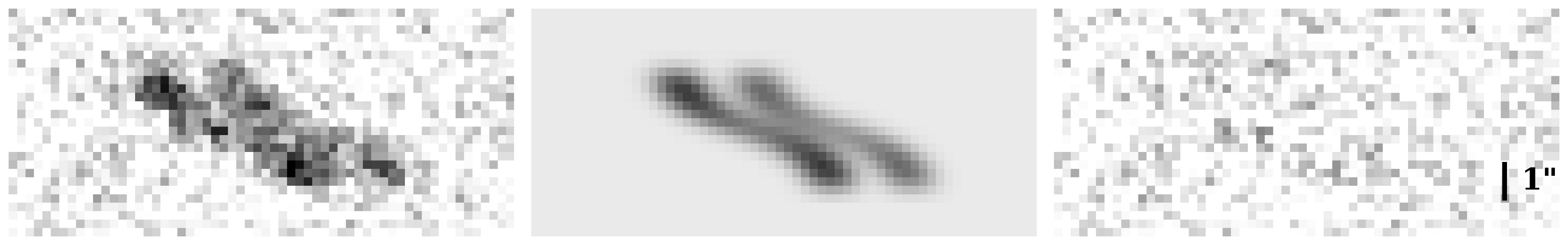}\\
   \medskip
    \includegraphics[width=0.14\hsize]{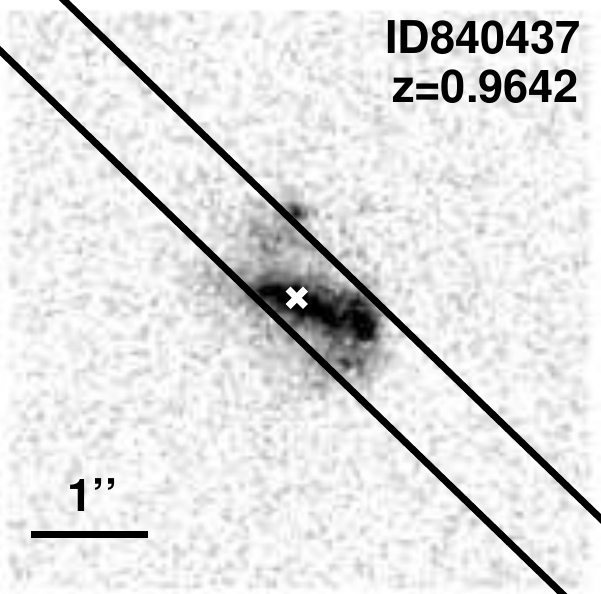}
   \includegraphics[width=0.68\hsize]{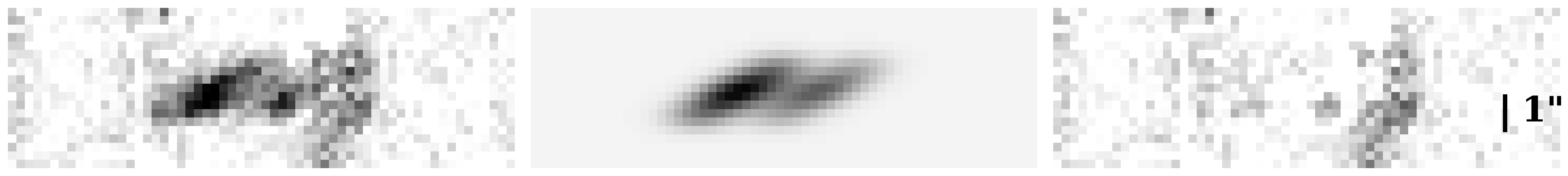}\\
   \medskip
    \includegraphics[width=0.14\hsize]{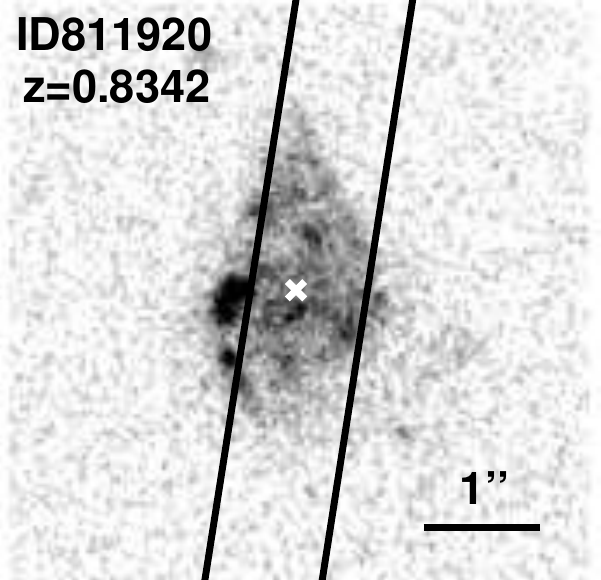}
   \includegraphics[width=0.68\hsize]{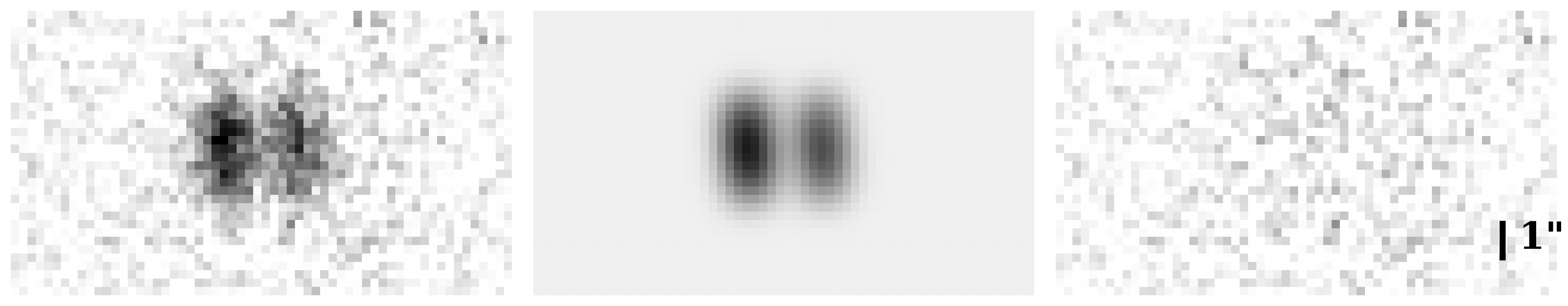}\\
   \medskip
    \includegraphics[width=0.14\hsize]{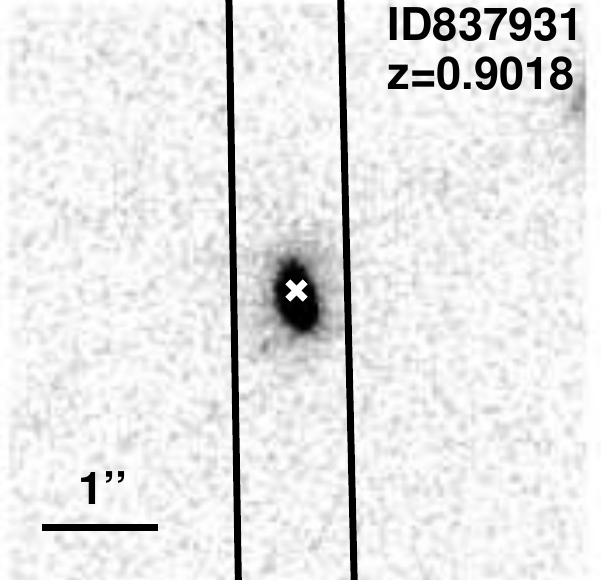}
   \includegraphics[width=0.68\hsize]{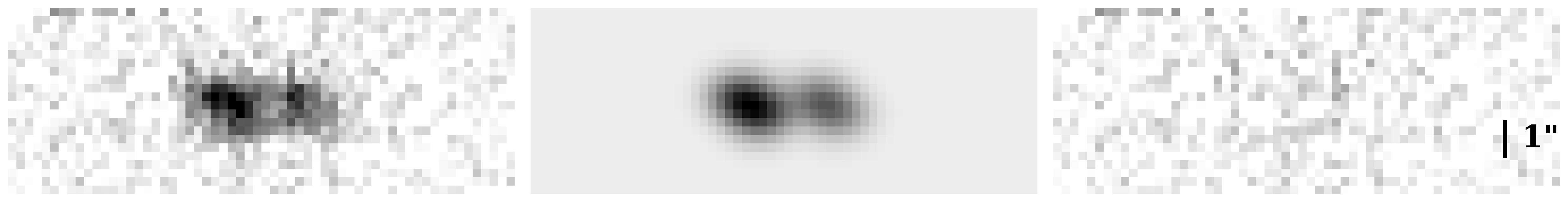}
      \caption{Examples of galaxies from our sample. For each galaxy we show the F814W \textit{HST}/ACS postage stamp with superimposed the 1$\arcsec$ width slit tilted to follow the galaxy major axis, the continuum-subtracted 2D spectrum, the best-fit kinematic model to the line emission and the residuals between the 2D spectrum and the best-fit model, on the same intensity scale as the 2D spectrum. In the spectra, the vertical and horizontal axes are the spatial position and wavelength, respectively. We show, both for the postage stamp and for the spectrum, the 1$\arcsec$ scale. The first 3 galaxies are classified as rotation-dominated, whereas the other 2 galaxies are dispersion-dominated.
       }
         \label{fig:model_data_rot}
   \end{figure*}

\subsection{Uncertainty estimation of the parameters} \label{subsec:uncetainties}
The fits are done by minimizing the weighted difference between the image of the 2D line emission and the 2D model of the same line. Each pixel is weighted by its inverse variance, which is computed by summing in quadrature the contribution from both the source and continuum-subtracted background. The first comes from the Poisson noise of the line-flux counts, and the second one is estimated directly from the RMS fluctuation in regions where there is no object.
These weights in the fit are translated into uncertainties on the parameters, computed from the covariance matrix.
To the 1-$\sigma$ formal error from the fits, we added  in quadrature to error on the parameter $V_t$ the uncertainties coming from the inclination (propagated from the uncertainties on $b/a$) and from the $PA$, since those parameters are kept fixed in the fitting process. The adopted $b/a$, from the Zurich Structure and Morphology Catalog, was measured with a software (GIM2D, see $\S$\ref{subsec:the_survey}), which gives morphological measurements that have been corrected for the instrumental point-spread function (PSF).  In $\S$\ref{appendix_incl}, we computed systematic uncertainties on the measurements of the axial ratio $b/a$ derived by small variation of the PSF and added them to the velocity error budget.   We measured the instrumental resolution from the skylines (see $\S$\ref{subsec:DataReduction}) and we found that it varies  from 2.58$\AA$ to 3.20$\AA$ with a distribution that peaks at 2.63$\AA$. Since we constructed our model using the nominal spectral resolution (3$\AA$) of the instrument,  we adjusted the value of the parameter $\sigma$ by applying a resolution correction using the value of the peak of the spectral FWHM distribution 2.63$\AA$, and then we added the dispersion of the distribution to the uncertainties on $\sigma$. In median, we added $-0.8\,km/s$ and $+4.7\,km/s$ to the formal error on $\sigma$ from the fit.
\begin{figure*}[!htb]
   \centering
   \includegraphics[width=0.8\textwidth]{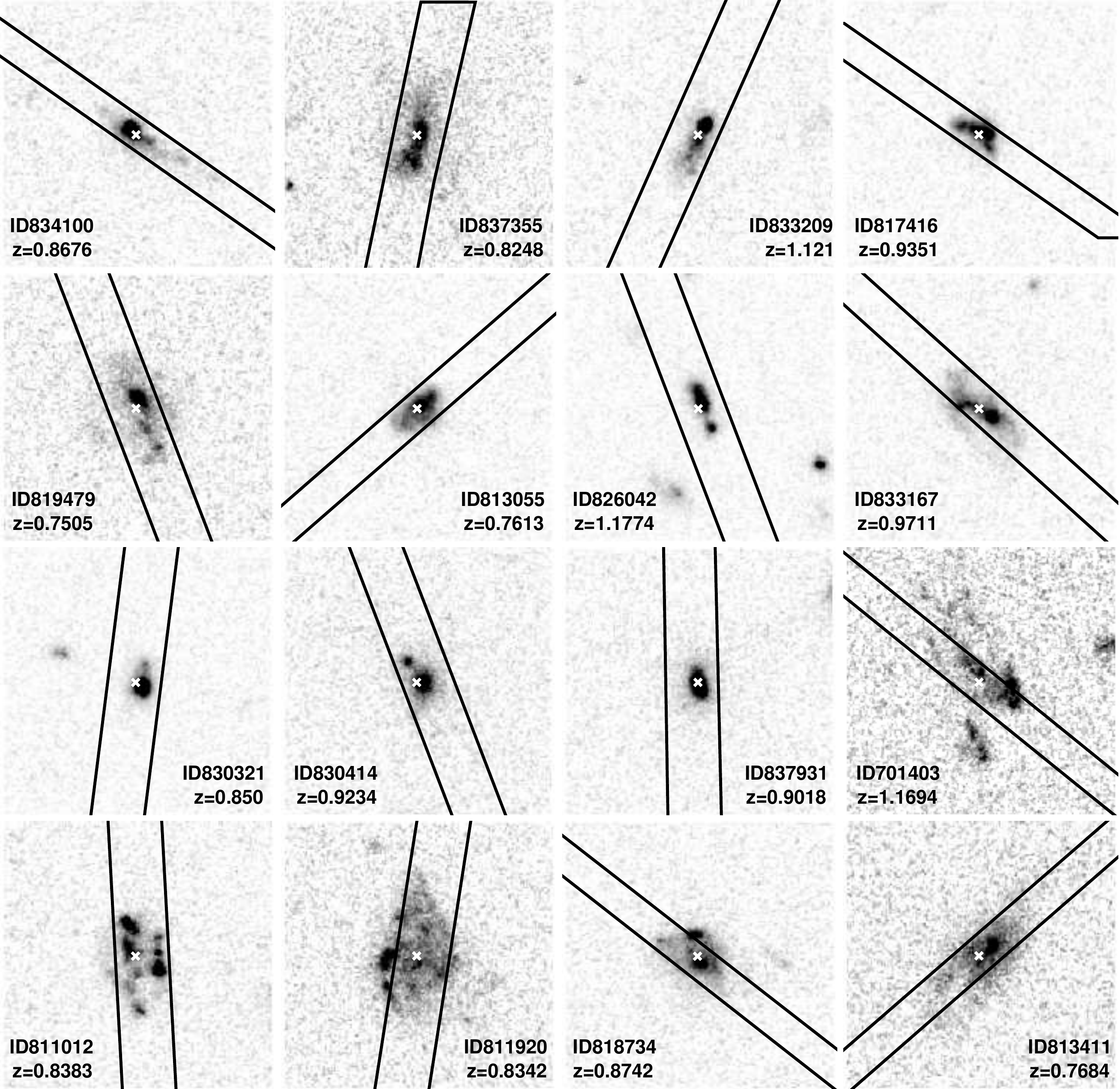}  
      \caption{F814W \textit{HST}/ACS postage stamps of our sub-sample of dispersion-dominated galaxies, with superimposed the 1$\arcsec$ width slit tilted to follow the galaxy major axis.
              }
              \label{fig:Disp_dom_gals}
\end{figure*}

\section{Results} \label{sec:results}
\subsection{Kinematic parameters} \label{subsec:kin_params}

The Levenberg–Marquardt least-squares fitting method is known to be sensitive to the local minima, thus it is important to choose the initial guesses as being as close as possible to the global minima. We accounted for this by measuring  the initial guesses as described in  $\S$\ref{subsec:fitting}. Furthermore, to explore the effect of the initial guesses on the resultant models, we adopted a Monte Carlo approach, which consists of perturbing the initial guesses by sampling a Gaussian distribution, obtaining 500 combinations of the free parameters and running the fit for every single combination. Amongst the fits that converged to a best-fit solution, we discarded the ones that have obviously wrong values for the parameters and extremely high values of the parameter errors, and we pre-selected the best-fit solutions in the lowest range of $\chi^2$ (within the 10\% of the lowest value). We then defined the best-fit model as the solution with the value of the parameter $V_t$ appeared most frequently at the most frequent $\chi^2$ value in the pre-selected $\chi^2$ distribution.  Our results do not change if we, instead, define the best-fit model as the solution with the most frequently occurring value of the parameter $V_t$ within the pre-selected $\chi^2$ distribution.

In this way, we obtained one best solution (when available) for each modeled rotation curve (exponential disc, flat model and arctangent function). For each galaxy, we visually inspected their residual images, the trace of the emission at each position (see $\S$\ref{subsec:fitting}) - both for the observed spectrum and the modeled one - and their rotation curves (see Figure \ref{fig:model}).
In general we chose the exponential disc solution, if available, by default. For our sample 60\% of the galaxies were modeled by an exponential disc rotation curve. If the exponential disc solution was not available, we adopted the flat model solution and, if not available, the arctangent function solution. This decision was motivated by the fact that for the 50\% of the sample, for which we got successfully a best-fit solution for each rotation curve model, the velocity V$_{2.2}$, that is the velocity used to derive the scaling relations (see later in this subsection), was always consistent within the uncertainties for all the rotation curve solutions.

The \mbox{$\chi^2$-minimization} fitting was run for 89 galaxies. Although seven of these presented noisy spectra (S/N<3), the fitting was attempted. However, from our visual check we found that no best-fit parameters could properly reproduce  the observations, so we discarded them, which resulted in a final sample of 82 galaxies. Examples of ours spectral data and best-fit model are shown in Figure \ref{fig:model_data_rot}.

To retrieve galaxy scaling relations and the dynamical masses, it is important to choose a fiducial radius at which the rotation velocity is measured. Past studies have shown that using different kinematic estimators (e.g., the maximal rotation velocity $V_{max}$, the plateau rotation curve velocity $V_{flat}$, the velocity $V_{80}$ measured at the radius containing $\sim$80$\%$ of the total  light, and the velocity $V_{2.2}$ measured at  $R_{2.2}=2.2\,R_d$, where $R_d$ is the disc scale-length estimated from broad band imaging) can lead to different results (see, e.g., \citealt{Verheijen2001, Pizagno2007}). In our sample, we adopted the $V_{2.2}$ estimator, which provides the tightest scatter in various galaxy scaling relations for bright galaxies and the best match to radio (21 cm) line widths for local galaxies \citep{Courteau1997}. The values of $R_{2.2}$ are computed as 2.2 times the galaxy disc scale length measured from the  \textit{HST}/ACS F814W images \citep{Scarlata2007} and included in the Zurich Structure and Morphology Catalog.

The model is not always able to reproduce the surface brightness profile of the observation, since a few discs are likely to be well described by an exponential profile, especially at high redshift, where a considerable number of galaxies shows clumpy star formation \citep{ForsterSchreiber2009,Wuyts2012}.  We  tested the reliability of rotation velocity parameters obtained from the fitting to the models by creating mock observations of galaxies (including the contribution of the background noise) with known rotation velocity and with diverse surface brightness profiles (generally described as a sum of two exponential functions with different $\Sigma_0$, $r_0$ and $r_{cen}$)  and we fitted them with our models, assuming an exponential disc surface brightness. We obtained that the values of the velocity best-fit parameters with different surface brightness profiles were always consistent within 1-$\sigma$ errors.  We also tested (see Appendix \ref{appendixA}) the correctness of the assumption of an exponential disc profile on the real data, by using an algorithm from \citet*{Scoville1983} (see Section 3 therein for a detailed description) to derive the  high-resolution spatial distribution of the galaxy emission that best matches the observed emission line profile shape. Including the profile derived in this way and re-fitting to the data, we found that the rotation velocity parameters were not affected by adopting this surface brightness profile. 
\begin{figure}[!htb]
   \centering
   \includegraphics[width=0.99\hsize]{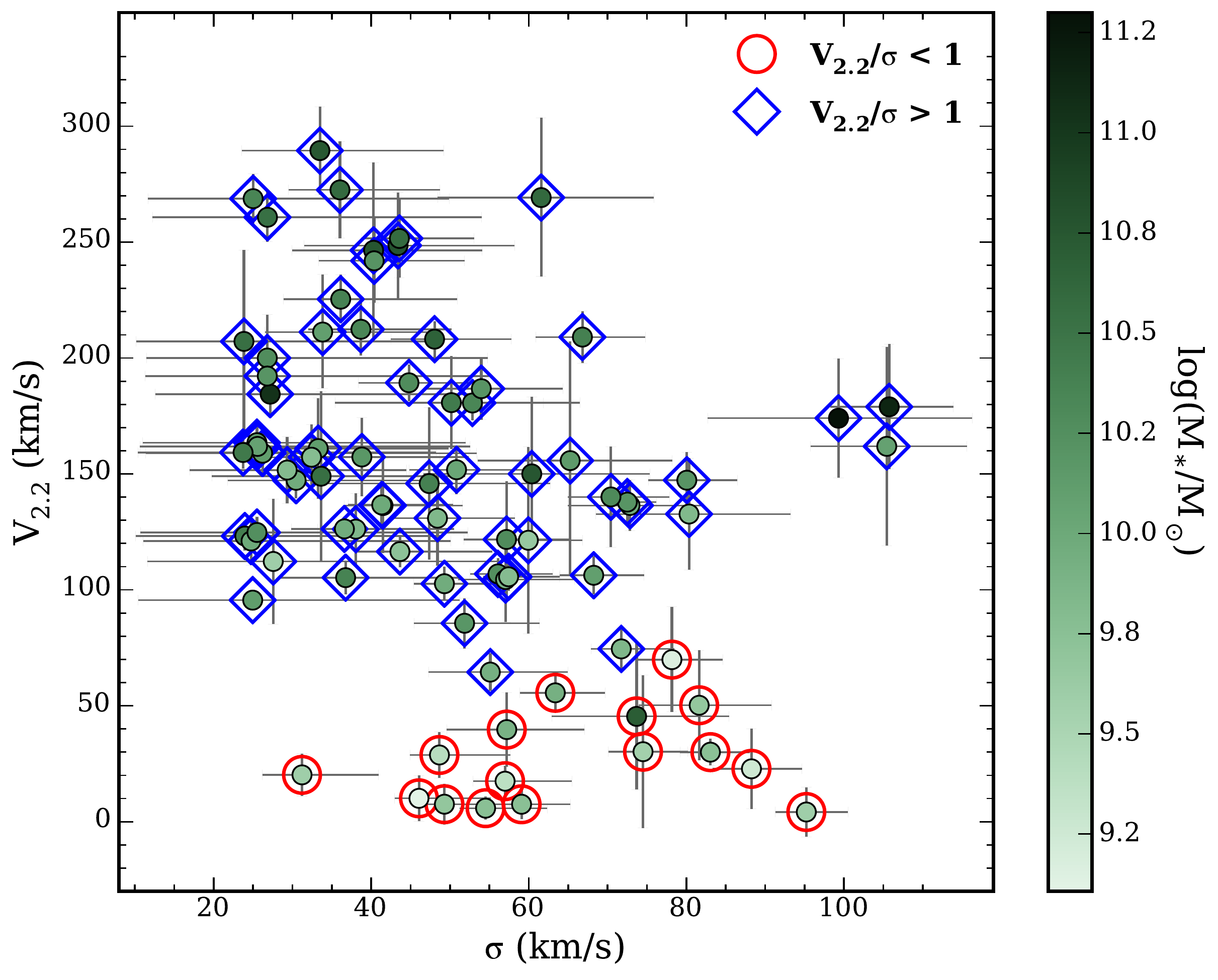}
      \caption{Rotation velocity $V_{2.2}$ as a function of the velocity dispersion $\sigma$. The points are color-coded according their stellar mass. Blue and red circles around the points identify rotation- and dispersion-dominated galaxies, respectively.
              }
         \label{fig:vrot_sigma}
   \end{figure}

The kinematic parameters of our sample are listed in the Table \ref{tab:results}. We found that in 82 successful fits, 66 galaxies (80$\%$) are formally rotation-dominated and 16 galaxies (20$\%$) are dispersion-dominated, by $V_{2.2}/\sigma>1$ and $V_{2.2}/\sigma<1$, respectively. We present in Figure \ref{fig:Disp_dom_gals} \textit{HST}/ACS images of our sample of dispersion-dominated galaxies.
 In Figure \ref{fig:vrot_sigma} we show a correlation between $V_{2.2}$ and $\sigma$ for dispersion-dominated and rotation-dominated galaxies. 
  Each point in Figure \ref{fig:vrot_sigma} is color-coded according to its stellar mass and we can see that the most massive galaxies amongst the rotation-dominated ones  have the largest values of  $V_{2.2}$.
 A similar result was found by \citet{Epinat2009}, who have studied the kinematics of a sample of star-forming galaxies at $\mathrm{1.2<z<1.6}$.
This would imply that the most massive star-forming galaxies are the most kinematically settled. Same result was presented by \citet{Kassin2012}, who studied the internal kinematics of 544 so called blue galaxies with stellar masses ranging $\mathrm{8.0 < logM_\ast(M_{\odot}) < 10.7}$ over $\mathrm{0.2 < z < 1.2}$, and found the most massive galaxies being the most evolved at any time. Another study from \citet{Simons2015} reports a transition mass in the smTF relation, $\mathrm{logM_\ast(M_{\odot})= 9.5}$, which they call the ``mass of disc formation'' $\mathrm{M_{df}}$, for a sample of emission line galaxies at  $\mathrm{0.1 < z < 0.375}$. This mass separates the galaxies that always form discs ($\mathrm{M_{\ast} > M_{df}}$) and are  settled on to the local smTF relation, from those which may or may not form discs ($\mathrm{M_{\ast} < M_{df}}$) and can either lie on the smTF relation or scatter off of it to low rotation velocity and higher disordered motions. We are aware, however, that the characterization of a galaxy as dispersion or rotatation-dominated is a strong function of the galaxy size. \citet{Newman2013_sins} compared the kinematic analysis of a sample of 81 star-forming galaxies at $\mathrm{z=1.0-2.5}$ using IFU data observed with both seeing-limited and adaptive optics (AO) mode,
 and found that   small galaxies are more likely to fall in the category of dispersion-dominated galaxies because of either insufficiently resolved rotation (especially with seeing-limited observations) or as a result of the almost constant values of velocity dispersion across all galaxy sizes while the values of rotation velocity linearly increase with size. They also found that many galaxies, which were considered dispersion-dominated from more poorly resolved data, actually showed evidence for rotation in higher resolution data but, in spite of this, they found that those galaxies  have different average properties than rotation-dominated galaxies. They tend to have lower stellar and dynamical masses, higher gas fractions, younger ages, and slightly lower metallicities. They suggest that these galaxies could be precursors of larger rotating galaxies, as they accrete more mass onto the outer regions of their discs.
We investigated any possible correlation between the size of the galaxies and the measured $V_{2.2}/\sigma$, by estimating the fraction of galaxies that have  the radius $r_{80}$, defined as the semi-major axis length of an ellipse encompassing 80\% of total light (from the Zurich Structure and Morphology Catalog), smaller than the measured seeing ($\S$\ref{subsec:DataReduction}). We found that this condition was true for 80\% of the dispersion-dominated galaxies, while 60\% of the total sample have $r_{80}$ smaller than the seeing. Therefore, although $r_{80}$ is only an approximation for the size of the galaxy, we found that the fraction of dispersion-dominated galaxies with small size is over-represented compared to the total fraction, suggesting that the classification as dispersion-dominated may be biased by the size of the galaxies.

We compared our kinematic classification to the galaxy morphologies as measured from the \textit{HST} images by \citet{Scarlata2007}, and we found that half of the dispersion-dominated sample is formed by galaxies defined as irregular, while 31\% have an intermediate bulge and only three galaxies (19\%) are classified as pure discs. The majority of the rotation-dominated galaxies, instead, are constituted of pure discs (42\%), while 29\% are formed by galaxies with intermediate bulge and 23\% are defined as irregular. The remaining 6\% is formed by three bulge-dominated galaxies and one early type. 
Three galaxies in Figure \ref{fig:vrot_sigma} do not follow the correlation and, even though they rotate quite fast, they  exhibit high velocity dispersion. For the two most massive galaxies a clear presence of a prominent bulge in the \textit{HST} images, or unresolved inner velocity gradient not described by simple models or intrinsic dispersion, may be the cause of the  high dispersion,  even though their discs are still rotating. The less massive one has a bright clumpy region in the periphery, which may have high dispersion and which, due to its relatively high brightness, may dominate the emission.

    \begin{figure*}[tb]
   \centering
   \includegraphics[width=0.7\hsize]{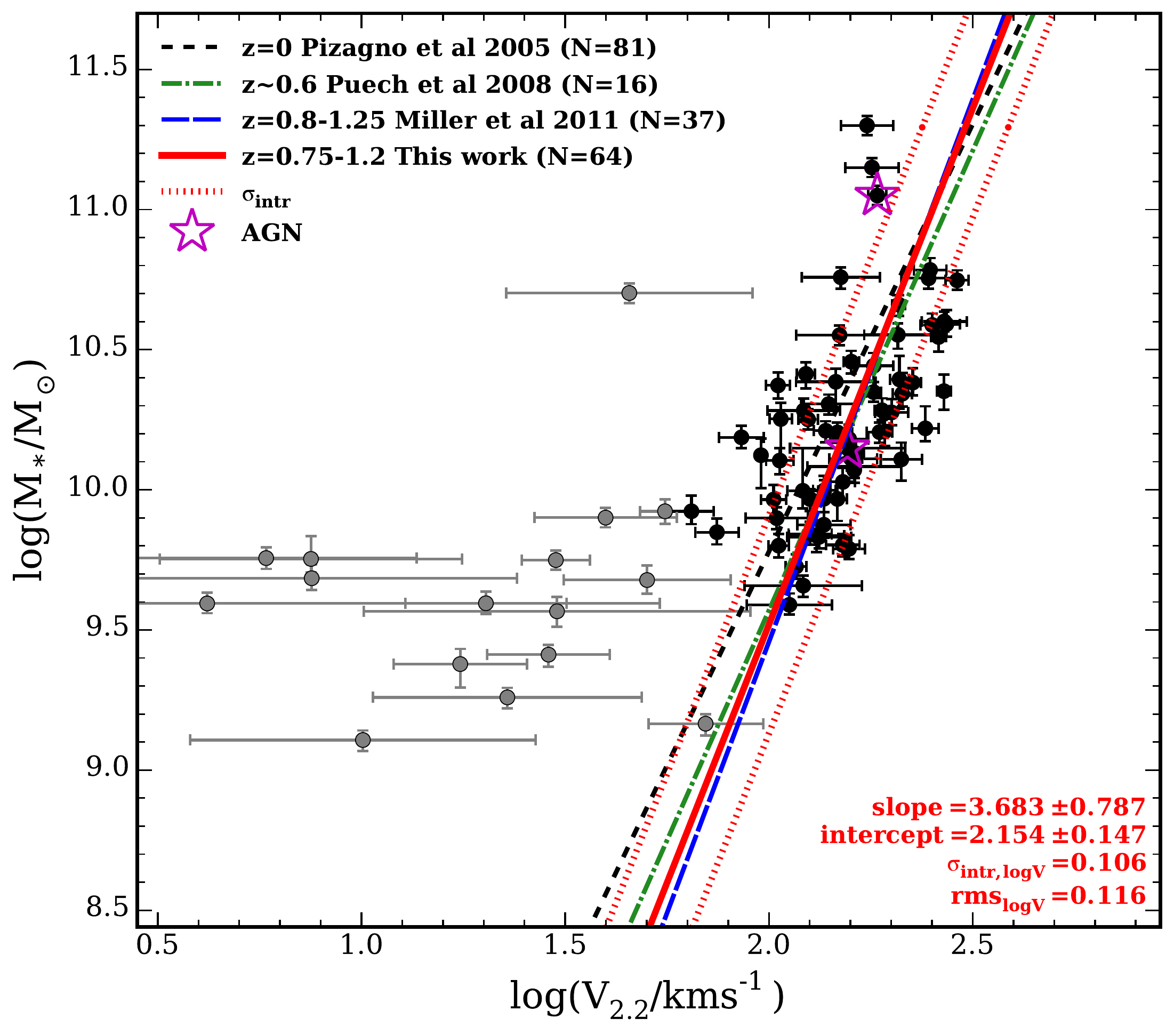}
      \caption{SmTF relation constrained for our sample of rotation-dominated galaxies (black circles). The gray points are the dispersion-dominated galaxies. The magenta stars show the AGNs. Our fit is represented with a solid red line. The dotted red lines show the intrinsic scatter $\sigma_{intr}$. We plot as references the relations at $\mathrm{z=0}$ \citep{Pizagno2005} with a short-dashed black line, $\mathrm{z\sim 0.6}$ \citep{Puech2008} with a dot-dashed green line and $\mathrm{z=0.8-1.25}$ \citep{Miller2011} with a long-dashed blue line.
              }
         \label{fig:smTF}
   \end{figure*}
Most of our kinematic sample is composed of [O{\small II}] emission line galaxies (79 out of 82) and, for those galaxies, we measured the line ratio $\mathrm{R_{[O{\small II}]}}$ from the model fitting. We show the distribution of $\mathrm{R_{[O{\small II}]}}$ along with its relationship to other galaxy parameters ($\sigma$, M$_{\ast}$, SFR) in Appendix \ref{appendixC}.  We measured the rest-frame equivalent width (EW) and computed the SFR from [O{\small II}] following \citet{Lemaux2014}. The choice of using EW([O{\small II}]) to derive the SFR instead of  [O{\small II}] flux was motivated by the lack of the absolute flux calibration for our spectroscopic observations (see $\S$ \ref{subsec:DataReduction}).  In Figure \ref{fig:SFR_sed_SFR_oii} we show the comparison between the SFR computed using the spectral energy distribution (SED) fitting and the [O{\small II}] emission. The values of $\mathrm{R_{[O{\small II}]}}$, $\mathrm{SFR_{SED}}$ and $\mathrm{SFR_{[O{\small II}]}}$ are listed in Table \ref{tab:oii_measurements}.

\begin{longtab}   
\tiny
\begin{longtable}{cccccccccccc}
\caption{Galaxy parameters}  \label{tab:results}\\
\hline\hline 
\noalign{\smallskip}       
ID  & RA & Dec & \textit{z} & PA & INCL & R$_{2.2}$ & V$_{2.2}$  & $\mathrm{\sigma}$& $V_{2.2}/\sigma$  & log(M$_\ast$/M$_\odot$) & log(M$_{\mathrm{dyn}}$(R$_{2.2}$)/M$_\odot$)\\[5pt]
 & deg & deg & & deg & deg & kpc & km/s  & km/s &  &  & \\  [5pt]
 (1) & (2)  & (3)  & (4) & (5) &(6)  & (7)  & (8) & (9) & (10) & (11) & (12) \\
\hline
\noalign{\smallskip} 
\endfirsthead
\caption{continued.}\\   
\hline\hline   
\noalign{\smallskip}     
ID  & RA & Dec & \textit{z} & PA & INCL & R$_{2.2}$ & V$_{2.2}$  & $\mathrm{\sigma}$& $V_{2.2}/\sigma$  & log(M$_\ast$/M$_\odot$) & log(M$_{\mathrm{dyn}}$(R$_{2.2}$)/M$_\odot$)\\[5pt]
 & deg & deg & & deg & deg & kpc & km/s  & km/s &  &  & \\  [5pt]
 (1) & (2)  & (3)  & (4) & (5) &(6)  & (7)  & (8) & (9) & (10) & (11) & (12) \\
\hline
\noalign{\smallskip} 
\endhead
\hline
\endfoot          
824384 & 150.200439 & 2.218330 & 0.8795 & $ -29.0_{-1.2}^{+1.2}$ & $ 52.7_{-6.8}^{+7.6}$ &   3.7 &    249$_{-30}^{+32}$ & $    43_{-12}^{+15}$ & $    5.7_{-1.7}^{+2.1}$ & $10.78_{-0.03}^{+0.04}$ & $10.76_{-0.10}^{+0.11}$\\ [2pt]
824508 & 150.173050 & 2.157201 & 0.8931 & $ +45.1_{-0.6}^{+0.6}$ & $ 62.8_{-2.5}^{+2.6}$ &  17.7 &    273$_{-21}^{+22}$ & $    36_{-7}^{+13}$ & $    7.6_{-1.5}^{+2.7}$ & $10.59_{-0.04}^{+0.05}$ & $11.50_{-0.07}^{+0.07}$\\ [2pt]
824658 & 150.142609 & 2.197006 & 0.8535 & $ +18.3_{-0.3}^{+0.4}$ & $ 78.3_{-2.7}^{+2.8}$ &  10.2 &    208$_{-8}^{+8}$ & $    48_{-6}^{+10}$ & $    4.3_{-0.5}^{+0.9}$ & $10.66_{-0.04}^{+0.03}$ & $11.06_{-0.03}^{+0.03}$\\ [2pt]
824791 & 150.110275 & 2.138679 & 0.8377 & $  +6.4_{-1.7}^{+1.5}$ & $ 49.1_{-7.3}^{+8.2}$ &   4.7 &    159$_{-18}^{+20}$ & $    26_{-15}^{+27}$ & $    6.1_{-3.5}^{+6.3}$ & $10.11_{-0.04}^{+0.04}$ & $10.46_{-0.10}^{+0.11}$\\ [2pt]
824847 & 150.097351 & 2.146748 & 0.9364 & $ -33.8_{-1.0}^{+1.3}$ & $ 64.3_{-4.1}^{+4.4}$ &   7.7 &    181$_{-21}^{+21}$ & $    50_{-15}^{+16}$ & $    3.6_{-1.1}^{+1.2}$ & $10.44_{-0.08}^{+0.05}$ & $10.83_{-0.09}^{+0.09}$\\ [2pt]
824675 & 150.137985 & 2.281988 & 0.8939 & $ -41.3_{-2.0}^{+1.6}$ & $ 62.1_{-5.5}^{+6.1}$ &   8.8 &     86$_{-12}^{+12}$ & $    52_{-6}^{+10}$ & $    1.7_{-0.3}^{+0.4}$ & $10.19_{-0.04}^{+0.04}$ & $10.43_{-0.08}^{+0.10}$\\ [2pt]
831229 & 150.200668 & 2.375840 & 0.7962 & $ -53.4_{-2.4}^{+2.1}$ & $ 41.4_{-7.3}^{+8.0}$ &   5.6 &    200$_{-31}^{+33}$ & $    27_{-15}^{+28}$ & $    7.5_{-4.4}^{+7.9}$ & $10.28_{-0.04}^{+0.05}$ & $10.73_{-0.13}^{+0.14}$\\ [2pt]
831256 & 150.194794 & 2.311815 & 0.8926 & $  -6.2_{-0.8}^{+0.6}$ & $ 72.9_{-2.6}^{+2.8}$ &  10.6 &    209$_{-11}^{+12}$ & $    67_{-6}^{+8}$ & $    3.1_{-0.3}^{+0.4}$ & $10.39_{-0.10}^{+0.08}$ & $11.12_{-0.04}^{+0.04}$\\ [2pt]
831534 & 150.136124 & 2.370541 & 0.9267 & $ -34.8_{-0.7}^{+0.8}$ & $ 52.1_{-3.9}^{+4.1}$ &   7.6 &    174$_{-27}^{+27}$ & $    99_{-17}^{+17}$ & $    1.8_{-0.4}^{+0.4}$ & $11.30_{-0.03}^{+0.03}$ & $10.96_{-0.10}^{+0.10}$\\ [2pt]
831675 & 150.107437 & 2.329219 & 0.9236 & $  -6.8_{-0.8}^{+1.0}$ & $ 50.5_{-3.7}^{+3.9}$ &  11.7 &    246$_{-40}^{+40}$ & $    40_{-10}^{+14}$ & $    6.1_{-1.9}^{+2.3}$ & $10.75_{-0.04}^{+0.03}$ & $11.24_{-0.13}^{+0.13}$\\ [2pt]
701002 & 149.998444 & 2.289785 & 0.9783 & $ +41.2_{-0.6}^{+0.6}$ & $ 67.2_{-4.2}^{+4.7}$ &   5.1 &    157$_{-17}^{+17}$ & $    39_{-10}^{+14}$ & $    4.1_{-1.2}^{+1.5}$ & $10.18_{-0.05}^{+0.05}$ & $10.52_{-0.09}^{+0.09}$\\ [2pt]
832385 & 149.987930 & 2.364365 & 0.9301 & $  +2.2_{-0.9}^{+0.8}$ & $ 69.7_{-3.2}^{+3.4}$ &  10.2 &    269$_{-12}^{+12}$ & $    25_{-13}^{+25}$ & $   10.7_{-5.8}^{+10.7}$ & $10.35_{-0.07}^{+0.06}$ & $11.24_{-0.04}^{+0.04}$\\ [2pt]
825474 & 149.982162 & 2.177024 & 0.8818 & $ +37.8_{-1.9}^{+1.7}$ & $ 44.9_{-5.6}^{+6.1}$ &   9.2 &    289$_{-32}^{+34}$ & $    33_{-10}^{+16}$ & $    8.6_{-2.7}^{+4.2}$ & $10.75_{-0.03}^{+0.04}$ & $11.27_{-0.09}^{+0.10}$\\ [2pt]
823045 & 150.478516 & 2.163036 & 0.8886 & $ +36.6_{-0.3}^{+0.3}$ & $ 73.6_{-2.4}^{+2.5}$ &   9.4 &    163$_{-7}^{+7}$ & $    25_{-14}^{+26}$ & $    6.4_{-3.7}^{+6.7}$ & $10.13_{-0.04}^{+0.04}$ & $10.79_{-0.04}^{+0.06}$\\ [2pt]
823323 & 150.423203 & 2.223220 & 0.8830 & $ -49.8_{-0.3}^{+0.4}$ & $ 70.5_{-2.2}^{+2.3}$ &   9.6 &    162$_{-4}^{+4}$ & $    26_{-15}^{+27}$ & $    6.3_{-3.7}^{+6.7}$ & $10.07_{-0.04}^{+0.05}$ & $10.79_{-0.03}^{+0.05}$\\ [2pt]
829955 & 150.444427 & 2.369805 & 0.8909 & $ +11.6_{-3.1}^{+2.4}$ & $ 33.4_{-8.0}^{+8.8}$ &   4.0 &    156$_{-52}^{+54}$ & $    65_{-12}^{+13}$ & $    2.4_{-0.9}^{+1.0}$ & $10.15_{-0.03}^{+0.03}$ & $10.49_{-0.21}^{+0.22}$\\ [2pt]
824317 & 150.212723 & 2.159449 & 0.9817 & $ +32.4_{-0.8}^{+0.5}$ & $ 63.2_{-4.0}^{+4.4}$ &   5.8 &     65$_{-8}^{+8}$ & $    55_{-8}^{+10}$ & $    1.2_{-0.2}^{+0.3}$ & $ 9.92_{-0.05}^{+0.06}$ & $10.16_{-0.09}^{+0.10}$\\ [2pt]
827096 & 149.657990 & 2.269869 & 0.8309 & $  +7.4_{-0.6}^{+0.6}$ & $ 59.4_{-3.2}^{+3.5}$ &   7.6 &    181$_{-9}^{+9}$ & $    53_{-5}^{+9}$ & $    3.4_{-0.4}^{+0.6}$ & $10.35_{-0.03}^{+0.04}$ & $10.83_{-0.04}^{+0.04}$\\ [2pt]
833862 & 149.698212 & 2.367727 & 0.7512 & $ +26.5_{-0.7}^{+0.5}$ & $ 67.9_{-3.8}^{+4.1}$ &   6.6 &    133$_{-24}^{+24}$ & $    80_{-12}^{+13}$ & $    1.7_{-0.4}^{+0.4}$ & $ 9.83_{-0.04}^{+0.05}$ & $10.69_{-0.11}^{+0.11}$\\ [2pt]
834100 & 149.655685 & 2.362452 & 0.8676 & $ +55.4_{-0.7}^{+0.7}$ & $ 66.4_{-4.9}^{+5.4}$ &   3.1 &     23$_{-17}^{+17}$ & $    88_{-5}^{+6}$ & $    0.3_{-0.2}^{+0.2}$ & $ 9.26_{-0.04}^{+0.03}$ & $10.10_{-0.05}^{+0.06}$\\ [2pt]
826948 & 149.687210 & 2.208733 & 0.9445 & $ +35.7_{-0.7}^{+0.9}$ & $ 59.6_{-3.4}^{+3.6}$ &   8.3 &    147$_{-13}^{+13}$ & $    80_{-5}^{+6}$ & $    1.8_{-0.2}^{+0.2}$ & $10.20_{-0.04}^{+0.04}$ & $10.84_{-0.05}^{+0.05}$\\ [2pt]
827050 & 149.666428 & 2.222278 & 0.8922 & $ +57.7_{-0.9}^{+1.1}$ & $ 42.8_{-6.7}^{+7.4}$ &   5.9 &    179$_{-34}^{+36}$ & $   106_{-7}^{+8}$ & $    1.7_{-0.3}^{+0.4}$ & $11.15_{-0.03}^{+0.03}$ & $10.89_{-0.10}^{+0.10}$\\ [2pt]
837355 & 150.392242 & 2.603468 & 0.8248 & $ -11.9_{-0.8}^{+0.8}$ & $ 66.3_{-3.6}^{+3.8}$ &   7.9 &     29$_{-10}^{+10}$ & $    49_{-4}^{+9}$ & $    0.6_{-0.2}^{+0.2}$ & $ 9.41_{-0.04}^{+0.04}$ & $10.04_{-0.07}^{+0.14}$\\ [2pt]
838455 & 150.189224 & 2.606611 & 1.0196 & $ +27.4_{-0.6}^{+0.7}$ & $ 72.0_{-3.5}^{+3.8}$ &   7.9 &    207$_{-40}^{+40}$ & $    24_{-14}^{+25}$ & $    8.7_{-5.3}^{+9.2}$ & $10.55_{-0.05}^{+0.04}$ & $10.91_{-0.16}^{+0.16}$\\ [2pt]
831223 & 150.202377 & 2.377197 & 0.7965 & $ +46.0_{-1.2}^{+1.3}$ & $ 59.2_{-4.9}^{+5.4}$ &   5.3 &    261$_{-15}^{+17}$ & $    27_{-15}^{+27}$ & $    9.7_{-5.3}^{+9.9}$ & $10.55_{-0.05}^{+0.05}$ & $10.93_{-0.05}^{+0.06}$\\ [2pt]
1254477 & 150.003080 & 2.400714 & 0.9447 & $  -6.9_{-1.7}^{+1.5}$ & $ 47.1_{-4.6}^{+4.9}$ &  10.0 &    212$_{-18}^{+19}$ & $    39_{-7}^{+11}$ & $    5.5_{-1.1}^{+1.7}$ & $10.34_{-0.05}^{+0.07}$ & $11.05_{-0.07}^{+0.07}$\\ [2pt]
840112 & 149.907379 & 2.529695 & 0.8908 & $ +38.4_{-2.5}^{+3.4}$ & $ 47.4_{-5.8}^{+6.1}$ &   6.7 &    150$_{-35}^{+35}$ & $    60_{-14}^{+15}$ & $    2.5_{-0.8}^{+0.9}$ & $10.76_{-0.04}^{+0.04}$ & $10.67_{-0.16}^{+0.16}$\\ [2pt]
840266 & 149.878723 & 2.601216 & 0.9587 & $ -32.1_{-1.3}^{+1.0}$ & $ 60.1_{-3.4}^{+3.6}$ &   8.7 &    107$_{-7}^{+7}$ & $    56_{-4}^{+7}$ & $    1.9_{-0.2}^{+0.3}$ & $10.25_{-0.15}^{+0.06}$ & $10.57_{-0.04}^{+0.06}$\\ [2pt]
840390 & 149.851700 & 2.519398 & 0.8452 & $ -24.6_{-0.4}^{+0.5}$ & $ 62.7_{-3.3}^{+3.6}$ &   6.8 &    161$_{-22}^{+22}$ & $    33_{-10}^{+17}$ & $    4.8_{-1.6}^{+2.5}$ & $10.11_{-0.04}^{+0.03}$ & $10.65_{-0.11}^{+0.11}$\\ [2pt]
833209 & 149.826096 & 2.382653 & 1.1210 & $ -23.8_{-0.7}^{+0.7}$ & $ 62.3_{-4.8}^{+5.2}$ &   4.2 &     56$_{-8}^{+8}$ & $    63_{-4}^{+6}$ & $    0.9_{-0.1}^{+0.2}$ & $ 9.92_{-0.04}^{+0.04}$ & $10.06_{-0.06}^{+0.07}$\\ [2pt]
811224 & 150.282700 & 1.932057 & 0.9137 & $  +8.3_{-7.1}^{+7.0}$ & $ 46.2_{-3.9}^{+4.2}$ &  13.7 &    269$_{-36}^{+37}$ & $    62_{-13}^{+14}$ & $    4.4_{-1.1}^{+1.2}$ & $10.60_{-0.04}^{+0.04}$ & $11.41_{-0.11}^{+0.11}$\\ [2pt]
811233 & 150.280426 & 1.921645 & 1.0099 & $ +33.7_{-0.8}^{+0.8}$ & $ 54.8_{-5.6}^{+6.2}$ &   5.1 &    123$_{-9}^{+10}$ & $    24_{-14}^{+25}$ & $    5.1_{-3.0}^{+5.4}$ & $10.41_{-0.05}^{+0.04}$ & $10.29_{-0.07}^{+0.09}$\\ [2pt]
817262 & 150.373260 & 2.086616 & 0.9294 & $ +24.8_{-0.4}^{+0.4}$ & $ 72.4_{-2.8}^{+3.0}$ &   7.3 &     96$_{-2}^{+2}$ & $    25_{-15}^{+26}$ & $    3.8_{-2.2}^{+4.0}$ & $10.12_{-0.12}^{+0.06}$ & $10.25_{-0.07}^{+0.12}$\\ [2pt]
817416 & 150.343170 & 2.057282 & 0.9351 & $ +55.0_{-0.6}^{+0.6}$ & $ 58.1_{-6.7}^{+7.6}$ &   3.2 &      4$_{-11}^{+11}$ & $    95_{-4}^{+5}$ & $    0.0_{-0.1}^{+0.1}$ & $ 9.59_{-0.03}^{+0.04}$ & $10.16_{-0.04}^{+0.05}$\\ [2pt]
818198 & 150.176285 & 2.111890 & 0.7872 & $ +52.2_{-0.6}^{+0.8}$ & $ 57.2_{-3.4}^{+3.6}$ &   8.0 &     75$_{-9}^{+9}$ & $    72_{-4}^{+7}$ & $    1.0_{-0.1}^{+0.2}$ & $ 9.85_{-0.04}^{+0.05}$ & $10.49_{-0.05}^{+0.07}$\\ [2pt]
811727 & 150.176453 & 1.955227 & 1.1494 & $ -42.1_{-1.6}^{+1.6}$ & $ 59.0_{-4.6}^{+5.0}$ &   6.2 &    146$_{-33}^{+33}$ & $    47_{-14}^{+15}$ & $    3.1_{-1.2}^{+1.2}$ & $10.39_{-0.10}^{+0.05}$ & $10.58_{-0.17}^{+0.17}$\\ [2pt]
818959 & 150.017838 & 2.049238 & 0.8724 & $ -59.6_{-1.7}^{+1.3}$ & $ 47.2_{-4.8}^{+5.2}$ &   7.1 &    131$_{-22}^{+22}$ & $    48_{-6}^{+10}$ & $    2.7_{-0.6}^{+0.7}$ & $ 9.84_{-0.06}^{+0.13}$ & $10.56_{-0.12}^{+0.12}$\\ [2pt]
819479 & 149.890701 & 2.089932 & 0.7505 & $ +20.5_{-1.3}^{+1.1}$ & $ 60.6_{-3.8}^{+4.0}$ &   7.3 &     17$_{-7}^{+7}$ & $    57_{-4}^{+8}$ & $    0.3_{-0.1}^{+0.1}$ & $ 9.38_{-0.08}^{+0.05}$ & $10.10_{-0.06}^{+0.12}$\\ [2pt]
819641 & 149.852707 & 2.100561 & 0.8957 & $ -49.8_{-1.8}^{+1.8}$ & $ 42.4_{-6.9}^{+7.5}$ &   5.7 &    149$_{-40}^{+41}$ & $    34_{-14}^{+18}$ & $    4.4_{-2.2}^{+2.7}$ & $10.55_{-0.04}^{+0.03}$ & $10.51_{-0.21}^{+0.22}$\\ [2pt]
819765 & 149.819427 & 2.093724 & 0.8543 & $  -1.6_{-0.5}^{+0.5}$ & $ 67.9_{-3.3}^{+3.5}$ &   6.6 &    136$_{-11}^{+11}$ & $    73_{-8}^{+9}$ & $    1.9_{-0.3}^{+0.3}$ & $ 9.97_{-0.05}^{+0.08}$ & $10.66_{-0.06}^{+0.06}$\\ [2pt]
812913 & 149.912201 & 1.923351 & 0.7747 & $ +43.3_{-0.6}^{+0.6}$ & $ 56.1_{-4.2}^{+4.5}$ &   6.5 &    184$_{-12}^{+12}$ & $    27_{-15}^{+27}$ & $    6.8_{-3.7}^{+6.8}$ & $11.05_{-0.03}^{+0.03}$ & $10.73_{-0.06}^{+0.07}$\\ [2pt]
813055 & 149.881760 & 1.910661 & 0.7613 & $ -48.5_{-0.7}^{+0.7}$ & $ 51.4_{-7.2}^{+8.2}$ &   3.4 &     50$_{-24}^{+24}$ & $    82_{-8}^{+9}$ & $    0.6_{-0.3}^{+0.3}$ & $ 9.68_{-0.05}^{+0.05}$ & $10.13_{-0.09}^{+0.10}$\\ [2pt]
824746 & 150.120987 & 2.208277 & 0.8467 & $ +33.7_{-0.9}^{+0.7}$ & $ 73.6_{-3.5}^{+3.8}$ &   8.6 &    126$_{-16}^{+16}$ & $    38_{-5}^{+12}$ & $    3.3_{-0.6}^{+1.1}$ & $ 9.84_{-0.04}^{+0.04}$ & $10.58_{-0.09}^{+0.10}$\\ [2pt]
824408 & 150.196198 & 2.267904 & 0.7496 & $ +22.1_{-0.8}^{+0.9}$ & $ 68.7_{-3.9}^{+4.2}$ &   5.3 &    117$_{-7}^{+7}$ & $    44_{-6}^{+12}$ & $    2.7_{-0.4}^{+0.7}$ & $ 9.73_{-0.04}^{+0.04}$ & $10.34_{-0.05}^{+0.07}$\\ [2pt]
831493 & 150.145020 & 2.336934 & 0.9360 & $ -40.7_{-0.5}^{+0.5}$ & $ 61.3_{-2.3}^{+2.4}$ &  12.3 &    252$_{-17}^{+18}$ & $    44_{-5}^{+9}$ & $    5.8_{-0.8}^{+1.3}$ & $10.59_{-0.03}^{+0.04}$ & $11.29_{-0.06}^{+0.06}$\\ [2pt]
832277 & 150.009109 & 2.371844 & 1.0299 & $ -23.1_{-1.5}^{+1.4}$ & $ 50.4_{-4.6}^{+5.0}$ &   7.4 &    159$_{-11}^{+12}$ & $    24_{-13}^{+24}$ & $    6.7_{-3.8}^{+6.9}$ & $10.46_{-0.04}^{+0.04}$ & $10.66_{-0.06}^{+0.07}$\\ [2pt]
832708 & 149.920853 & 2.306864 & 0.9272 & $ +26.5_{-0.6}^{+0.6}$ & $ 61.3_{-3.9}^{+4.3}$ &   6.2 &    122$_{-25}^{+25}$ & $    57_{-5}^{+8}$ & $    2.1_{-0.5}^{+0.5}$ & $10.28_{-0.04}^{+0.04}$ & $10.50_{-0.13}^{+0.13}$\\ [2pt]
825250 & 150.024475 & 2.190769 & 0.7587 & $ +36.9_{-1.0}^{+1.0}$ & $ 54.2_{-3.7}^{+4.0}$ &   7.9 &    225$_{-13}^{+14}$ & $    36_{-7}^{+15}$ & $    6.2_{-1.3}^{+2.6}$ & $10.38_{-0.05}^{+0.05}$ & $11.00_{-0.05}^{+0.05}$\\ [2pt]
825269 & 150.021332 & 2.178634 & 0.7954 & $ +34.6_{-0.7}^{+0.7}$ & $ 57.2_{-3.5}^{+3.7}$ &   7.6 &    157$_{-15}^{+15}$ & $    32_{-8}^{+16}$ & $    4.8_{-1.2}^{+2.5}$ & $ 9.79_{-0.04}^{+0.05}$ & $10.68_{-0.08}^{+0.08}$\\ [2pt]
823909 & 150.304321 & 2.269336 & 0.9256 & $ -37.7_{-0.9}^{+0.9}$ & $ 49.8_{-6.7}^{+7.4}$ &   4.8 &    242$_{-27}^{+29}$ & $    40_{-7}^{+11}$ & $    6.0_{-1.2}^{+1.8}$ & $10.22_{-0.05}^{+0.08}$ & $10.84_{-0.09}^{+0.10}$\\ [2pt]
824079 & 150.259933 & 2.292139 & 0.9871 & $  +4.2_{-0.3}^{+0.4}$ & $ 71.4_{-3.5}^{+3.8}$ &   5.8 &    106$_{-8}^{+8}$ & $    68_{-4}^{+6}$ & $    1.6_{-0.2}^{+0.2}$ & $10.10_{-0.05}^{+0.04}$ & $10.46_{-0.04}^{+0.05}$\\ [2pt]
831094 & 150.228714 & 2.316025 & 0.8972 & $ -15.7_{-0.2}^{+0.2}$ & $ 66.6_{-3.4}^{+3.7}$ &   5.6 &    138$_{-10}^{+10}$ & $    73_{-2}^{+4}$ & $    1.9_{-0.1}^{+0.2}$ & $10.21_{-0.04}^{+0.03}$ & $10.60_{-0.04}^{+0.04}$\\ [2pt]
826042 & 149.868927 & 2.130142 & 1.1774 & $ +21.2_{-0.3}^{+0.3}$ & $ 79.0_{-4.4}^{+5.0}$ &   4.0 &     30$_{-6}^{+6}$ & $    83_{-4}^{+5}$ & $    0.4_{-0.1}^{+0.1}$ & $ 9.75_{-0.03}^{+0.03}$ & $10.16_{-0.04}^{+0.05}$\\ [2pt]
826065 & 149.864700 & 2.209825 & 0.8941 & $ -14.6_{-1.6}^{+1.2}$ & $ 56.0_{-4.7}^{+5.1}$ &   6.7 &    104$_{-19}^{+19}$ & $    57_{-6}^{+9}$ & $    1.8_{-0.4}^{+0.4}$ & $ 9.90_{-0.04}^{+0.04}$ & $10.45_{-0.10}^{+0.11}$\\ [2pt]
826091 & 149.857941 & 2.137394 & 0.8937 & $ -31.0_{-1.1}^{+1.0}$ & $ 57.9_{-3.6}^{+3.8}$ &  10.3 &    152$_{-12}^{+12}$ & $    51_{-6}^{+9}$ & $    3.0_{-0.4}^{+0.6}$ & $10.03_{-0.07}^{+0.05}$ & $10.84_{-0.06}^{+0.06}$\\ [2pt]
833167 & 149.836288 & 2.301956 & 0.9711 & $ +47.7_{-0.4}^{+0.4}$ & $ 72.7_{-3.4}^{+3.7}$ &   5.8 &      8$_{-6}^{+6}$ & $    59_{-3}^{+6}$ & $    0.1_{-0.1}^{+0.1}$ & $ 9.75_{-0.05}^{+0.08}$ & $10.01_{-0.04}^{+0.09}$\\ [2pt]
833707 & 149.731277 & 2.329328 & 0.7841 & $ +44.9_{-0.5}^{+0.5}$ & $ 63.2_{-2.8}^{+3.0}$ &   8.4 &    187$_{-14}^{+14}$ & $    54_{-7}^{+10}$ & $    3.5_{-0.5}^{+0.7}$ & $10.21_{-0.04}^{+0.03}$ & $10.90_{-0.06}^{+0.06}$\\ [2pt]
827090 & 149.659027 & 2.224841 & 0.9205 & $  +1.0_{-0.6}^{+0.6}$ & $ 65.5_{-2.7}^{+2.9}$ &  11.1 &    136$_{-21}^{+21}$ & $    42_{-5}^{+10}$ & $    3.3_{-0.6}^{+0.9}$ & $ 9.88_{-0.05}^{+0.04}$ & $10.76_{-0.11}^{+0.11}$\\ [2pt]
830321 & 150.383362 & 2.371992 & 0.8500 & $  -7.1_{-1.0}^{+1.0}$ & $ 60.8_{-8.6}^{+10.4}$ &   1.8 &     70$_{-23}^{+23}$ & $    78_{-4}^{+6}$ & $    0.9_{-0.3}^{+0.3}$ & $ 9.17_{-0.04}^{+0.03}$ & $ 9.88_{-0.08}^{+0.09}$\\ [2pt]
830414 & 150.369095 & 2.394953 & 0.9234 & $ +20.8_{-2.0}^{+2.5}$ & $ 43.9_{-8.2}^{+9.2}$ &   4.4 &     45$_{-32}^{+32}$ & $    74_{-11}^{+12}$ & $    0.6_{-0.4}^{+0.4}$ & $10.70_{-0.04}^{+0.03}$ & $10.15_{-0.14}^{+0.15}$\\ [2pt]
837931 & 150.287567 & 2.477325 & 0.9018 & $  +0.7_{-0.6}^{+0.5}$ & $ 66.0_{-7.1}^{+8.3}$ &   2.3 &     40$_{-16}^{+16}$ & $    57_{-8}^{+10}$ & $    0.7_{-0.3}^{+0.3}$ & $ 9.90_{-0.03}^{+0.03}$ & $ 9.67_{-0.11}^{+0.14}$\\ [2pt]
837433 & 150.378189 & 2.530785 & 0.7963 & $ -22.9_{-0.5}^{+0.4}$ & $ 70.3_{-2.2}^{+2.3}$ &  12.0 &    192$_{-4}^{+4}$ & $    27_{-15}^{+28}$ & $    7.2_{-4.1}^{+7.5}$ & $10.21_{-0.05}^{+0.05}$ & $11.03_{-0.03}^{+0.04}$\\ [2pt]
837491 & 150.367355 & 2.520236 & 0.9940 & $ +11.9_{-0.6}^{+0.5}$ & $ 68.3_{-2.8}^{+2.9}$ &   8.2 &    106$_{-6}^{+6}$ & $    57_{-3}^{+6}$ & $    1.8_{-0.2}^{+0.2}$ & $ 9.80_{-0.04}^{+0.04}$ & $10.54_{-0.04}^{+0.05}$\\ [2pt]
837613 & 150.341934 & 2.568233 & 0.8182 & $ -20.8_{-0.3}^{+0.3}$ & $ 79.4_{-2.6}^{+2.8}$ &   7.7 &    105$_{-7}^{+7}$ & $    37_{-9}^{+14}$ & $    2.9_{-0.7}^{+1.1}$ & $10.37_{-0.05}^{+0.05}$ & $10.40_{-0.06}^{+0.09}$\\ [2pt]
831296 & 150.187332 & 2.414294 & 0.8517 & $ -12.7_{-0.4}^{+0.4}$ & $ 61.4_{-4.3}^{+4.7}$ &   5.0 &    147$_{-9}^{+9}$ & $    30_{-9}^{+17}$ & $    4.8_{-1.4}^{+2.7}$ & $ 9.97_{-0.08}^{+0.06}$ & $10.44_{-0.05}^{+0.06}$\\ [2pt]
831848 & 150.082047 & 2.458241 & 0.9363 & $ -41.6_{-0.4}^{+0.4}$ & $ 76.9_{-3.1}^{+3.3}$ &   6.4 &    162$_{-43}^{+43}$ & $   105_{-10}^{+10}$ & $    1.5_{-0.4}^{+0.4}$ & $10.08_{-0.04}^{+0.05}$ & $10.87_{-0.13}^{+0.13}$\\ [2pt]
832184 & 150.026810 & 2.403100 & 0.9468 & $ -31.7_{-1.4}^{+1.9}$ & $ 45.3_{-5.9}^{+6.4}$ &   6.5 &    121$_{-15}^{+16}$ & $    25_{-14}^{+25}$ & $    4.9_{-2.8}^{+5.0}$ & $10.00_{-0.06}^{+0.15}$ & $10.39_{-0.10}^{+0.13}$\\ [2pt]
839193 & 150.058121 & 2.562992 & 0.8887 & $ -54.2_{-1.5}^{+1.2}$ & $ 46.0_{-3.8}^{+4.1}$ &  10.4 &    125$_{-10}^{+10}$ & $    25_{-15}^{+27}$ & $    4.9_{-2.9}^{+5.2}$ & $10.25_{-0.04}^{+0.04}$ & $10.61_{-0.07}^{+0.10}$\\ [2pt]
839379 & 150.026962 & 2.589348 & 0.7469 & $ -44.3_{-1.6}^{+1.5}$ & $ 52.0_{-4.9}^{+5.3}$ &   5.9 &    112$_{-28}^{+28}$ & $    28_{-16}^{+29}$ & $    4.1_{-2.6}^{+4.4}$ & $ 9.59_{-0.03}^{+0.04}$ & $10.29_{-0.20}^{+0.22}$\\ [2pt]
840437 & 149.842422 & 2.570137 & 0.9642 & $ +46.5_{-0.8}^{+0.7}$ & $ 61.0_{-4.0}^{+4.4}$ &   6.2 &    121$_{-40}^{+40}$ & $    60_{-4}^{+7}$ & $    2.0_{-0.7}^{+0.7}$ & $ 9.66_{-0.04}^{+0.04}$ & $10.51_{-0.19}^{+0.19}$\\ [2pt]
701403 & 149.895996 & 2.410341 & 1.1694 & $ +51.2_{-2.7}^{+2.8}$ & $ 54.1_{-5.1}^{+5.5}$ &   8.3 &     30$_{-33}^{+33}$ & $    75_{-4}^{+6}$ & $    0.4_{-0.4}^{+0.4}$ & $ 9.57_{-0.06}^{+0.05}$ & $10.39_{-0.08}^{+0.09}$\\ [2pt]
811012 & 150.331024 & 1.879742 & 0.8383 & $  +2.6_{-0.9}^{+0.8}$ & $ 47.2_{-4.0}^{+4.2}$ &   7.0 &     10$_{-10}^{+10}$ & $    46_{-3}^{+9}$ & $    0.2_{-0.2}^{+0.2}$ & $ 9.11_{-0.04}^{+0.03}$ & $ 9.88_{-0.06}^{+0.17}$\\ [2pt]
811108 & 150.309006 & 1.916707 & 0.8963 & $ -55.8_{-0.9}^{+0.9}$ & $ 54.1_{-3.4}^{+3.6}$ &   8.7 &    140$_{-22}^{+22}$ & $    70_{-5}^{+7}$ & $    2.0_{-0.4}^{+0.4}$ & $10.31_{-0.04}^{+0.03}$ & $10.79_{-0.09}^{+0.10}$\\ [2pt]
817640 & 150.296600 & 1.969438 & 0.9357 & $ +23.9_{-1.1}^{+1.0}$ & $ 43.6_{-6.2}^{+6.8}$ &   5.4 &    211$_{-30}^{+32}$ & $    34_{-7}^{+13}$ & $    6.2_{-1.6}^{+2.6}$ & $10.11_{-0.08}^{+0.06}$ & $10.77_{-0.12}^{+0.13}$\\ [2pt]
817426 & 150.339340 & 2.100803 & 0.8667 & $  +9.9_{-0.7}^{+0.6}$ & $ 63.4_{-3.3}^{+3.5}$ &   6.9 &    103$_{-8}^{+8}$ & $    49_{-4}^{+9}$ & $    2.1_{-0.2}^{+0.4}$ & $ 9.97_{-0.08}^{+0.05}$ & $10.40_{-0.05}^{+0.07}$\\ [2pt]
818113 & 150.192520 & 2.019528 & 1.0123 & $ +16.2_{-0.9}^{+0.8}$ & $ 61.8_{-3.7}^{+4.0}$ &   6.5 &    137$_{-9}^{+9}$ & $    41_{-4}^{+9}$ & $    3.3_{-0.4}^{+0.8}$ & $10.00_{-0.03}^{+0.04}$ & $10.53_{-0.05}^{+0.06}$\\ [2pt]
811920 & 150.138062 & 1.881366 & 0.8342 & $  -9.5_{-1.0}^{+1.0}$ & $ 65.4_{-2.9}^{+3.1}$ &   8.3 &      6$_{-5}^{+5}$ & $    55_{-3}^{+8}$ & $    0.1_{-0.1}^{+0.1}$ & $ 9.76_{-0.04}^{+0.04}$ & $10.09_{-0.05}^{+0.12}$\\ [2pt]
818734 & 150.074448 & 2.038712 & 0.8742 & $ +52.3_{-1.5}^{+1.8}$ & $ 41.9_{-7.2}^{+7.9}$ &   5.1 &      8$_{-9}^{+9}$ & $    49_{-3}^{+8}$ & $    0.2_{-0.2}^{+0.2}$ & $ 9.68_{-0.04}^{+0.05}$ & $ 9.80_{-0.06}^{+0.14}$\\ [2pt]
813128 & 149.862106 & 1.936515 & 0.7887 & $ -13.7_{-0.6}^{+0.8}$ & $ 61.9_{-2.8}^{+2.9}$ &   9.6 &    126$_{-6}^{+6}$ & $    37_{-7}^{+14}$ & $    3.5_{-0.7}^{+1.3}$ & $ 9.97_{-0.04}^{+0.04}$ & $10.62_{-0.04}^{+0.06}$\\ [2pt]
830282 & 150.388718 & 2.309891 & 0.7432 & $ -39.6_{-0.6}^{+0.4}$ & $ 62.8_{-4.2}^{+4.6}$ &   4.9 &    152$_{-15}^{+15}$ & $    29_{-12}^{+15}$ & $    5.2_{-2.2}^{+2.7}$ & $ 9.80_{-0.04}^{+0.04}$ & $10.46_{-0.08}^{+0.09}$\\ [2pt]
831655 & 150.112289 & 2.387432 & 0.7573 & $ +22.0_{-0.3}^{+0.3}$ & $ 74.1_{-2.1}^{+2.2}$ &   9.4 &    189$_{-8}^{+8}$ & $    45_{-6}^{+8}$ & $    4.2_{-0.6}^{+0.8}$ & $10.28_{-0.03}^{+0.04}$ & $10.94_{-0.04}^{+0.04}$\\ [2pt]
813411 & 149.793304 & 1.927607 & 0.7684 & $ -48.9_{-1.2}^{+1.3}$ & $ 59.0_{-4.3}^{+4.7}$ &   7.0 &     20$_{-9}^{+9}$ & $    31_{-5}^{+10}$ & $    0.6_{-0.3}^{+0.4}$ & $ 9.60_{-0.04}^{+0.04}$ & $ 9.61_{-0.13}^{+0.24}$\\ [2pt]
\end{longtable}
\begin{flushleft}
\justify
(1) Source HR-COSMOS identification number, (2) and (3) right ascension and declination J2000 coordinates, (4) HR-COSMOS spectroscopic redshift, (5) morphological galaxy position angle, defined as the angle measured counterclockwise (East of North) between the North direction in the sky and the galaxy major axis, from Zurich Structure and Morphology catalog \citep{Sargent2007}, (6) inclination, defined as the angle between the line of sight and the normal to the plane of the galaxy ($i$ = 0 for face-on galaxies), from the \textit{HST}/ACS F814W images  \citep[Zurich Structure and Morphology catalog,][]{Sargent2007}, (7) characteristic radius 2.2 times the galaxy disc scale length \citep[Zurich Structure and Morphology catalog,][]{Scarlata2007}, (8) rotation velocity measured from the kinematic models at $R_{2.2}$, (9) velocity dispersion from the kinematic models, (10) ratio between $V_{2.2}$ and the velocity dispersion $\sigma$, (11) log of the stellar mass computed as described in $\S$ \ref{subsec:stellarmass}, (12) log of the dynamical mass.
\end{flushleft}
\end{longtab}

\subsection{Stellar Mass Tully-Fisher Relation at z$\,\sim\,$0.9} \label{subsec:smTF}
We present the stellar mass Tully-Fisher (smTF) relation obtained at $\mathrm{z\sim 0.9}$ in the COSMOS field. 
This is shown in Figure~\ref{fig:smTF} and takes the form:
\begin{equation} \label{eq:smTF}
\\ \\  \qquad \mathrm{logM_\ast}= a \,(\mathrm{logV_{2.2}-logV_{2.2, 0}}) + b \, ,
\end{equation}
where $a$ and $b$ are the slope and the y-intercept of the relation, respectively, and  $\mathrm{logV_{2.2, 0}}$ is chosen to be equal to 2.0 dex to minimize the correlation between the errors on $a$ and $b$ \citep{Tremaine2002}.
Since the smTF relation is known to be valid for rotating galaxies, we decided to fit the relation for the rotation-dominated sub-sample, shown in Figure \ref{fig:smTF} with black circles, whereas the dispersion-dominated galaxies are plotted as gray points. We decided not to include the two galaxies known to be NL AGNs in the fit of the smTF relation, since we cannot tell if the emission is dominated by the AGN or by the host. We note that, for those galaxies, the stellar mass measurements may not be correct since AGNs were not taken into account in the SED-fitting process. 
The relation in Eq. \ref{eq:smTF} is obtained using \mbox{MPFITEXY} routine \citep*{William2010}, which adopts a least-squares approach accounting for the uncertainties in both coordinates and incorporates the measurement of the intrinsic scatter $\sigma_{intr}$ on the velocity variable, added in quadrature to the overall error budget. The MPFITEXY routine depends on the MPFIT package \citep{Markwardt2009}.
Following previous analyses of the Tully–Fisher relation  \citep[see e.g.,][]{Verheijen2001, Pizagno2005, Pizagno2007, Miller2011, Reyes2011}, we fitted an \textit{inverse} linear regression to our data, where  velocity is treated as the dependent variable. The relation is then inverted to compare the fitted parameters to other works, which show the \textit{forward} best-fit parameters of the smTF relation. MPFITEXY automatically handles the inversion of the results and the propagation of errors.
Forward and Inverse fitting is only symmetric if there is no intrinsic scatter ($\sigma_{intr}$=0) \citep{Tremaine2002}, but generally this is not the case. 
It has been shown in previous works \citep[e.g.,][]{Willick1994, Weiner2006_2} that there is a significant bias in the slope of the forward best fit relation introduced by the sample selection limits. Therefore the inverse relationship is usually preferred, where the galaxy parameters that are more subject to the selection effects (magnitude, stellar mass) are treated as the independent variable.

We determined the 1-$\sigma$ errors on $a$, and $b$, by repeating the fit for 100 bootstrap sub-samples of 82 galaxies from the full rotation-dominated galaxy sample and taking the dispersion of the distribution of bootstrap estimated parameters as the uncertainty in the same parameters. 
Our best-fit parameters are shown in Table~\ref{table:smTFparam}.
\begin{table*}[!tb]
\begin{center}
\large
\caption{Stellar Mass Tully-Fisher parameters}             
\label{table:smTFparam}              
\begin{tabular}{c c c c c c c c} 
\hline\hline     
\noalign{\smallskip}   
$a$ & $b$ & $\sigma_{intr,\,\mathrm{logV}}$\tablefootmark{i} & $RMS_{\mathrm{logV}}$\tablefootmark{i} & $\delta_{med,\,\mathrm{logV}}$\tablefootmark{i} & $\sigma_{intr,\, \mathrm{logM_\ast}}$\tablefootmark{ii} & $RMS_{\mathrm{logM_\ast}}$\tablefootmark{ii} & $\delta_{med,\,\mathrm{logM_\ast}}$\tablefootmark{ii}\\ 
\hline         
\noalign{\smallskip}   
   {\small 3.683$\pm$0.787 }   &   {\small 2.154$\pm$0.147} & {\small 0.106 } & {\small 0.116} & {\small 0.043} & {\small 0.389}  &{\small  0.425} & {\small 0.043}\\
 \hline    
\end{tabular}
\end{center}
{\footnotesize Best-fit parameters of the smTF relation, expressed as $\mathrm{logM_\ast}= a \,(\mathrm{logV_{2.2}-logV_{2.2, 0}}) + b$, where $\mathrm{logV_{2.2, 0} =2.0\; dex}$ is the `pivot' value adopted to minimize the correlation between the errors on $a$ and $b$. The forward parameters are obtained by inverting the best-fit parameters from  the inverse fit. The slope of the relation is $a$ and $b$ is the y-intercept. $\sigma_{intr}$, $RMS$ and $\delta_{med}$ are the intrinsic scatter, the total scatter and the median error, respectively, for both the velocity and stellar mass in logaritmic units. \tablefoottext{i} { dex in V$_{2.2}$}, \tablefoottext{ii} { dex in M$_{\ast}$}. }
\end{table*}

To investigate the evolution of the smTF relation with redshift, we plotted the relations at $\mathrm{z=0}$ from \citet{Pizagno2005} and at $\mathrm{z\sim 0.6}$ from \citet{Puech2008}. A smTF relation in similar redshift range to our sample ($\mathrm{0.8<z<1.25}$) was presented by \citet{Miller2011} for 37 galaxies in the GOODS fields and we also plotted it for comparison. Our result shows no significant evolution with redshift and it is in good agreement with the result obtained by \citet{Miller2011}. We also fitted the smTF relation, fixing the slope to the $\mathrm{z=0}$ relation form \citet{Pizagno2005} and we quantified an offset of $\Delta M_\ast= - 0.1\,dex$ that is within 1-$\sigma$ error on the y-intercept and is consistent with the no significant evolution observed by \citet{Miller2011} ($\Delta M_\ast= -0.037\,dex$) and by the predictions from the cosmological simulations up to z=1 \citep{Portinari2007}. Conversely, \citet{Puech2008} found an offset of the smTF relation at $\mathrm{z\sim 0.6}$  equal to $\Delta M_\ast= - 0.36\,dex$ with respect to the local relation, suggesting an evolution of the smTF relation. Small differences in the slope that we measured compared to \citet{Pizagno2005} and  \citet{Puech2008} may arise from different adopted techniques used to compute galaxy stellar mass and, in particular, from the IMF assumption. Indeed, the former assumed a Kroupa IMF \citep{Kroupa2001} and the latter a ``diet-Salpeter'' IMF \citep{BelldeJong2001, Bell2003}. \citet{Miller2011}, conversely, assumed a Chabrier IMF, as we did.

A recent work on the TF relation at $\mathrm{z\sim 1}$ from the KROSS survey, using the multi-object IFU KMOS, has been presented by \citet{Tiley2016}. They constrained the smTF and K-band TF relations for a sample, denoted as ``disky'', of 56 rotation-dominated galaxies (out of 210 galaxies with well-measured rotation velocity) at $\mathrm{z=0.8-1.0}$, applying the strict selection criterion $\mathrm{V_{80}/\sigma>3}$. They computed stellar masses assuming Chabrier IMF. They found evidence of significant evolution of the smTF relation towards lower masses, $\Delta M_\ast= - 0.41\,dex$, while no evolution in the K-band TF relation. This result is very different from the one we found both on the slope (their relation has a slope of a value 4.7) and on the evidence of evolution of the relation, despite the fact that we also selected galaxies dominated by the rotation to constrain the smTF relation but with a less stringent cut ($\mathrm{V_{2.2}/\sigma>1}$). \citet{Tiley2016} stress the importance of their result, obtained by selecting  strictly rotation-dominated galaxies, to make the comparison with  TF relations for $z\sim0$ late-type galaxies that are dominated by the rotation. They argue that the difference with past works, which found no evolution of the smTF relation \citep{Conselice2005, Miller2011}, is due to the inclusion of galaxies with low ratios of rotation-to-pressure support. We find that the cut applied by \citet{Tiley2016} to keep galaxies only supported by the rotation is overly strict and proba\-bly it introduces a bias in their ``disky'' sample which removes all the galaxies with velocity smaller than $\mathrm{\sim 120 \; km/s}$ (except for one galaxy with 84 km/s) for stellar masses between $10^{8.9} \, M_\odot$ and $10^{11} \, M_\odot$. In our kinematic sample, conversely, 12 galaxies have $\mathrm{V_{2.2}< 120 \; km/s}$ with a minimum value of $\mathrm{64 \; km/s}$. Some of the galaxies used by these authors have been observed in the COSMOS field and we may have some targets in common but, unfortunately, we were not able to recover the coordinates of these galaxies to make a direct comparison.

Our smTF relation shows an intrinsic scatter of 0.106 dex in velocity. This is smaller, than the value obtained by \citeauthor{Puech2008} (0.12 dex), but it is larger than the scatter obtained by \citet{Pizagno2005} and \citet{Miller2011} ($\sim\,$0.05~dex for both). An important factor, which influences the scatter, is the measurement of the errors since $\sigma_{intr}$ is computed assuming that the uncertainty estimation is correct. Therefore the details of how the analysis is actually carried out have a very strong effect on the result. Another factor that definitely affects the scatter around the smTF relation is the sample selection.  \citet{Kannappan2002} have shown that the scatter around the TF relation increases by broadening their spiral sample, for which the relation is computed, to include all morphologies. They also argue that local studies tend to weed out kinematically irregular galaxies; therefore if a more representative sample of spirals is selected, the scatter could actually be much higher. At intermediate redshifts, small irregularities in rotation curves are harder to detect, owing to limited spatial resolution, and so a higher measured scatter might reasonably be expected \citep{Moran2007}. We emphasize that our sample was not subject to a pre-selection that aims to prefer some galaxy morphologies rather than others, and that our selection on the inclination and the $PA$ was driven only by the necessity to create  the observing conditions that favor the extraction of the kinematics information. Moreover, the selection based on the GINI parameter ($\S$ \ref{subsec:the_survey}) was  applied to a small fraction of our sample (17$\%$), for which we did not have prior spectroscopic information. In fact, the whole sample spans a wide range of GINI parameters from 0.29 to 0.60.
Conversely, \citet{Pizagno2005} applied a strict morphological cut in their sample, selecting galaxies with a disc-to-total luminosity ratio greater than 0.9, and \citet{Miller2011} made a morphological selection favoring disc-like structures, even though they attempted to avoid to favor ``well-behaved'' spirals and included a more morphologically disturbed population in their sample.
We conclude that we do not have the conditions to investigate the evolution of the scatter with redshift. Therefore, rather than comparing the scatter with other studies that have  different selections, we opt to internally compare our own data from the overall sample as a function of redshift in a future work.

\begin{figure*}[!htb]
   \centering
   \includegraphics[width=0.47 \hsize]{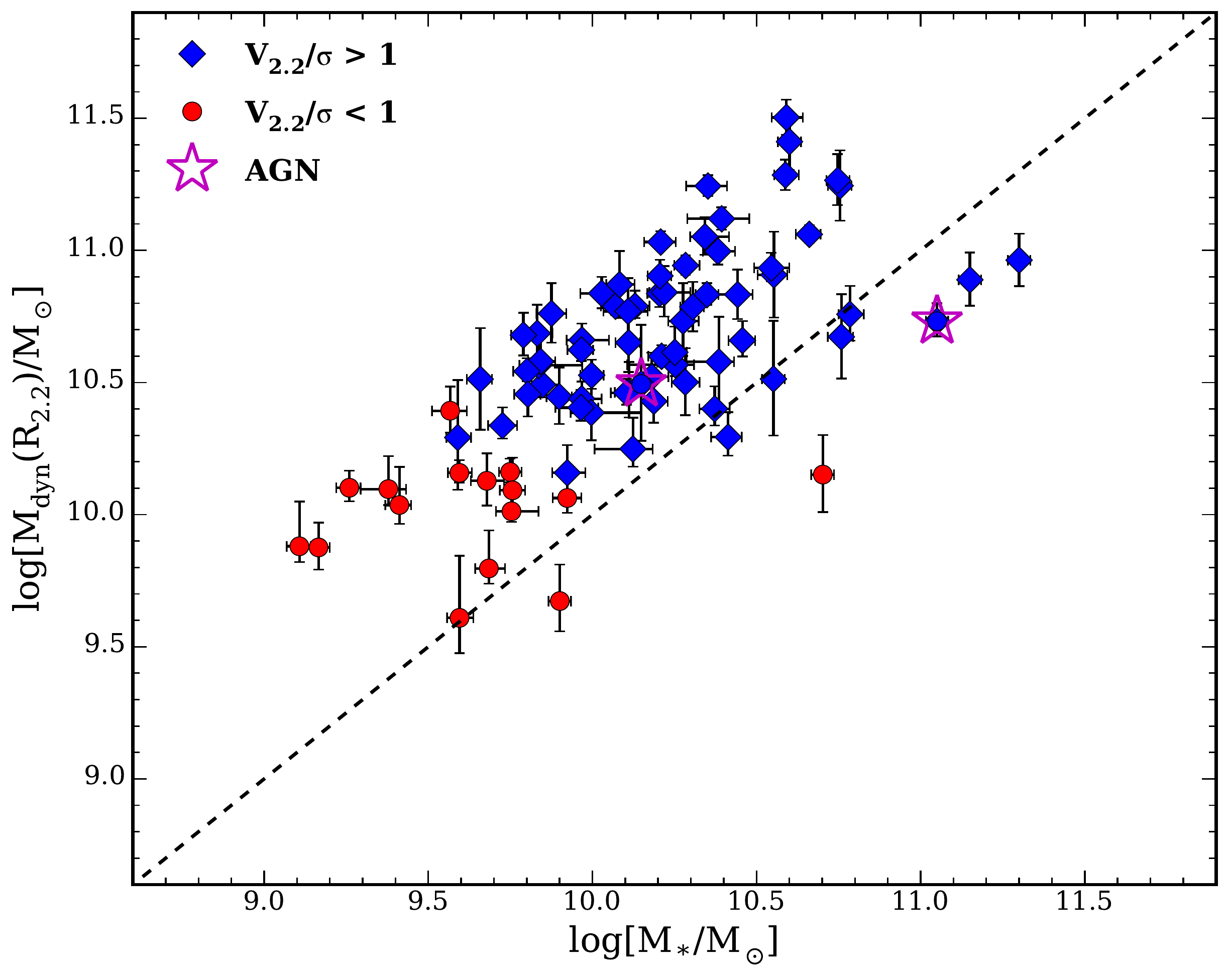}  $\;$
    \includegraphics[width=0.47 \hsize]{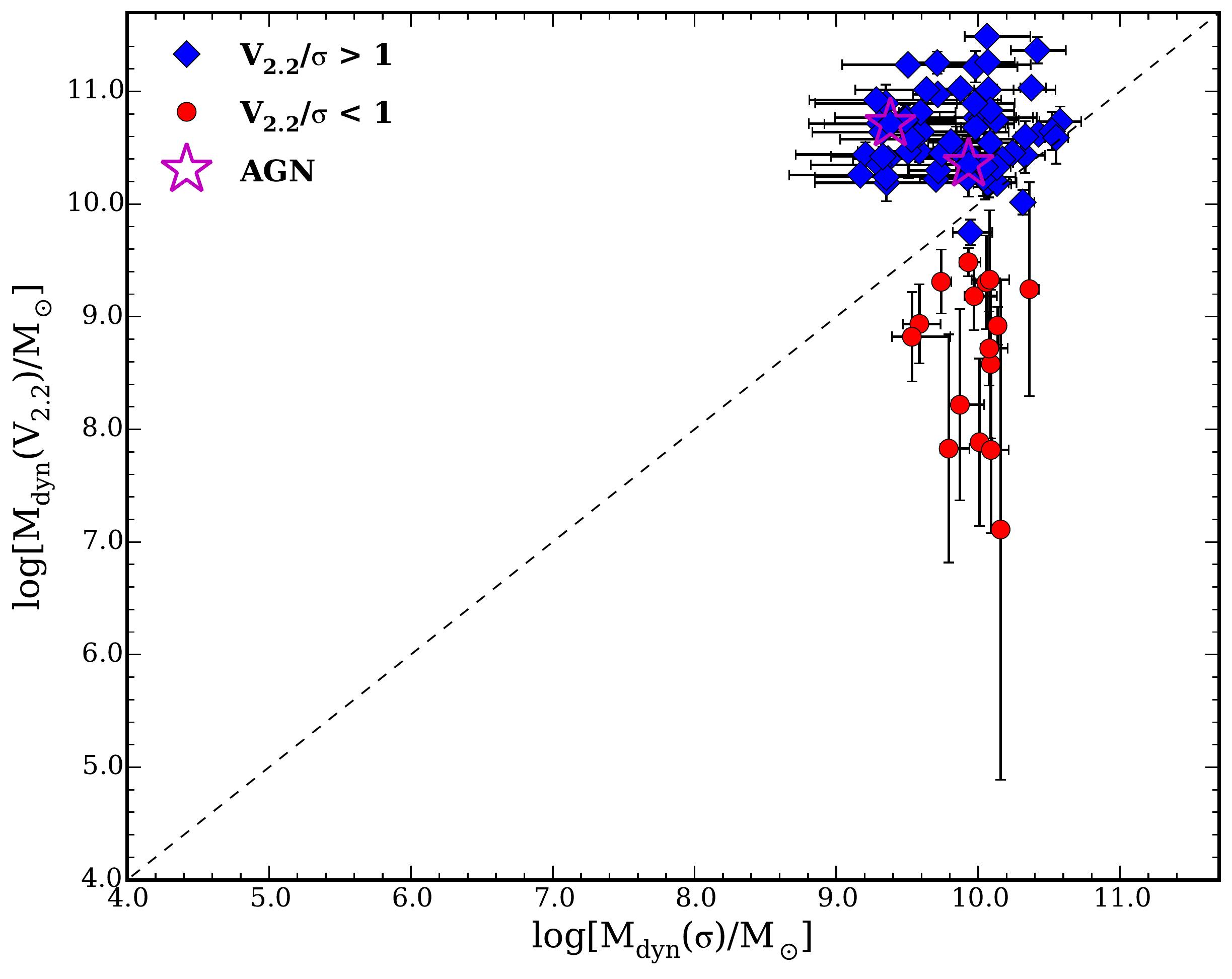}
      \caption{\textit{Left panel:} Comparison between dynamical ($y$-axis) and stellar ($x$-axis) masses in logarithmic units. Blue diamonds and red circles represent the rotation- and dispersion-dominated galaxies, respectively. The magenta stars show the AGNs. 
 The dashed line represents the relation 1:1.  \textit{Right panel:} Rotation versus dispersion contributions to the M$_{dyn}$. Blue diamonds and red circles are the rotation- and dispersion-dominated galaxies, respectively. The magenta stars show the AGNs. The dashed line represents the relation 1:1. }
              \label{fig:Mdyn_Mstar}
   \end{figure*}
Knowing that M$_\ast$ and V$_{2.2}$ are not measured at the same radius, to ensure consistency of our smTF relation and that the comparison with other works is fair, we repeated the fit of the relation following the technique adopted by \citet{Miller2011}.
In one case, we constrained the relation for stellar masses and velocities, both measured at the radius R$_{2.2}$. To that end, we applied an aperture correction to the stellar mass by estimating the fraction of light contained within R$_{2.2}$ for an exponential profile galaxy (see $\S$\ref{subsec:massdynamic}). In the other case, we derived the smTF relation using the total stellar mass of our galaxies and the velocity measured at the radius R$_{3.2}$, also called optical radius, which contains 83\% of the total integrated light and, for an exponential disc, is equal to 3.2 times the disc-scale length. In both cases, the smTF relation is in good ($\la2\sigma$) agreement with the relation presented in Figure \ref{fig:smTF} for the best-fit value of slope and $y$-intercept. We find, moreover, that the relation computed at the radius R$_{2.2}$ exhibit the same intrinsic scatter ($\sigma_{intr}=0.106$~dex) as our original relation, while the relation derived at R$_{3.2}$ shows a larger intrinsic scatter ($\sigma_{intr}=0.128$~dex), most likely due to the fact that the emission of most of our galaxies does not extend that far out and the measurements of the velocity at R$_{3.2}$ are the result of an extrapolation at that radius. We are therefore confident that the comparison with other works of the smTF relation derived in our study is not subject to bias introduced by the different radii at which M$_\ast$ and V$_{2.2}$ are measured.

\subsection{Dynamical mass measurements} \label{subsec:massdynamic}
In this section, we present the dynamical masses computed for our sample of star-forming galaxies at $\mathrm{0.75<z<1.2}$. The dynamical mass of disc galaxies is defined by the maximum rotation velocity following the formula
\begin{equation} \label{eq:Mdyn_1}
\\ \\  \qquad M_{dyn}(R)= \dfrac{V_{ROT}^2\,R}{G} \, ,
\end{equation}
where $R$ is the radius at which we measure the maximum veloci\-ty and $G$ is Newton's gravitational constant. However, this formula is valid for galaxies with total mass that is supported mainly by the rotation and, as we have shown in $\S$ \ref{subsec:kin_params}, our sample is 20$\%$ composed of dispersion-dominated galaxies, for which the measured slow rotation is not enough to dynamically support the galaxy mass. For these galaxies, to take into account the pressure support, we apply an ``asymmetric drift correction'' term \citep{Meurer1996},  which involves radial gradients of the gas surface density, the gaseous velocity dispersion, and the disc scale-height. We assumed that i) the galaxy kinematics is axisymmetric, ii) the gas velocity dispersion is isotropic, iii) the velocity dispersion and the scale-height of the galaxy disc are constant with radius, and iv) the gas surface density profile follows the exponential galaxy surface brightness, assumed in our models (Eq. \ref{eq:surface_bright}).
Adapting Eq. 2 from \citet{Meurer1996} for an exponential gas surface density profile, we computed the asymmetric drift correction, which leads to the following expression of the dynamical mass within R$_{2.2}$:
\begin{equation} \label{eq:Mdyn_2}
 M_{dyn}(R_{2.2})= M_{dyn}(V_{2.2}) + M_{dyn}(\sigma) =\dfrac{V_{2.2}^2\,R_{2.2}^{}}{G} + \dfrac{2.2\sigma^2R_{2.2}^{}}{G}.
\end{equation}
The two terms of Eq. \ref{eq:Mdyn_2}, M$_{dyn}(V_{2.2})$ and M$_{dyn}(\sigma)$, refer to the rotation and dispersion contribution to the total mass, respectively. The first term, M$_{dyn}(V_{2.2})$, represents the mass enclosed within the radius R$_{2.2}$, the second term, M$_{dyn}(\sigma)$,  represents the asymmetric drift correction at the same radius.
The dynamical mass of each galaxy is given in Col. (12) of Table \ref{tab:results}, the uncertainties are estimated by the propagation of the errors on V$_{2.2}$ and $\sigma$. The values range between $2.5\times 10^{9}\,M_{\odot}$ and $2.3\times 10^{10}\,M_{\odot}$ with a median value of $1.0\times 10^{10}\,M_{\odot}$ for the dispersion-dominated galaxies, and  between $1.3\times 10^{10}\,M_{\odot}$ and $3.1\times 10^{11}\,M_{\odot}$ with a median value of $4.5\times 10^{10}\,M_{\odot}$ for rotation-dominated galaxies.

In Figure \ref{fig:Mdyn_Mstar} (left panel), we present the comparison between the computed dynamical mass and the stellar mass ($\S$ \ref{subsec:stellarmass}). We find that our independent measurements of stellar and dynamical mass have highly significant correlation. The dynamical mass gives us a measure of the total mass of the galaxy within the radius R$_{2.2}$, including all its components (stars, gas, and dark matter). Therefore, as expected, M$_{dyn} (R_{2.2})$ is generally equal to or greater than M$_{\ast}$.  Within their uncertainties, a few galaxies have M$_{\ast}$ > M$_{dyn}(R_{2.2})$. 

Knowing that both masses, M$_{\ast}$ and M$_{dyn}$, are measured at different radii (M$_{\ast}$ is evaluated at larger radius), to make a comparison with previous works,  we computed an aperture correction by estimating the fraction of light contained within R$_{2.2}$ for an exponential profile galaxy. 
We found that 65\% of the total F814W light is included in R$_{2.2}$ and we have applied this  correction to M$_{\ast}$ measurements.
We computed the stellar-to-dynamical mass fraction within R$_{2.2}$ for our sample, and we found a median value of 0.2, which means that the contribution from gas+dark matter masses is 80$\%$ of the total dynamical mass within R$_{2.2}$.
Our stellar-to-dynamical mass fraction is consistent with the fraction measured by~\citet{Stott2016}  within R$_{2.2}$ for star-forming galaxies at $\mathrm{z=0.8-1.0}$ from the KROSS survey. They found that, on average, 78\% of the total mass within R$_{2.2}$ is composed of non-stellar material. \citet{Miller2011} obtained a value of stellar-to-dynamical mass fraction within R$_{2.2}$, across all their sample at $\mathrm{0.2 < z < 1.3}$, equal to~$\sim\,$0.3 with a considera\-ble scatter ($\mathrm{\sigma_{int}=0.25}$), while \citet{Wuyts2016} found a median fraction equal to 0.32 within the H-band half-light radius for a sample of  star-forming discs from the KMOS$^{\mathrm{3D}}$ survey, spanning a wide redshift range of $\mathrm{0.6 < z < 2.6}$.
 We note that the stellar-to-dynamical mass fraction is dependent on the IMF assumed when computing the stellar mass. Assuming, for exam\-ple, a Salpeter IMF \citep{Salpeter1955}, the stellar-to-dynamical mass fraction would have been equal to $\sim\,$0.33, since  galaxy stellar masses calculated assuming a Chabrier IMF are $\sim\,$0.6 times smaller than the ones calculated assuming a Salpeter IMF. For the comparison with previous works, the studies discussed here have computed stellar masses assuming Chabrier IMF. 

 Figure \ref{fig:Mdyn_Mstar} (right panel) presents the rotation versus the dispersion contributions to the M$_{dyn}$. 
  Here we see, even more clearly than in Figure \ref{fig:vrot_sigma}, the separation between rotation- and dispersion-dominated galaxies.

\subsection{SmTF relation as a function of the environment}  \label{subsec:environment}
Another important question we are trying to answer is: does the smTF relation have any dependence on the environment? Usual\-ly galaxy parameters vary according to the environment in which they are located \citep{Baldry2006, Bamford2009, Capak2007,  Sobral2011, Scoville2013}, hence we expect to observe this dependence reflecting in the galaxy scaling relations.
To investigate possible variations of the smTF relation we first defined environments by using the local surface density measurements by  \citet{Scoville2013} in the COSMOS field. They used the two-dimensional Voronoi tessellation technique \citep{Ebeling1993} to measure the local density ($\mathrm{\Sigma_{vor}}$) associated with each galaxy, and to map and visuali\-ze coherent the large-scale structure for a $K_S\leq24$ sample of $155\,954$ galaxies out to $\mathrm{z=3}$, using extremely well-constrained photometric redshifts. \citet{Darvish2015} have shown, with extensive simulations on evaluating the performance of  different density estimators, that the Voronoi tessellation technique (along with adaptive kernel smoothing technique) outperforms other methods and thus it should be considered as more reliable and robust than the widely-used nearest-neighbor (usually 5$^{th}$ or 10$^{th}$ NN) or count-in-cell (CC) techniques.

To attempt other environment measurements we searched in group catalogs available for the COSMOS field to determine if an appreciable number of galaxies in our sample had been classified as group members. We looked at a catalog of X-ray selected groups from \citet{George2011} and found that only one galaxy in our sample was defined as a member of a poor (N$_{\mathrm{mem}}$<3) group. The percentage of our selected galaxies that are classified as group members is 1$\%$, hence smaller than the 3$\%$ found for the galaxies part of the X-ray selected groups cata\-log at $\mathrm{0.75<z<1.2}$. Another group catalog was built by \citet{Knobel2012}, based on spectroscopic redshift measurements in the zCOSMOS-bright survey. We found that only 18 galaxies of our sample are present in their catalog, of which 13 have a significant  probability of being associated to a group. We note that the work of \citet{George2011} presents inconsistencies with \citet{Knobel2012}. Owing to a low number of our galaxies in group catalogs and to the inconsistency between these two works, we use the local density measurements to define the environment and we defer to future works to proper define other environment measurements.

\begin{figure}[!htb]
   \centering
   \includegraphics[width= 0.99\hsize]{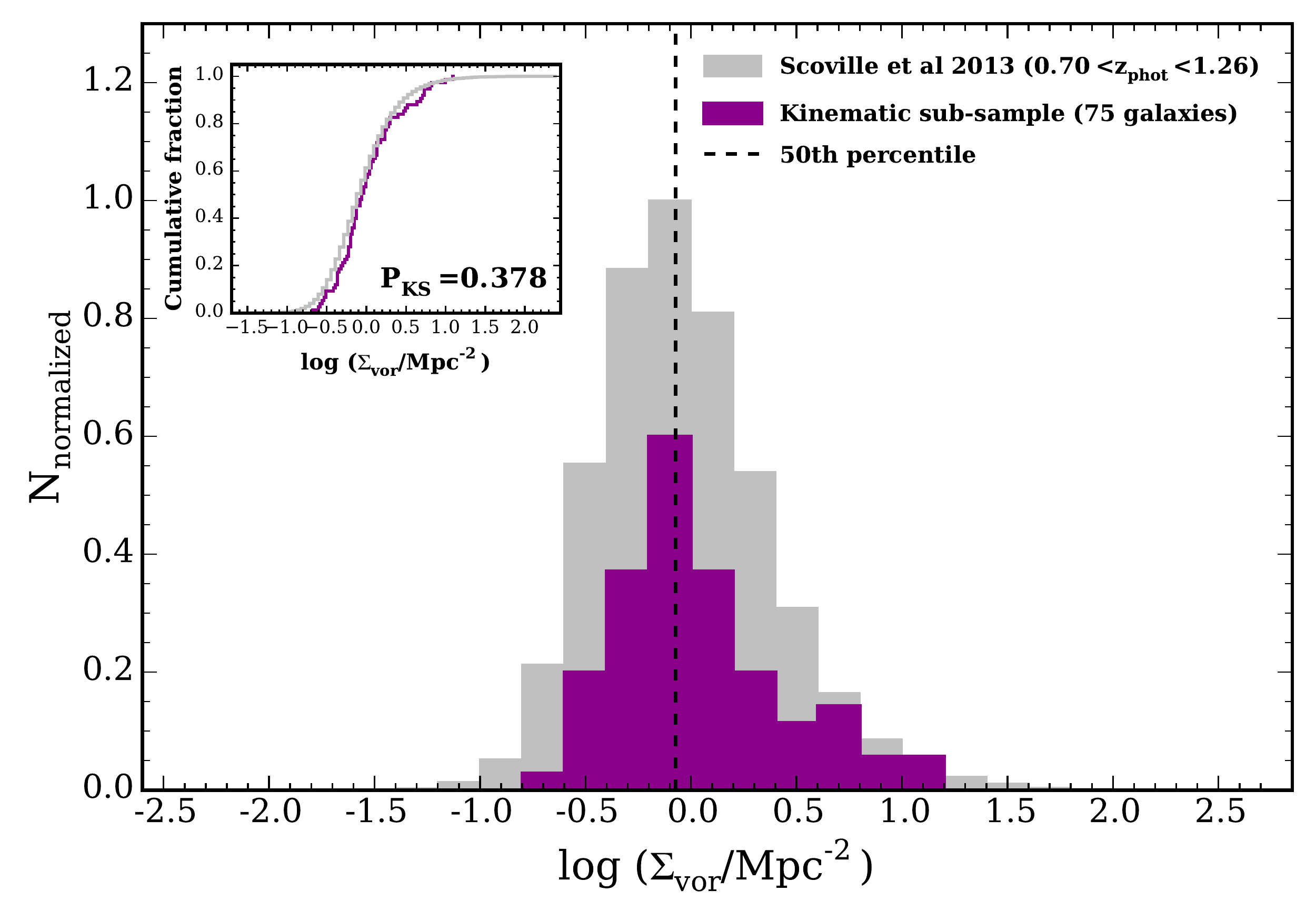}
      \caption{Local density distributions. The gray histogram represents the distribution of $\mathrm{log_{10}(\Sigma_{vor}/Mpc^{2})}$ for the parent galaxy sample from \citet{Scoville2013} at $\mathrm{0.75 < z < 1.2}$. The purple histogram shows the density distribution of our sample. Both histograms are re-normalized for a better visual comparison of their spreads and peaks, therefore their normalization does not reflect the actual scale. The vertical line splits in half the parent local density distribution, to define the low- (to the left) and the high- (to the right) density environments. The plot embedded shows the comparison of the two cumulative distributions of both samples with the resulting value from the K-S test.
}
\label{fig:dens_distr}
   \end{figure}

In Figure \ref{fig:dens_distr}, we show the density distribution of the parent galaxy sample from \citet{Scoville2013} at $\mathrm{0.75<z<1.2}$ and the density distribution of our galaxy sample. In the parent sample, seven of our galaxies do not have density measurements and, therefore, will not contribute to our investigation.  Since we are not interested in comparing the normalization of the distributions, we re-normalized the histograms to better visually compare their spreads and peaks (their normalization does not reflect the actual scale). To quantify any bias between the two samples we performed a \textit{Kolmogorov–Smirnov} (K-S) test.  We defined a null hypothesis that both samples are drawn from the same distribution and we rejected the null hypothesis if the \mbox{p-value} $\mathrm{P_{KS}}$ is below the value 0.05. We find that both samples are consistent with having been drawn from the same distribution ($\mathrm{P_{KS}=0.38}$). 

We defined low- and high-density environments such that the parent local density distribution in the redshift range of our interest was split in half. A first attempt to divide the overall distribution in three environmental density bins -- low, moderate, higher density -- have been made based on projected density quartiles and interquartile, but we have too low statistical significance in two of the three bins. Therefore we decided to increase the statistic significance and to consider  two  environment classes. The low-density environment has  values of $\mathrm{log_{10}(\Sigma_{vor}/Mpc^{2})}$ between $\mathrm{-0.678}$ and $\mathrm{-0.074}$ and the high-density environment has values between $- 0.074$  and $1.113$ $\mathrm{log_{10}(\Sigma_{vor}/Mpc^{2})}$.
As we see in Figure \ref{fig:dens_distr}, the local density distribution of our sample does not reach the lowest nor the highest value of the parent distribution. Our galaxies do not probe the least dense environments (void-like regions) in the COSMOS field nor the most dense ones typical to cluster galaxies, hence our sample is characterized by a local density distribution that goes from a slightly less dense environment than field galaxies to group galaxies.

 \begin{figure}[!htb]
   \centering
   \includegraphics[width= 0.79\hsize]{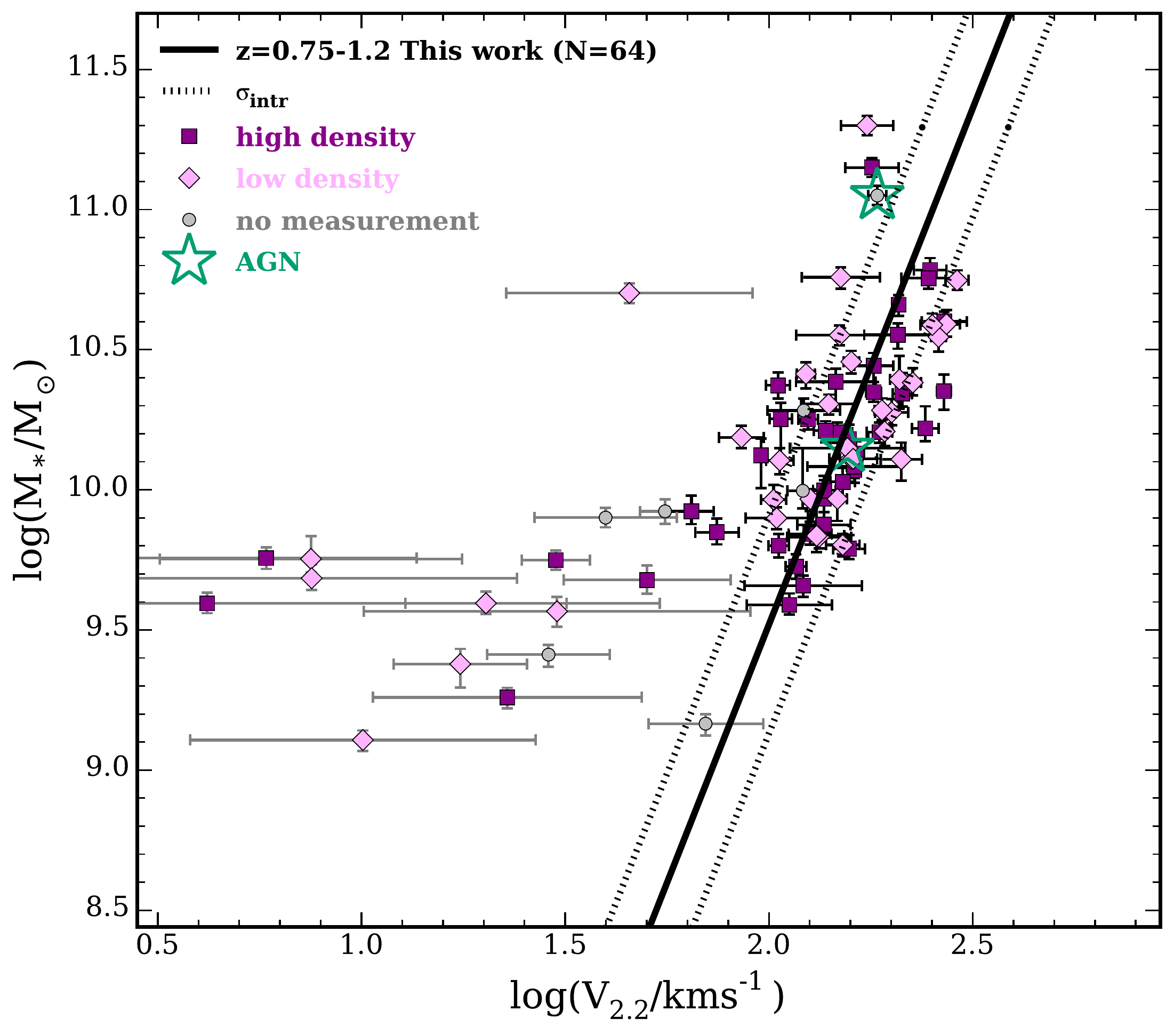}
    \includegraphics[width=0.79\hsize]{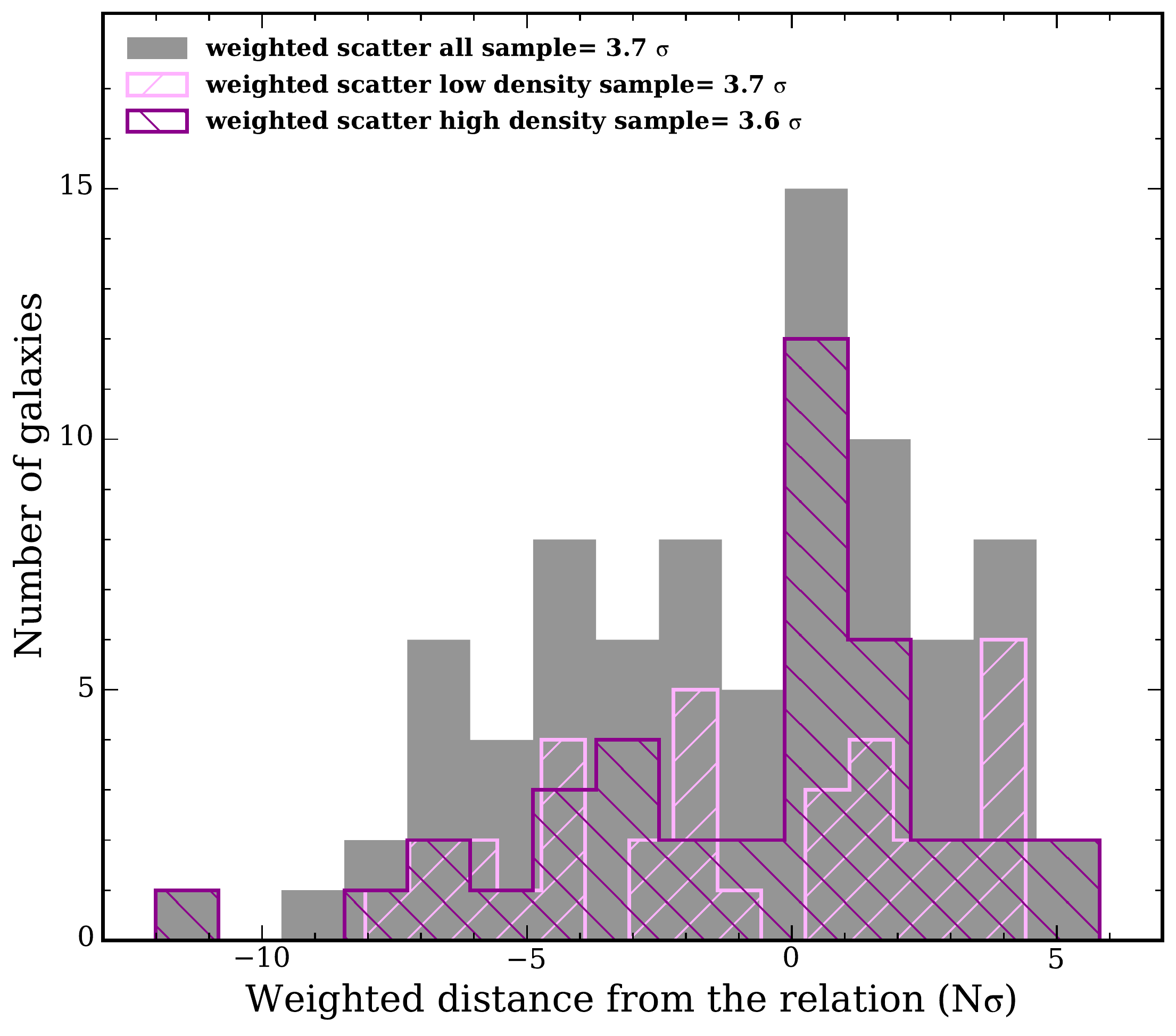}
   \includegraphics[width=0.79\hsize]{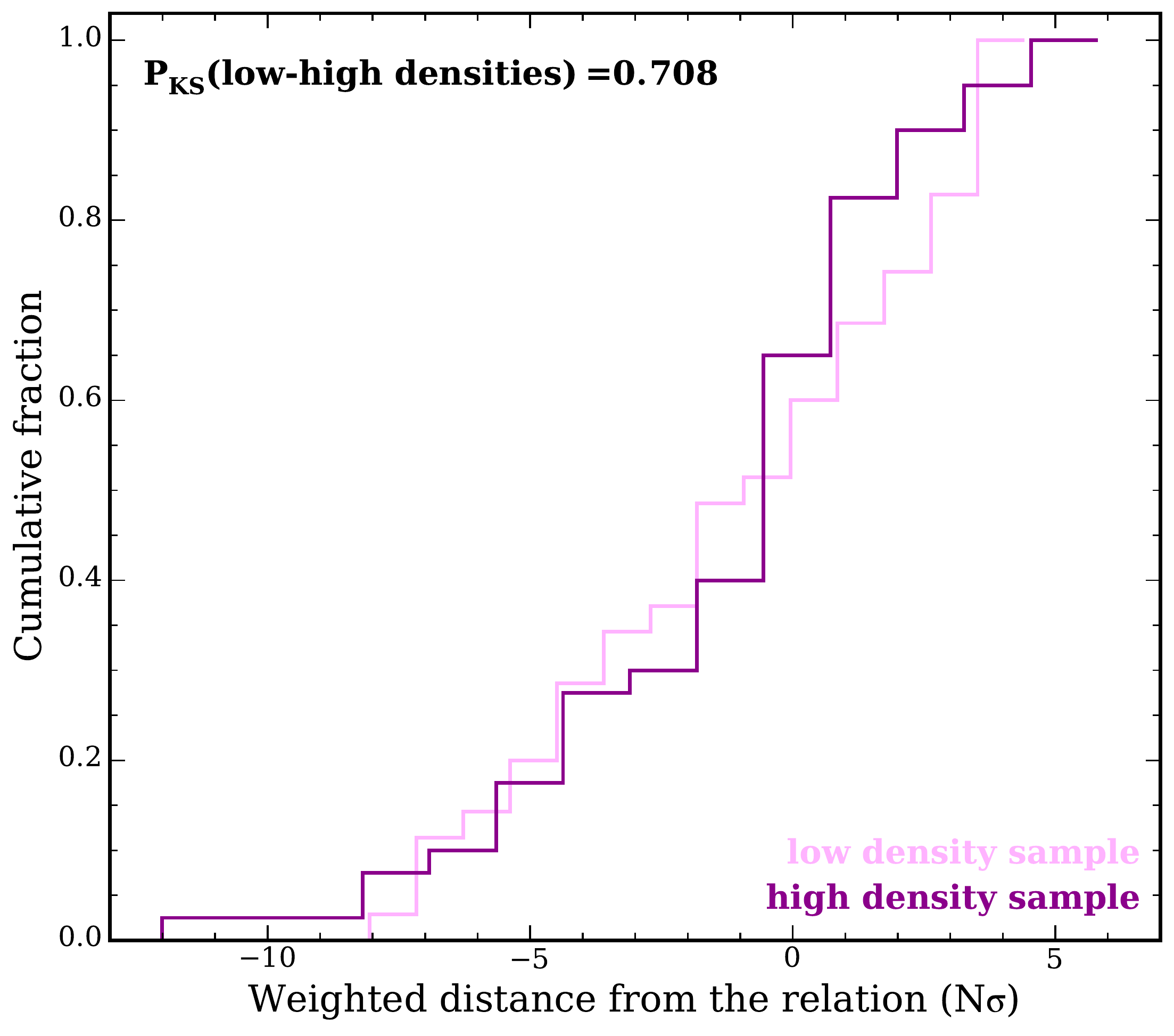}
      \caption{\textit{Top panel:} Stellar mass $M_{\ast}$ versus rotation velocity $V_{2.2}$ for our sample at $\mathrm{0.75<z<1.2}$ in different environments. The pink diamonds and the purple squares are for galaxies in lower and higher density environment, respectively. The gray circles are the galaxies for which we do not have density measurements. The green stars identify the AGNs. The black solid and dotted lines show our smTF relation with the intrinsic dispersion. \textit{Middle panel:} Distribution of the 1-$sigma$ error weighted distance $d_w$ of the data points from the relation for the entire sample in gray solid histogram, and for the galaxies in low- and high-density environments in pink and purple hashed histograms, respectively. The value of the weighted scatter around the relation for the whole sample and the two sub-samples is also shown.\textit{ Bottom panel:} Cumulative distribution of $d_w$ for low (pink) and high (purple) density environment and the result of the K-S test.}
         \label{fig:environment}
   \end{figure} 
We plotted  our smTF relation for galaxies in lower (pink diamonds) and higher (purple squares) density environments in the top panel of Figure~\ref{fig:environment}. Here we investigate the possibility that any correlation exists between the scatter of the galaxies around the relation (even for the dispersion-dominated ones) and their environment, in order to understand if the environment can influence the galaxies so that they deviate from the scaling relation.  Visually we do not detect any difference in the spread of the two galaxy sub-samples around the relation computed for the rotation-dominated galaxies ($\S$ \ref{subsec:smTF}). To properly quantify  the difference in scatter between the galaxies in both density environments, we define a quantity $d_w$ as the 1-$\sigma$ error weighted shortest distance to the relation between our data points and the smTF relation. We introduced this quantity to investigate environmental effect on both rotation velocity and stellar mass.  We explain in detail how $d_w$ is computed in Appendix \ref{appendixB}. Figure \ref{fig:environment} (middle panel) presents the distribution of the distances $d_w$ for the full sample and for both sub-samples at low- and high-density. We see that galaxies in the two environments have remarkably similar distributions to that for the full sample. In the bottom panel of Figure \ref{fig:environment}, we compare the cumulative distributions of the galaxies in  low- and high-density environments. A K-S test between them tells us that the probability that the two samples are consistent with having been drawn from the same distribution is $\sim$71$\%$. This implies no difference owing to the environment in the scatter around the smTF relation. 
Furthermore, we explored whether any effect of the environment exists in constraining the smTF relation and we again fitted the relation, separately, for rotation-dominated galaxies in low- and high-density environments, with the slope set to the value found for our whole rotation-dominated sample. We find no significant offset of the relation ($\Delta M_\ast=-0.04\,dex$ for the sub-sample at low-density, $\Delta M_\ast=+0.02\,dex$ for the sub-sample at high-density), which is consistent with no environmental effect.

\subsubsection{Previous studies}
In this study, we present our investigation of the dependence of the stellar mass Tully-Fisher relation with the environment at $\mathrm{z>0}$, for the first time, using local density measurements obtained with the 2D Voronoi tessellation technique that identified large scale structure in the $\sim$2 deg$^2$ COSMOS field. This approach for tracing the environment  makes use of the local number density of galaxies as a proxy for their host region. The conventional method to define the environment as two extreme regions in the density distribution of galaxies, i.e. galaxy cluster and general field, is an overly coarse binning of the full dynamical range of the density field at these redshifts. There are, indeed, intermediate environments, such as galaxy groups, outskirts of clusters, and filaments that are equally important  \citep{Fadda2008, Porter2008, Tran2009, Coppin2012,  Darvish2014, Darvish2015}. 
Previous works on constraining the B-band TF relation for cluster and field galaxies at $\mathrm{0<z<1}$ \citep[e.g.][]{Ziegler2003, Nakamura2006, Jaffe2011, Bosch2013} have found no dependence of the TF relation with environment, while other studies by \citet{Bamford2005} and \citet{Moran2007}  have found this kind of change. \citet{Bamford2005} derived a B-band TF relation for 58 field galaxies at $\mathrm{0\lesssim z\lesssim 1}$ and for 22 cluster galaxies at $\mathrm{0.3\lesssim z\lesssim 0.85}$ and have found that the cluster galaxies relation is offset from the one for the field galaxies by $\Delta M_B = 0.7\pm 0.2$ mag. \citet{Moran2007} computed the scatter around the K$_{s}$-band (proxy of the stellar mass) and V-band TF relations at $\mathrm{0.3\lesssim z\lesssim 0.65}$  for 40 cluster galaxies and 37 field galaxies, showing that cluster galaxies are more scattered than the field ones.
As already pointed out, the most dense environment that our galaxies sampled is a group-scale one, therefore we cannot assert  agreement or disagreement with those previous works, since they focus on comparison between field and cluster galaxies. \citet{Jaffe2011} included group galaxies in their sample and have found no correlation between the scatter around the B-band TF and the environment, nevertheless they did not explore the environmental effect on the smTF relation. 

In the local Universe ($\mathrm{z\ll 0.1}$), \citet{Torres-Flores2010, Torres-Flores2013} compared the kinematics of a galaxy sample composed of nearby groups from the sample of \citet{Hickson1993} and concluded that the behaviour of most compact group galaxies on the different TF relations (B-band, K-band, stellar and
baryonic) do not fundamentally differs from those shown by field galaxies. They argue, however, that the larger scatter around the relation for the compact group galaxies  and the presence of some outliers indicate subtle, but intrinsic, differences between compact group and field galaxies, which may be linked to transient evolutionary phenomena owing to the dense environment of compact group of galaxies.
We expect, therefore, that a galaxy that lies in a dense environment would be affected by the gravitational potential created by the surrounding galaxies and have its gas and its kinematics disturbed. 
In our high-density sub-sample, we find no such evidence of kinematic disturbances relative to the low-density sub-sample. This may be an effect of the limited dynamic range of densities probed by our sample and by the COSMOS field as a whole, an effect of combining galaxies in field and group environments in the high-density bin, a true similarity between the two populations, or some combination of the three.
Within nearby cluster galaxies, it has been shown by \citet{Amram1993} that the kinematics of cluster members is not perturbed until the optical radius, while \citet{Torres-Flores2014} have found that in structures only as dense as local compact groups, a high fraction of galaxies do present significant kinematic disturbance. \citet{Jaffe2011}, at \mbox{0.36 < z < 0.75}, also computed the fraction of kinematically disturbed  galaxies $\mathrm{f_{K}}$ in field, group and cluster environments and they have obtained a higher $\mathrm{f_{K}}$ for cluster galaxies than for field and group ones.

We conclude that a more complete study, extending to more dense environments, is needed to better understand the effect of the environment on the galaxy kinematics.

\section{Conclusions}  \label{sec:conclusion}
We have presented our new survey HR-COSMOS aimed to obtain the first statistical and representative sample to study the kinematics of star-forming galaxies in the COSMOS field at intermediate redshift.
The COSMOS field is a unique treasury of multi-wavelength photometric information, morphological and spectroscopic parameters highly optimized for the study of galaxy evolution and environments in the early Universe.
We collected a large sample (766 galaxies) of morphologically blind galaxies at $\mathrm{0<z<1.2}$, observed with the multi-slit spectrograph ESO-VLT/VIMOS in HR mode. We have aligned each spectral slit along the galaxy major axis, using the information from the high spatial resolution \textit{HST}/ACS F814W imaging data. 
In this paper, we focused our analysis on the galaxies in the highest redshift range, $\mathrm{0.75<z<1.2}$. We estimated stellar masses making use of the latest COSMOS photometric catalog \citep{Laigle2016} derived from the deep-ground and space-based imaging in 30 broad, intermediate, and narrow bands, including the latest data releases from UltraVISTA and \textit{Spitzer}. 
We modeled the rotation velocity of the galaxies taking into account the instrumental contribution to the observations, such as the seeing and the wavelength resolution. We obtained the rotation velocity at the characteristic radius R$_{2.2}$ to constrain the stellar mass Tully-Fisher relation at $\mathrm{z\sim 0.9}$. Our main results are summarized below.
\begin{itemize}
\item We obtained velocity rotation measurements of 82 gala\-xies at $\mathrm{z\simeq 0.9}$, of which 80$\%$ being rotation-dominated and 20$\%$ dispersion-dominated. For the rotation-dominated galaxies, thanks to the relatively large number of galaxies in our sample, we were able to fit the smTF relation without the necessity to set the slope to a local relation, as has often been done in previous works. \\

\item  Our smTF best-fit parameters (\mbox{slope=3.68$\pm$0.79}, \mbox{y-intercept=2.15$\pm$0.15}) are formally consistent with previous  works in a similar redshift range, like \citet{Miller2011} with spectra obtained with the spectrograph Keck/DEIMOS in the GOODS fields and \citet{DiTeodoro2016}, who use 3D spectra of a small selected sample of galaxies from ESO-VLT/KMOS. However, our results largely differ from \citet{Tiley2016}, who constrained the smTF relation for a sample of rotation-dominated galaxies ($\mathrm{V_{80}/\sigma>3}$) from the KROSS survey.\\

\item No apparent evolution of the smTF relation with redshift was detected when comparing our relation to previous works at lower redshift, such as \citet{Pizagno2005} at $\mathrm{z=0}$ and \citet{Puech2008} at $\mathrm{z\sim 0.6}$. We also fitted the smTF relation setting the slope to the $\mathrm{z=0}$ relation form \citet{Pizagno2005} and we found an offset of $\Delta M_\ast=-0.1\,dex$ that is within 1-$\sigma$ error on the y-intercept and consistent with no significant evolution predictions from cosmological simulations \citep{Portinari2007}. 
Nevertheless, since the sample selection and the details of how the analysis is actually carried out have a  strong effect on the results, we will extend our kinematic study to the full sample over $\mathrm{0<z<1.2}$ to investigate, in a consistent manner, any possible evolution of the smTF relation with redshift.\\

\item The scatter around our smTF relation is smaller than the one obtained by \citet{Puech2008} at $\mathrm{z\sim 0.6}$, and larger than the scatter obtained by \citet{Pizagno2005} at $\mathrm{z=0}$ and \citet{Miller2011} at $\mathrm{z\sim 0.9}$. We argue that the sample selection plays an important role in the determination of the scatter around the smTF relation, and therefore an internal comparison of the scatter   within our own full sample at $\mathrm{0<z<1.2}$ will be presented in a future work.\\

\item We presented our dynamical mass measurements within the radius R$_{2.2}$ for both rotation- and dispersion-dominated galaxi\-es. To account for the fact that some galaxies are not dynamically supported by the rotation, we introduced an ``asymmetric drift correction'' term \citep{Meurer1996}, which adds a pressure contribution to the dynamical mass. In comparing the dynamical and stellar masses within R$_{2.2}$ at $\mathrm{z\simeq 0.9}$, we find a median stellar-to-dynamical mass ratio equal to 0.20 (for a Chabrier IMF), which means that gas+dark matter masses contribute for  80$\%$ of the total mass. 
Our result is consistent with the stellar-to-dynamical mass fraction measured by~\citet{Stott2016}  within R$_{2.2}$ for star-forming galaxies from the KROSS survey at $\mathrm{z=0.8-1.0}$, and by \citet{Miller2011} within R$_{2.2}$ across their full sample at $\mathrm{0.2 < z < 1.3}$. \\

\item We finally presented our investigation of the dependence of the smTF relation with the environment at $\mathrm{z > 0}$, for the first time using the local surface density measurement by \citet{Scoville2013}, obtained with the 2D Voronoi tessellation technique. We have set lower density and higher density environments for the our sample (dispersion-dominated galaxies included) and we computed the scatter around the relation for galaxies in the two environmental bins. We find no dependence of the scatter on environment. We explored whether any effect of the environment exists in constraining the smTF relation and we again fitted the relation, separately, for rotation-dominated galaxies in low- and high-density environments. Still we found no significant offset of the relation consistent with no environmental effect. We note that our sample does not probe the extremes of the COSMOS local density distribution, hence our galaxies in the high-density environmental bin are most likely sitting in a group scale environment. We argue that the gravitational potential created by a group scale environment is possibly not strong enough to kinematically perturb the gas in galaxies. Extending this kinematic study to galaxies in denser environment (cluster-like), is needed to better explore the effect of the environment on the kinematics.
\end{itemize}

\noindent Our analysis demonstrates the power of systematic multi-slit spectroscopy over deep redshift surveys in treasury fields like COSMOS. Even though recent integral field spectroscopy combined with adaptive optics (AO) facilities enable us to retrieve precise 3D velocity map of galaxies, these facilities remain limited in terms of galaxy selection and they require huge amount of telescope time to acquire large representative samples. Multi-slit spectroscopy remains a good complementary alternative to study kinematics of galaxies over deep cosmological fields allowing us to observe, in the case of VIMOS, $\sim$120 galaxies per exposure in a wide field of view, $4\times \, 7\arcmin \times 8\arcmin$, much larger than the $\sim 2\arcmin \times 2\arcmin$ and $1\arcmin \times 1\arcmin$ fields of view provided by the last generation of IFUs as KMOS and MUSE, respectively. 

\begin{acknowledgements}
We wish to thank the ESO staff at Paranal Observatory and especially the VIMOS team at ESO for their support during observations. We thank Bodo Ziegler for useful discussion, comments, and suggestions during the preparation of the observations and Katarina Kova\v{c}  for kindly providing us with their environment measurements. D.P. and L.T. wish to thank Alisson Michel and Bruno Ribeiro for their help with programming codes. We thank Marco Scodeggio for his assistance during the preparation of the observations and the data reduction. D.P. gratefully acknowle\-dges Lori Lubin and UC Davis hospitality during the last phases of the project. We thank Alfred Tiley for providing us with their non-public measurements.
This paper is based on data products from observations made with ESO Telescopes at the La Silla Paranal Observatory under ESO programme ID 179.A-2005 and on
data products produced by TERAPIX and the Cambridge Astronomy Survey
Unit on behalf of the UltraVISTA consortium.
\end{acknowledgements}
\bibliographystyle{aa} 
\bibliography{DP_LT_et_al2016} 

\appendix
\section{Uncertainties on the axial ratio \textit{b/a}} \label{appendix_incl}
The adopted $b/a$ is included in the Zurich Structure and Morphology Catalog and was measured with a software (GIM2D, see $\S$\ref{subsec:the_survey}) that gives morphological measurements corrected for the instrumental PSF. We used the axial ratio to derive the inclination of the galaxies of our sub-sample at z$\, \sim \,$0.9. 

\begin{figure}[!htb]
   \centering
    \includegraphics[width=0.99 \hsize]{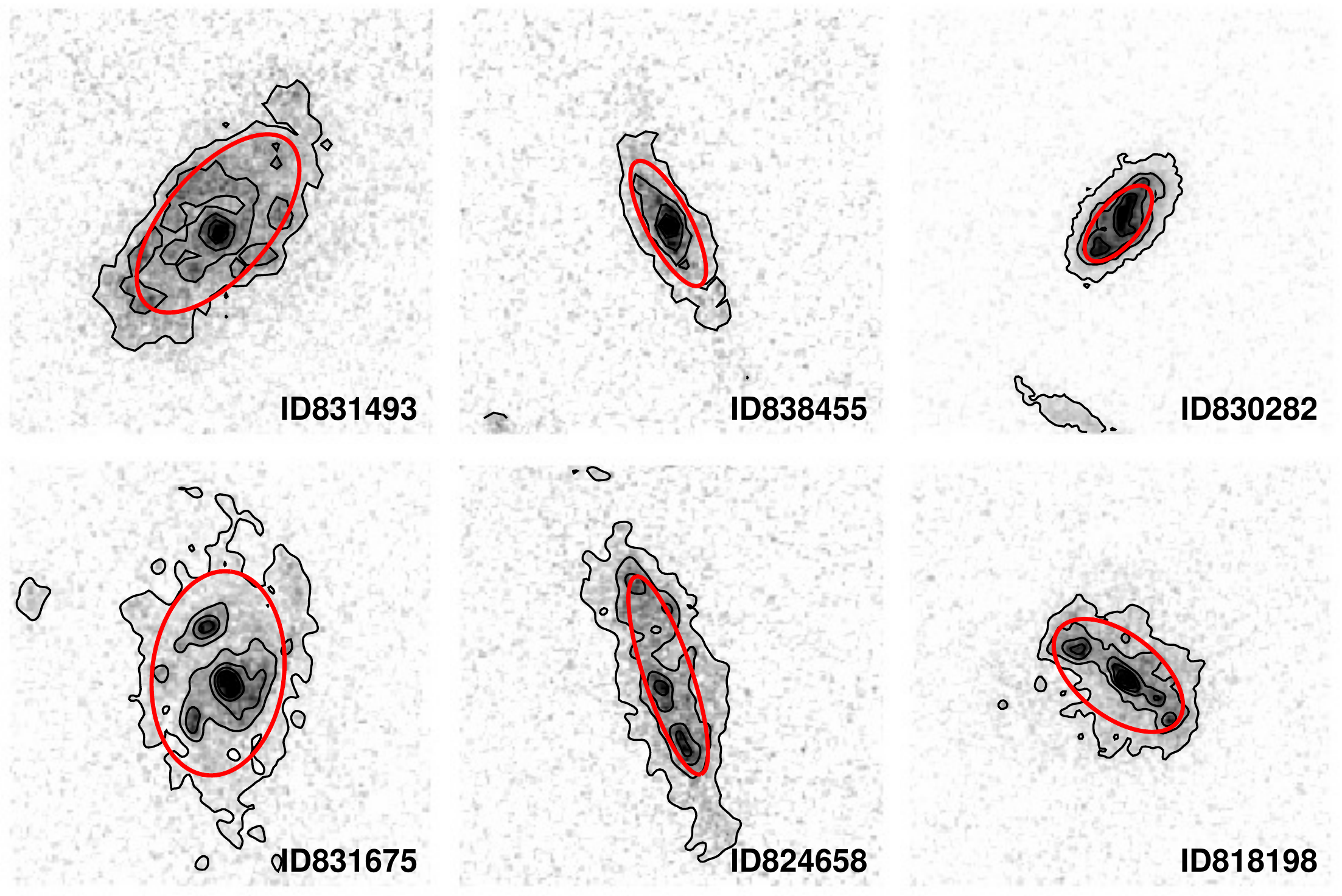}
      \caption{Example of the visual check of the correctness of the inclination adopted in our kinematic analysis, by placing ellipses (in red) with the axial ratio $b/a$ from the Zurich Structure and Morphology Catalog on \textit{HST}/ACS F814W galaxy images. The contours begin at 1-$\sigma$ and continue with 2$\sigma$ step. }
              \label{fig:ellipses}
   \end{figure}

The inclination is an important parameter in our kinematic analysis since it scales the observed velocity to obtain the galaxy intrinsic velocity rotation (see Equation $\S$\ref{eq:velocity_short}). Therefore, we visually checked the correctness of this parameter by overlapping  an ellipse with the adopted $b/a$ on the \textit{HST}/ACS F814W image of each galaxy. We plot this visualization for a randomly chosen subset of six galaxies drawn from our main sample in Figure \ref{fig:ellipses}.
 Since  the GIM2D measurements in the catalog only included the value of the ellipticity ($1- b/a$), but not those of $a$ and $b$, we adopted as major axis a quantity called $r80$ from the Zurich Structure and Morphology Catalog. This quantity is defined as the semi-major axis length of an ellipse encompassing 80\% of total light, with the adopted minor axis being, therefore, equal to $r80$ scaled by $b/a$.
This inspection allowed us to ensure that the inclinations employed in our kinematic analysis are reasonably measured and that they do not bias the rotation velocity derived. 

\begin{figure}[!htb]
   \centering
    \includegraphics[width=0.99\hsize]{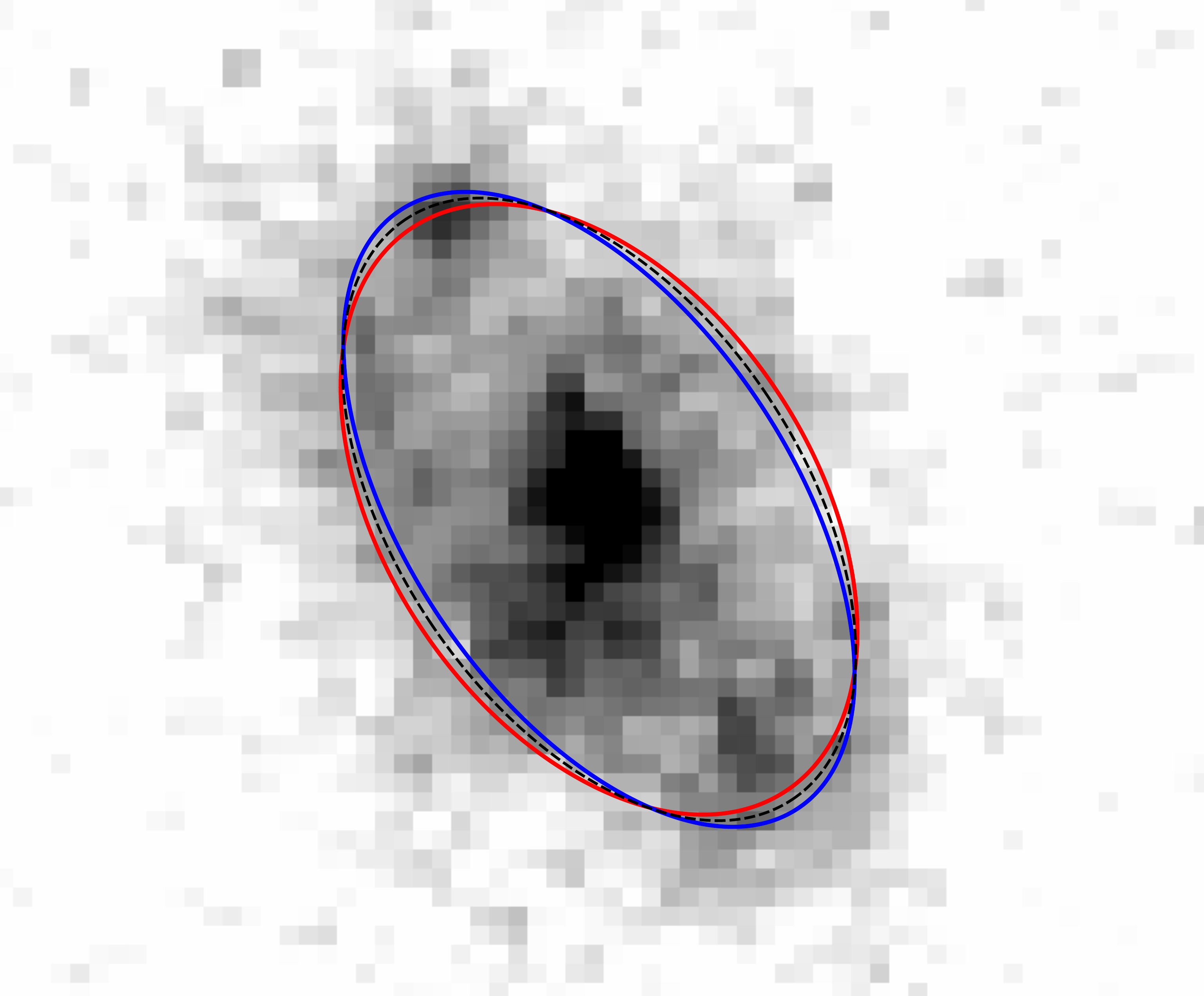} $\;$
      \caption{Example of the ellipse that overlaps the \textit{HST}/ACS image with $b/a$ as measured in the Zurich Structure and Morphology Catalog (black dashed ellipse) and as it appears using the lower (red solid ellipse) and upper (blue solid ellipse) value of $b/a$ measured  in the Equation \ref{eq:b_a_minmax}.  }
              \label{fig:ellipse_minmax}
   \end{figure}

However, systematic uncertainties on the axial ratio measurements due to  small variations of the instrument PSF must be taken into account in the velocity error budget. To that end, following \citep{Epinat2009}, we estimated upper and lower limits for the axial ratio $b/a$. As before, we adopted $a = r_{80}$ and $b = r_{80} \times b/a$. We obtain that $b/a$, as a result of the small variation of the PSF, which we assume to be the same for both major and minor axes and equal to one third of the \textit{HST}/ACS PSF (where the ACS PSF is $\sim\,$0.1$\arcsec$), can vary in the following range:
\begin{equation}\label{eq:b_a_minmax}
\qquad \qquad \qquad\qquad\dfrac{b-\Delta}{a+\Delta}\leq\frac{b}{a}\leq\dfrac{b+\Delta}{a-\Delta}
\end{equation}
with $\Delta=0.03$. In Figure \ref{fig:ellipse_minmax}, we show an example of how the ellipse  overlapped to the \textit{HST}/ACS image changes when using the upper and lower limit of the axial ratio.
   
We quantified that the relative uncertainties coming from the small variations of the instrument PSF go from a minimum of 6\% to a maximum of 30\%. These uncertainties are propagated to the inclination and the rotation velocity, the values of which are presented in the Table \ref{tab:results}.

\label{appendix}
\section{Model of the galaxy surface brightness} \label{appendixA}
\begin{figure}[!htb]
   \centering
    \includegraphics[width=0.98 \hsize]{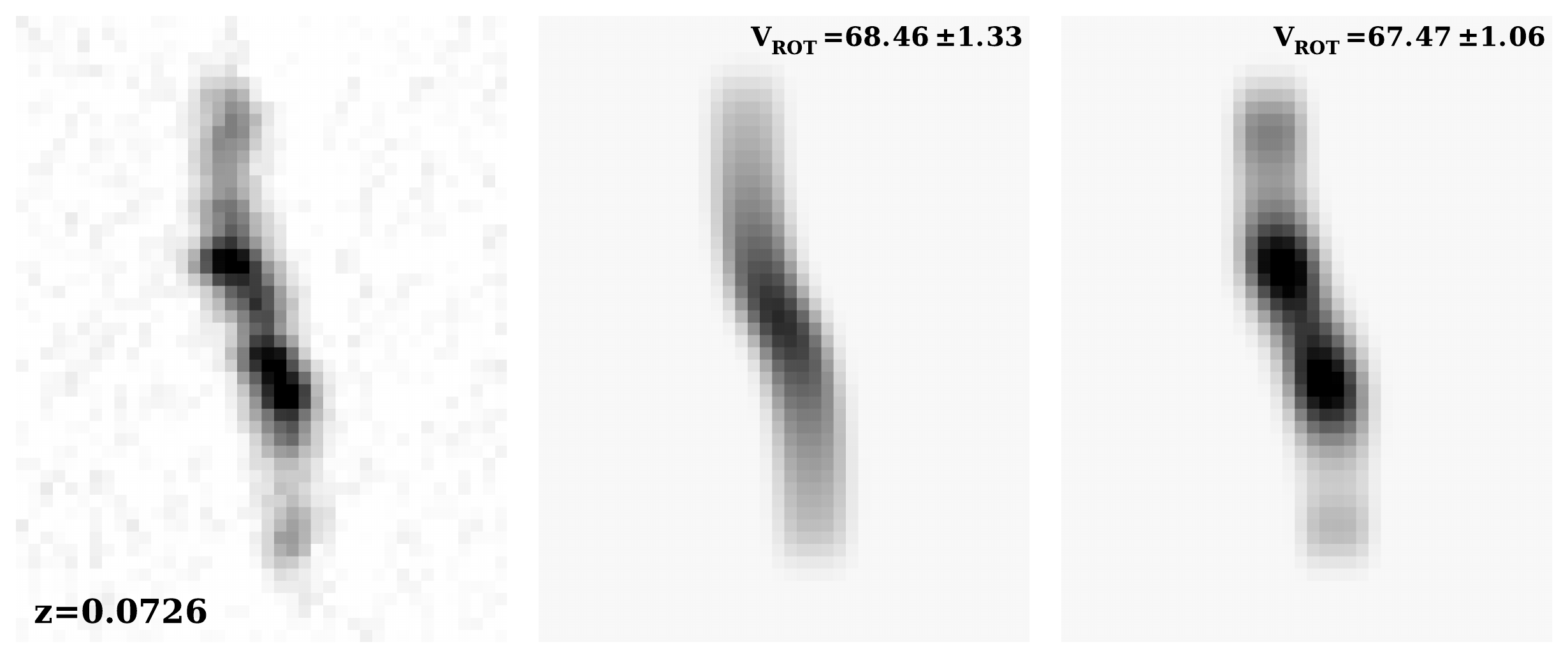}
    \includegraphics[width=0.98 \hsize]{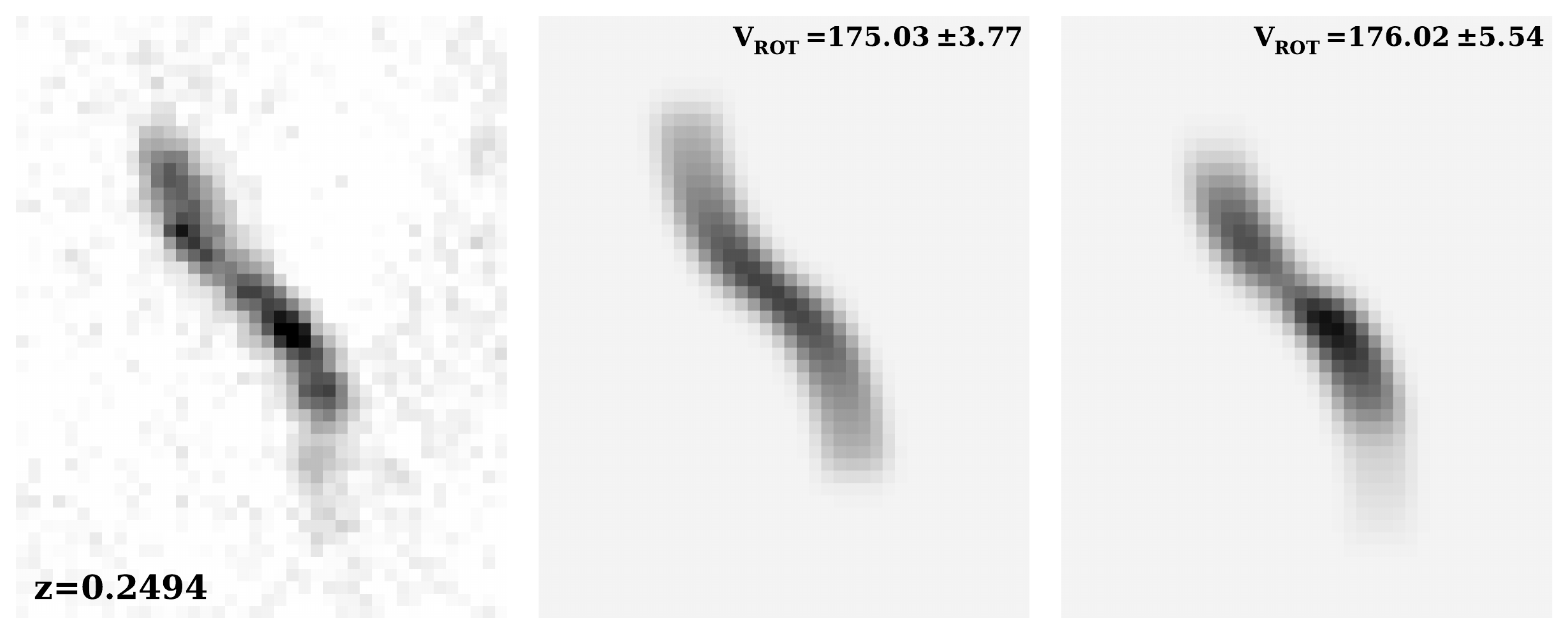}
    \includegraphics[width=0.98 \hsize]{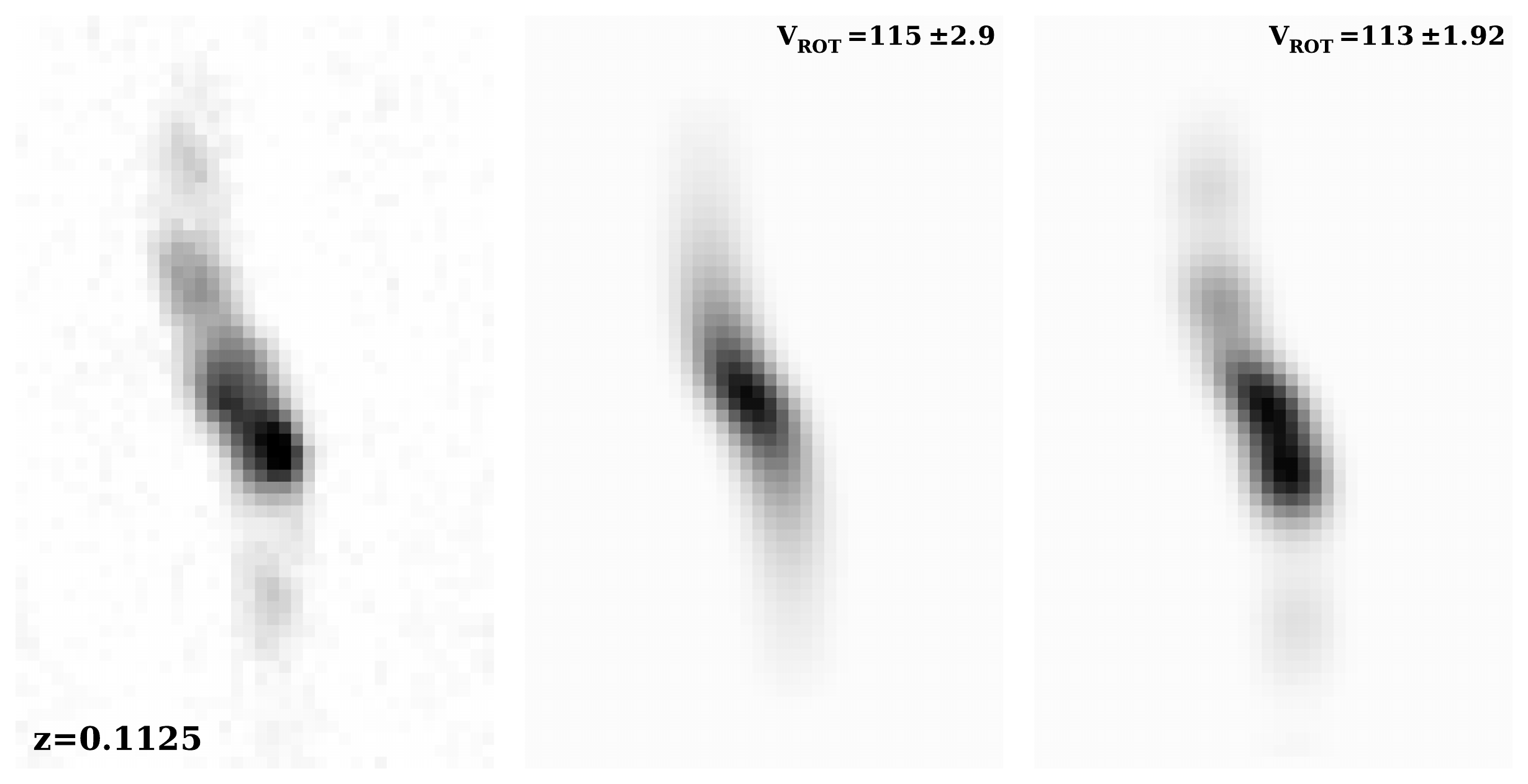}
      \caption{Comparison between two modeling techniques for three cases with the H$\alpha$ emission line observed at z=0.0726, z=0.2494 and z=0.1125.
       \textit{Left panels:} Continuum subtracted 2D spectra centered at H$\alpha$. \textit{Middle panels:} Best-fit models to the emission line with an exponential disc light profile. \textit{Right panels:} Best-fit models to the emission line with the light profile estimated from the algorithm from \citet*{Scoville1983}. The measurement of the velocity best-fit parameter within its uncertainties is given in the upper right corners. The images scale is: 1pixel=0.205$\arcsec$ in the spatial direction and 1pixel=0.6 $\AA$ in the spectral direction.
              }
              \label{fig:comp_nickcode}
   \end{figure}
We showed in $\S$\ref{sec:results} that our models are not always able to reproduce the observed spectra light profile. These models are particularly ineffective at high redshift where galaxies are known to exhibit clumpy star formation. 

Here we test the effect of our assumption about exponential disc profile on our derived rotation velocities.  To that end, we used an algorithm from \citet*{Scoville1983} (see Section 3 therein for a detailed description), which derives the spatial line emission distribution (even non-axisymmetric) at high-resolution that best matches the observed emission line profile. 
Since the algorithm was not implemented to work with doublet emission lines, we made our test on galaxies with  H$\alpha$ emission line spectra from our sample at lower redshift. For those galaxies we fitted the observed spectrum with kinematic models obtained with both an exponential surface brightness profile and the profile derived by the algorithm. The comparison between the two best-fit models for three cases at redshift z=0.0726, z=0.2494 and z=0.1125 is shown in Figure \ref{fig:comp_nickcode}.

We find that the models with the light profile derived with the algorithm from \citet*{Scoville1983} (right panels) reproduce much better the observed emission line than the models obtained with the exponential profile (middle panels). If we compare the rotation velocity $\mathrm{V_{ROT}}$ estimated with both models (values given in the upper right corners with their associated uncertainties), we find that they are always consistent within the uncertainties. We claim, therefore, that there is no effect from the modeled surface brightness profile in the derivation of the galaxy rotation velocity.

\section{[O{\small II}] doublet measurements} \label{appendixC}
We provide our measurements of  the ratio $\mathrm{R_{[O{\small II}]}}$, the rest-frame EW, and the SFR derived from both SED-fitting and the \mbox{[O{\small II}] EW}. They are listed in Table \ref{tab:oii_measurements}.
\begin{figure}[!htb]
   \centering
    \includegraphics[width=0.9 \hsize]{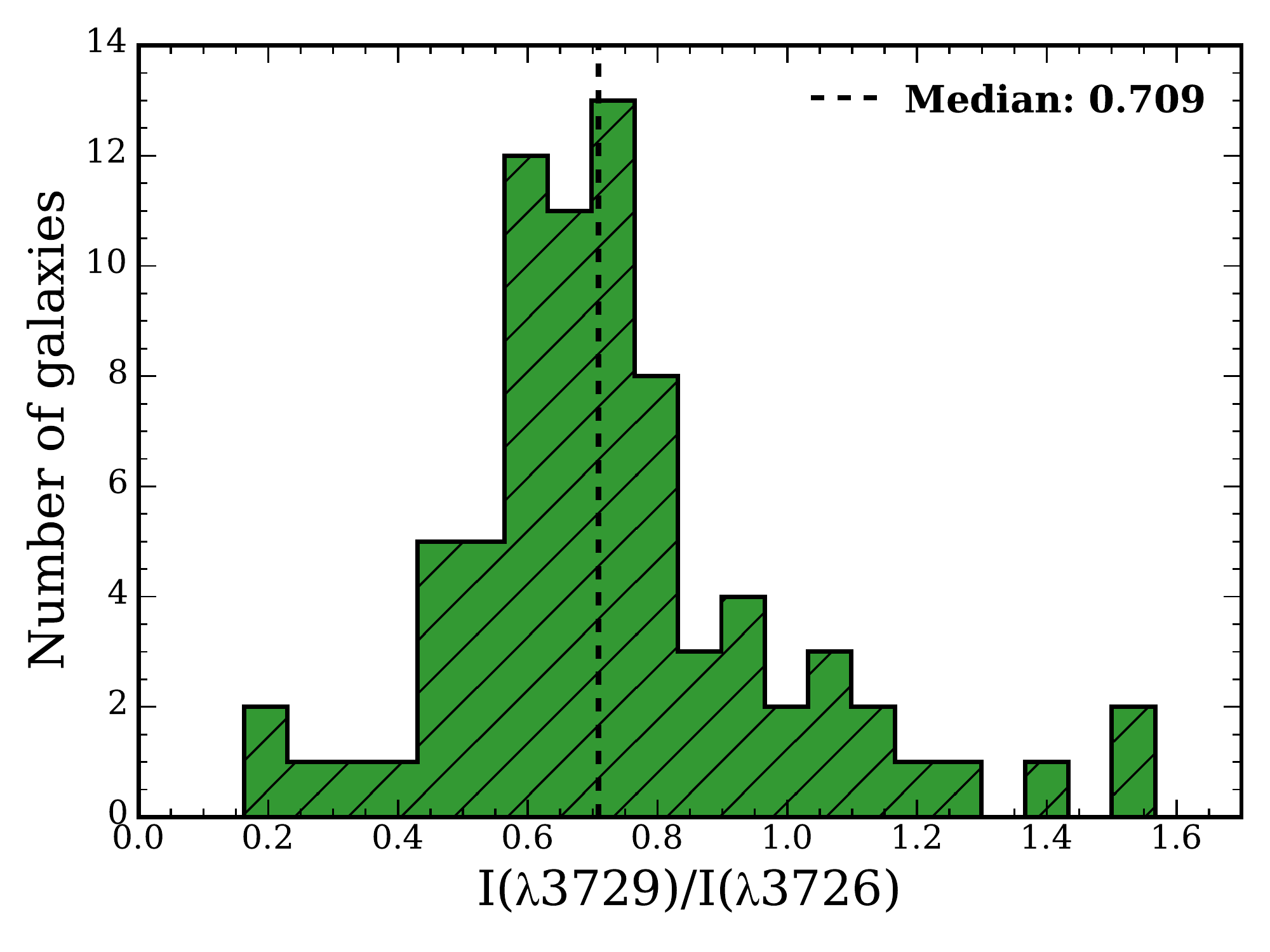}
      \caption{Intensity ratio of the [O{\small II}] doublet distribution of our galaxy sample at $\mathrm{z\simeq 0.9}$. The vertical dotted line lies at the median value of the distribution.
              }
              \label{fig:OII_dist}
   \end{figure}
The value of $\mathrm{R_{[O{\small II}]}}$ was computed during our modeling of the galaxy kinematics (see $\S$ \ref{subsec:fitting}), and is defined as the flux ratio between the line at longer wavelength ($\lambda=3729\,\AA$) and the one at shorter wavelength ($\lambda=3726\,\AA$). Figure \ref{fig:OII_dist} shows  $\mathrm{R_{[O{\small II}]}}$ distribution for our sample at $\mathrm{0.75<z<1.2}$. It is consistent with the range of values computed by \citet{Osterbrock1989}, from $\mathrm{R_{[O{\small II}]}=0.35}$ in the limit of  high electronic density ($\mathrm{N_e\rightarrow \infty}$) to $\mathrm{R_{[O{\small II}]}=1.5}$ in the limit of low electronic density ($\mathrm{N_e\rightarrow 0}$) for temperatures typical of star-forming regions ($\mathrm{T\sim 10^4 K}$).
\begin{figure}[!tb]
   \centering
    \includegraphics[width=0.9 \hsize]{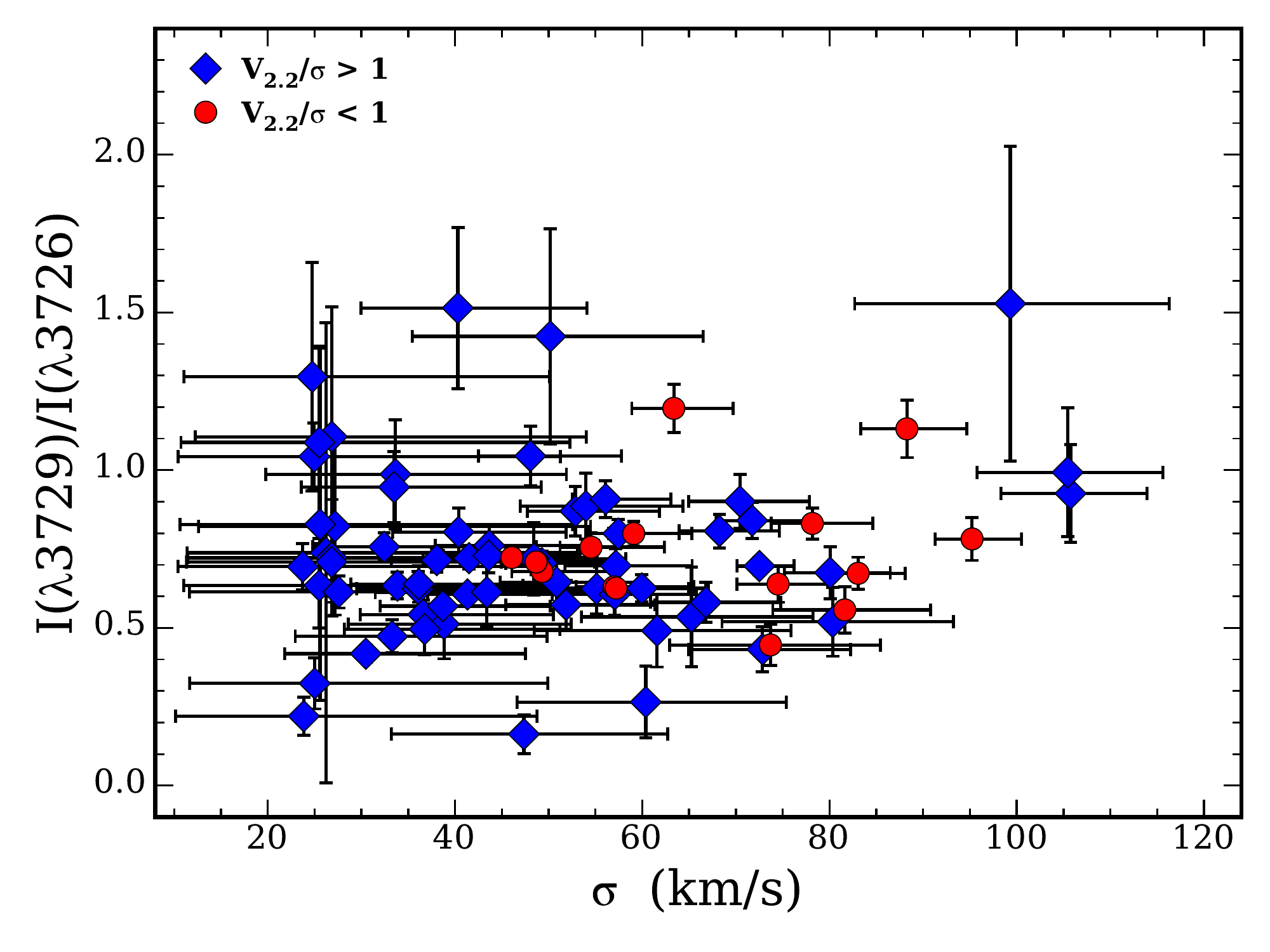}
    \includegraphics[width=0.9 \hsize]{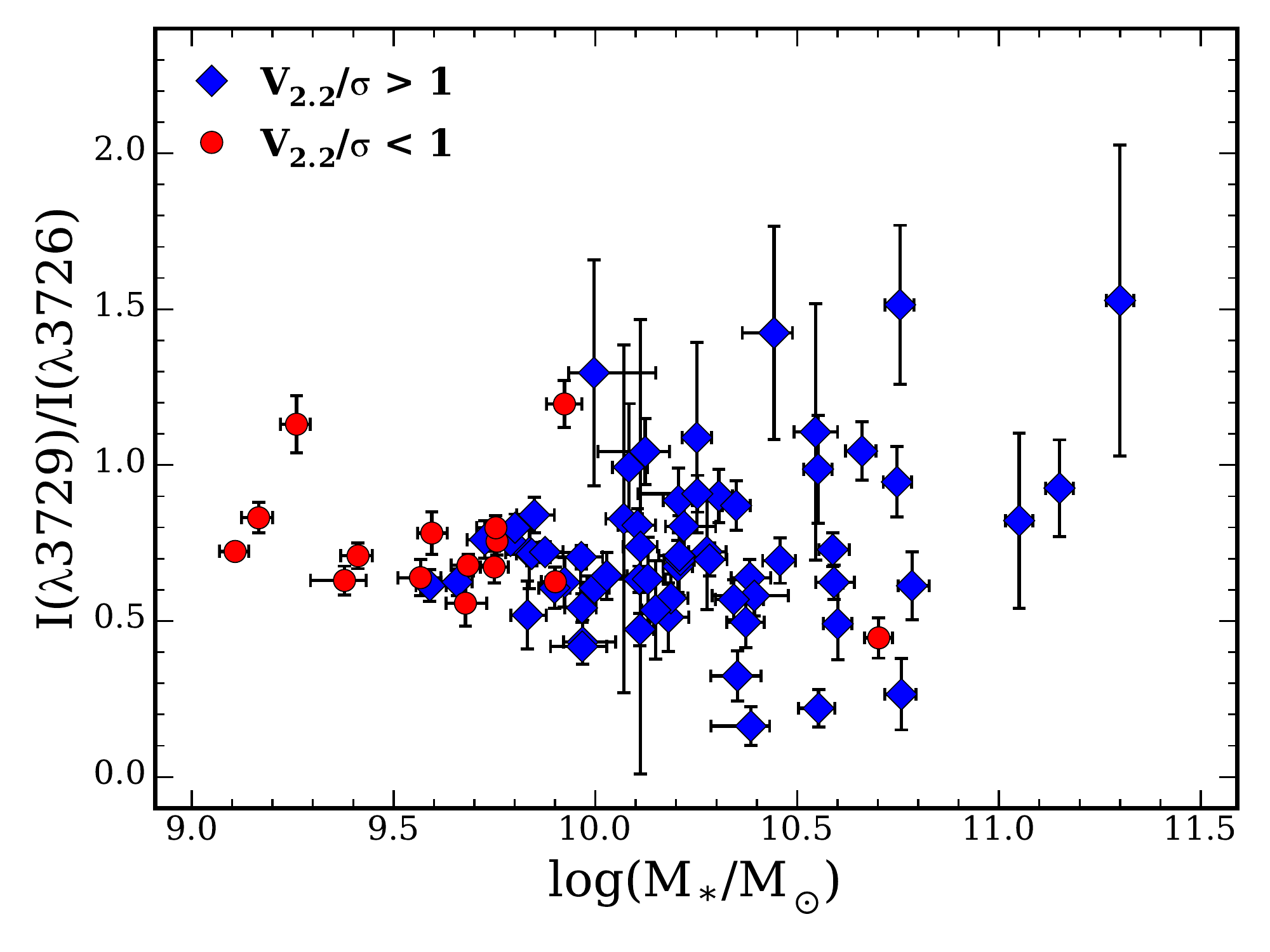}
     \includegraphics[width=0.9 \hsize]{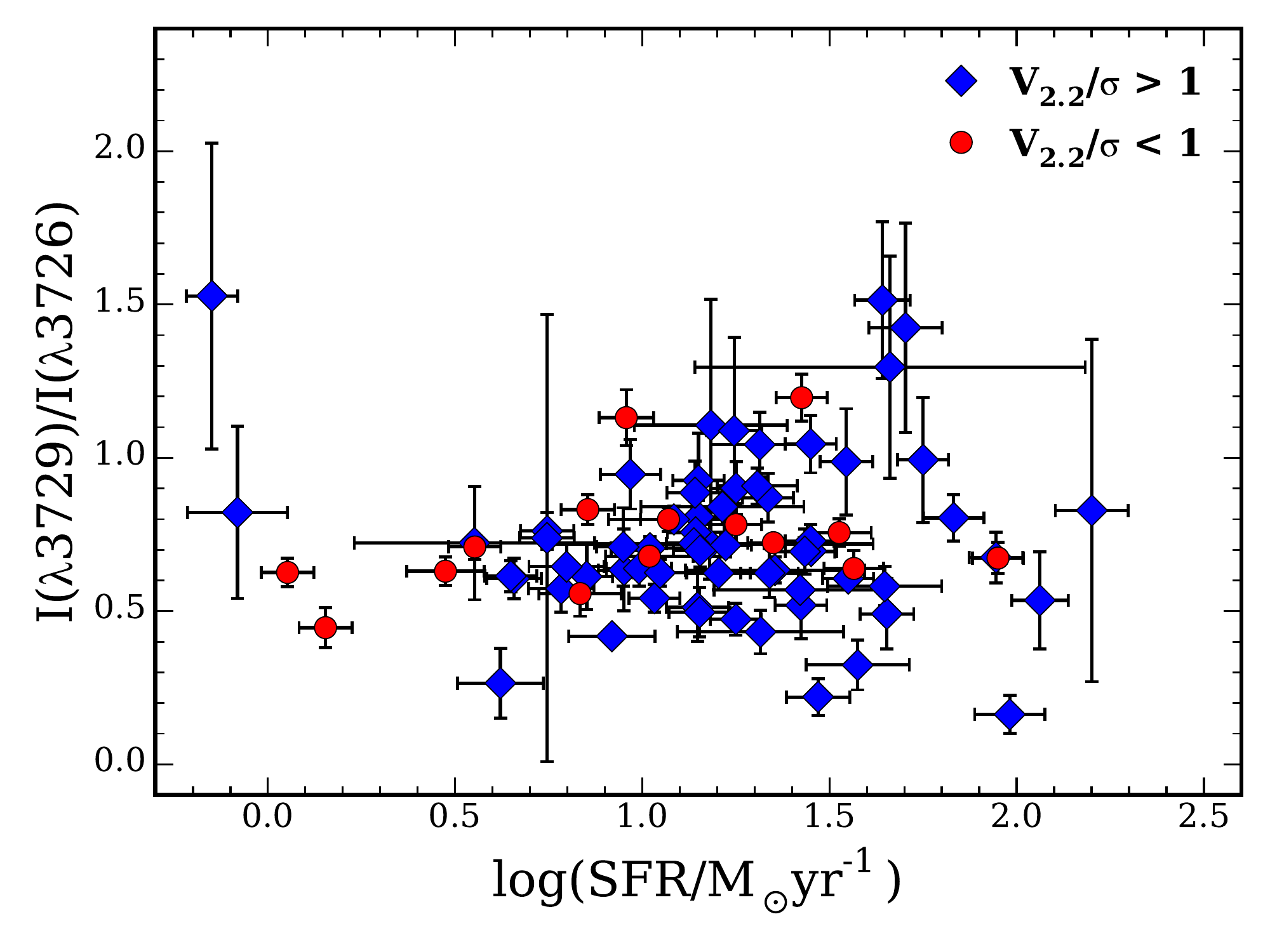}
      \caption{Intensity ratio of the [O{\small II}] doublet  versus velocity dispersion (top), stellar mass (middle) and $\mathrm{SFR_{SED}}$ (bottom). The blue diamonds and the red circles show the rotation- and dispersion-dominated galaxies, respectively.
              }
              \label{fig:OII_vs_sig_star_sfr}
   \end{figure}

 Figure \ref{fig:OII_vs_sig_star_sfr} present $\mathrm{R_{[O{\small II}]}}$ versus the velocity dispersion (top panel), the stellar mass (middle panel) and the $\mathrm{SFR_{SED}}$ (bottom panel). We find that the electronic density (probed by $\mathrm{R_{[O{\small II}]}}$) for those star-forming disc galaxies does not depend on the velocity dispersion (both for dispersion- and rotation-dominated galaxies), nor on the stellar mass or the star formation rate. We note that the spread of the electronic density values appears larger in massive galaxies ($\mathrm{M_{\ast}>10^{10} M_{\odot}}$).   
\begin{figure}[t]
   \centering
    \includegraphics[width=0.9 \hsize]{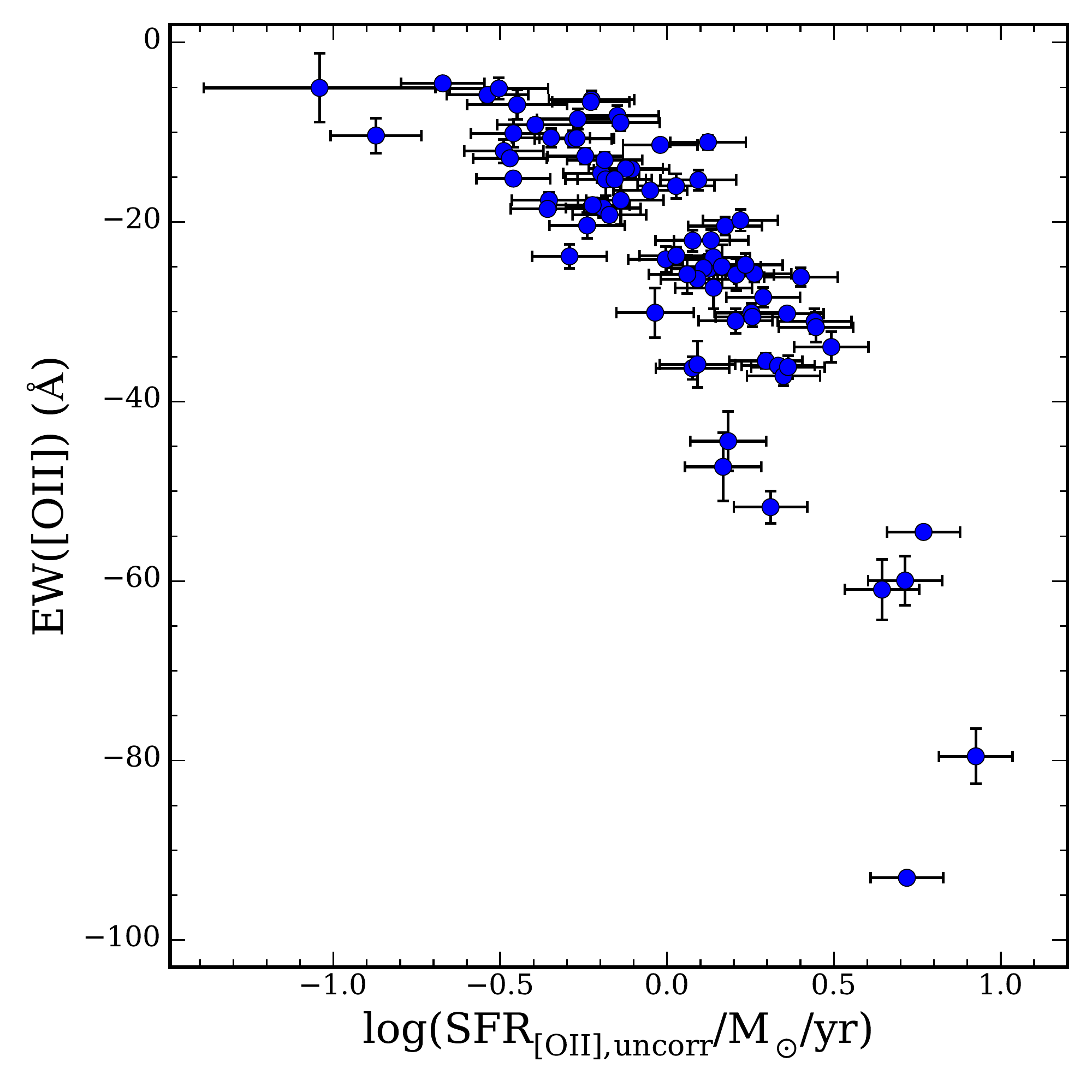}
      \caption{[O{\small II}] equivalent width versus log SFR computed from [O{\small II}] emission line and uncorrected for interstellar extinction.
              }
              \label{fig:EW_sfr_uncorr}
   \end{figure}
   
The EW([O{\small II}]) is computed by defining two ``continuum'' bandpasses, slightly blueward and redward of the [O{\small II}] doublet, which are used to estimate  the stellar continuum across the spectral feature. An additional ``feature'' bandpass is defined to include the spectral line. The stellar continuum is computed as the median value over the two continuum bandpasses. The bandpasses were chosen by eye for each galaxy spectrum to avoid possible contaminating features near the spectral lines of interest.
The EW is defined as
 \begin{equation} \label{eq:ew}
\\ \\  \qquad \mathrm{EW(\AA)}=\sum_{i=0}^{n} \dfrac{C_i - F_i}{C_i}\, \Delta \lambda_i ,
\end{equation}
where $F_i$ is the flux in the $i$th spectral pixel in the feature bandpass, $C_i$ is the continuum flux in the $i$th spectral pixel over the same bandpass, and $\Delta \lambda_i$ is the pixel scale of the spectrum ($\AA$/pixel). Errors in the EW were derived using the Poisson errors on the spectral feature. Given this definition of EW, the convention adopted in this work is for negative EWs to correspond to spectral lines observed in emission.
\begin{figure}[tb]
   \centering
    \includegraphics[width=0.9 \hsize]{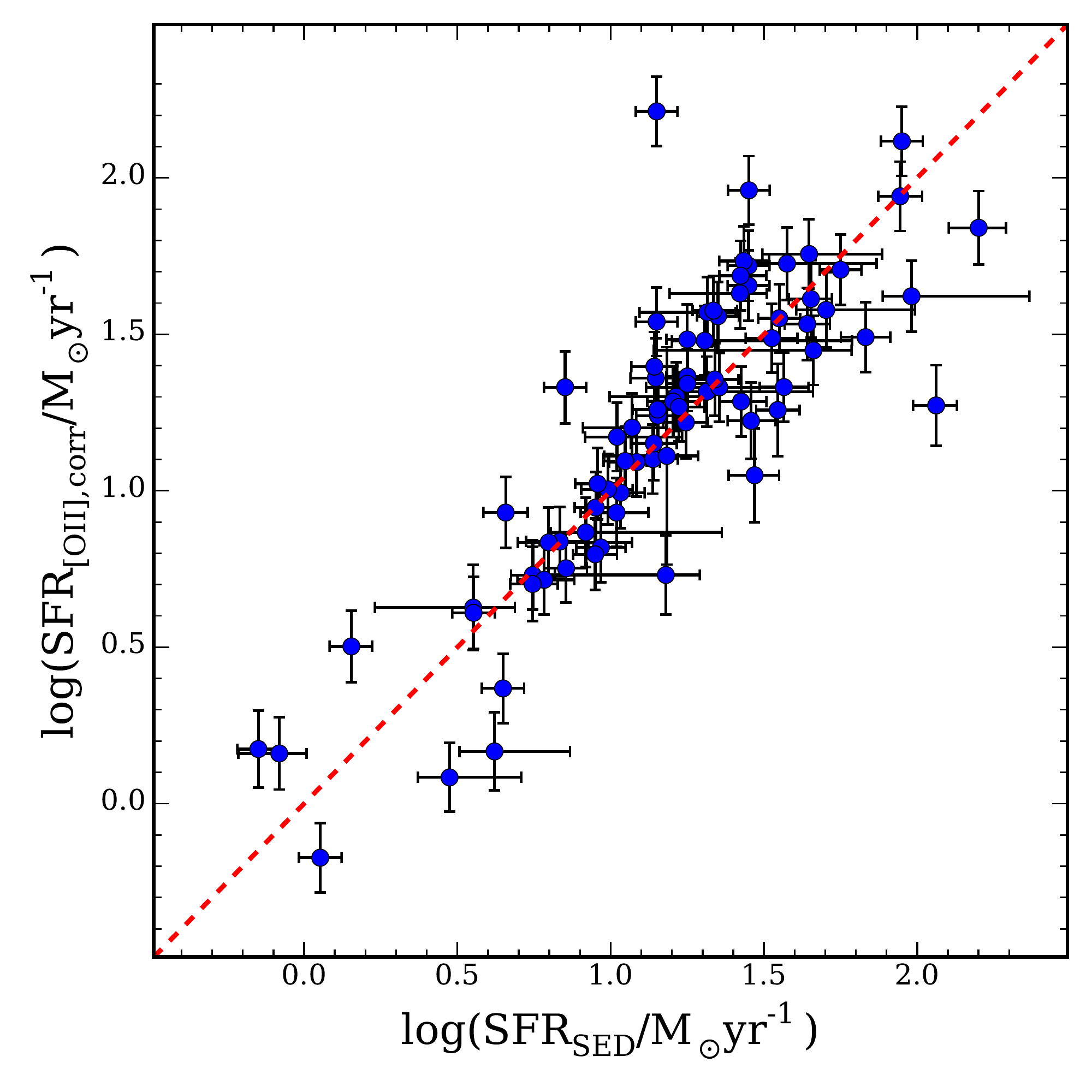}
      \caption{Comparison of SFR computed from SED fitting process and [O{\small II}] emission line corrected for internal extinction in logarithmic units. The red dashed line represents the 1:1 relation.
              }
              \label{fig:SFR_sed_SFR_oii}
   \end{figure}
The $\mathrm{SFR_{SED}}$ is computed making use of the Le Phare software \citep{Arnouts2002, Ilbert2006} following the same recipes applied to compute the stellar masses ($\S$ \ref{subsec:stellarmass}).  The $\mathrm{SFR_{[O{\small II}]}}$ is computed using the equation (5) from \citet{Lemaux2014} based on SFR formula of Kewley et al. (2004) and adapted to substitute the [O{\small II}] flux with the measurement of EW([O{\small II}]). The choice of using EW([O{\small II}]) instead of  [O{\small II}] flux was motivated by the lack of the absolute flux calibration for our spectroscopic observations (see $\S$ \ref{subsec:DataReduction}).  We show  the relation between the computed $\mathrm{SFR_{[O{\small II}]}}$ uncorrected for the internal extinction and the EW([O{\small II}] in Figure \ref{fig:EW_sfr_uncorr}.
The correction for the internal extinction was applied using a prescription proposed by \citet{Wuyts2013}, based on the stellar continuum reddening from the SED fitting, modified to take into account an extra extinction expected from the HII regions.

In the Figure \ref{fig:SFR_sed_SFR_oii}, we show the comparison between the $\mathrm{SFR_{[O{\small II}]}}$ and $\mathrm{SFR_{SED}}$. Our two SFR estimations are in a very good agreement.

\section{Scatter around the smTF relation} \label{appendixB}
Physical scatter around the smTF relation can be due to variations in the stellar mass fraction (stellar-to-total mass ratio), or differences in how the observed rotation velocity relates to the total mass \citep{Kannappan2002}, or a combination of the two. We decided, therefore, to geometrically compute the scatter around the smTF relation as the shortest distance $d_w$ of the data from the relation weighted by the uncertainties, since $d_w$ carries information on the physical scatter coming from both the stellar mass and the rotation velocity. We used this approach in our investigation of the dependence of the smTF relation with the environment ($\S$ \ref{subsec:environment}).
\begin{figure}[!htb]
   \centering
    \includegraphics[width=0.97 \hsize]{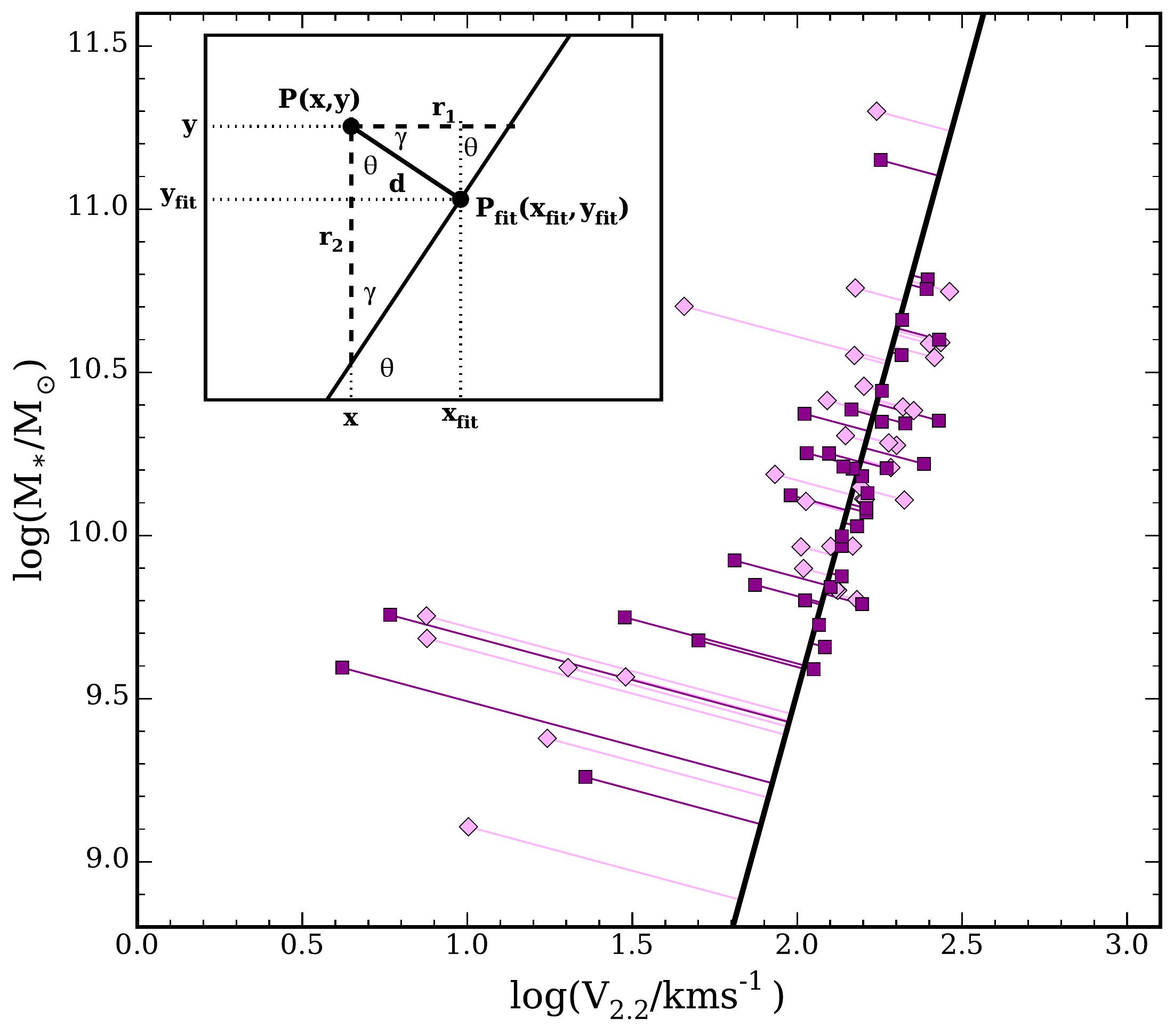}
      \caption{Distances $d_w$ of the data points from the smTF relation. Pink diamonds and purple squares with relative lines refer to galaxies in lower and higher density environment, respectively. The embedded plot in the upper left corner show a simplified scheme of the geometry used to compute $d_w$ (see text $\S$ \ref{appendixB}).
              }
               \label{fig:dist_fromTf}
   \end{figure}
In Figure \ref{fig:dist_fromTf} we show the distances computed from our smTF relation, color coded according to our environment definition (purple squares for higher density and pink diamonds for lower density environment). In the upper left corner of the plot we show a scheme of the geometry used to compute the scatter. Following the trigonometry, we defined the distance $d$ between the data point P(x,y) and the point $\mathrm{P_{fit}(x_{fit},y_{fit})}$ on the relation as
\begin{equation} \label{eq:scatter_env}
\\ \\  \qquad d=\dfrac{\mid r_1\,\sin\theta \mid + \mid r_2\,\cos\theta \mid }{2} ,
\end{equation}
where $\theta$ is the angle between the relation and the positive direction of the $x $-axis, and is expressed as $\mathrm{\theta =\arctan(slope)}$, and $r_1$ (or $r_2$) is the distance between the $x$ value (or the $y$ value) of the data points and the relation at fixed value of $y$ (or $x$). We chose to define $d$ as in Equation \ref{eq:scatter_env}, so that we can include the uncertainties from both the $x$ and $y$ variables in the error budget. Therefore the scatter error is expressed as
\begin{equation} \label{eq:err_scatter_env}
\\ \\  \qquad \delta_d=\dfrac{\sqrt{(\delta_{r_1}\,\sin\theta)^2 + (\delta_{r_2}\,\cos\theta)^2} }{2} .
\end{equation}
We then computed the weighted $d_w$ as
\begin{equation} \label{eq:d_weighted}
\\ \\  \qquad  d_{w} = d/\delta_d .
\end{equation}
In $\S$ \ref{subsec:smTF}, we did not use this approach to compute the intrinsic scatter $\sigma_{intr}$ of the rotation-dominated galaxy sample around the smTR relation to ease the comparison of our results with previous works. In the study of possible dependencies of the smTF relation with the environment, instead, we made a comparison  internal  to our sample between scatters for galaxies in two different environments, and the distance $d_w$ was a more appropriate quantity for this purpose.
\setcounter{table}{0}
\renewcommand{\thetable}{B\arabic{table}}
\newpage
\onecolumn
\begin{longtab}   
\begin{longtable}{cccccc}
\caption{[O{\small II}] measurements} \label{tab:oii_measurements}\\
\hline\hline
\noalign{\smallskip} 
ID  & $\mathrm{R_{[O{\small II}]}}$ & $\mathrm{EW([O{\small II}])}$ & $\mathrm{E_s(B-V)_{SED}}$ & $\mathrm{log(SFR_{SED}/ M_\odot\, yr^{-1})}$ & $\mathrm{log(SFR_{[O{\small II}], corr}/ M_\odot\, yr^{-1})}$\\[5pt] 
&  & $\AA$ & &  & \\  [5pt] 
(1) & (2)  & (3)  & (4) & (5) & (6) \\
\hline
\noalign{\smallskip} 
\endfirsthead
\caption{continued.}\\   
\hline\hline   
\noalign{\smallskip}     
ID  & $\mathrm{R_{[O{\small II}]}}$ & $\mathrm{EW([O{\small II}])}$ & $\mathrm{E_s(B-V)_{SED}}$ & $\mathrm{log(SFR_{SED}/ M_\odot\, yr^{-1})}$  & $\mathrm{log(SFR_{[O{\small II}], corr}/ M_\odot\, yr^{-1})}$\\[5pt] 
&  & $\AA$ & &  & \\  [5pt] 
(1) & (2)  & (3)  & (4) & (5) & (6)  \\
\hline
\noalign{\smallskip} 
\endhead
\hline
\endfoot       
701002 &  0.51$\pm$ 0.11 & -17.57$\pm$ 2.70 & 0.4 & $ 1.15_{-0.08}^{+0.08}$ &  1.36$\pm$ 0.13\\ [2pt]
701403 &  0.64$\pm$ 0.06 & -33.92$\pm$ 1.70 & 0.2 & $ 1.57_{-0.08}^{+0.08}$ &  1.33$\pm$ 0.11\\ [2pt]
811012 &  0.72$\pm$ 0.02 & -93.06$\pm$ 0.59 & 0.2 & $ 1.35_{-0.07}^{+0.07}$ &  1.56$\pm$ 0.11\\ [2pt]
811108 &  0.90$\pm$ 0.09 & -20.45$\pm$ 0.99 & 0.3 & $ 1.25_{-0.07}^{+0.07}$ &  1.37$\pm$ 0.11\\ [2pt]
811224 &  0.49$\pm$ 0.12 & -8.16$\pm$ 1.12 & 0.5 & $ 1.65_{-0.07}^{+0.07}$ &  1.61$\pm$ 0.12\\ [2pt]
811233 &  4.84$\pm$ 7.38 & -5.84$\pm$ 0.75 & 0.5 & $ 1.46_{-0.08}^{+0.08}$ &  1.22$\pm$ 0.12\\ [2pt]
811727 &  0.16$\pm$ 0.06 & -11.11$\pm$ 0.80 & 0.4 & $ 1.98_{-0.09}^{+0.38}$ &  1.62$\pm$ 0.11\\ [2pt]
811920 &  0.76$\pm$ 0.04 & -35.48$\pm$ 0.82 & 0.3 & $ 1.53_{-0.09}^{+0.08}$ &  1.49$\pm$ 0.11\\ [2pt]
812913 &  0.82$\pm$ 0.28 & -10.75$\pm$ 0.93 & 0.1 & $-0.08_{-0.13}^{+0.09}$ &  0.16$\pm$ 0.12\\ [2pt]
813055 &  0.56$\pm$ 0.07 & -17.57$\pm$ 0.88 & 0.3 & $ 0.83_{-0.11}^{+0.09}$ &  0.84$\pm$ 0.11\\ [2pt]
813128 &  0.54$\pm$ 0.05 & -14.56$\pm$ 1.08 & 0.3 & $ 1.03_{-0.07}^{+0.08}$ &  0.99$\pm$ 0.11\\ [2pt]
817262 &  1.04$\pm$ 0.11 & -25.19$\pm$ 1.61 & 0.3 & $ 1.31_{-0.13}^{+0.35}$ &  1.32$\pm$ 0.11\\ [2pt]
817416 &  0.78$\pm$ 0.07 & -60.95$\pm$ 3.35 & 0.2 & $ 1.25_{-0.07}^{+0.07}$ &  1.48$\pm$ 0.11\\ [2pt]
817426 &  0.70$\pm$ 0.04 & -36.01$\pm$ 0.78 & 0.2 & $ 1.02_{-0.10}^{+0.21}$ &  1.17$\pm$ 0.11\\ [2pt]
817640 &  0.63$\pm$ 0.04 & -23.93$\pm$ 0.65 & 0.3 & $ 1.35_{-0.24}^{+0.13}$ &  1.33$\pm$ 0.11\\ [2pt]
818113 &  0.61$\pm$ 0.03 & -30.20$\pm$ 0.60 & 0.3 & $ 1.55_{-0.07}^{+0.07}$ &  1.55$\pm$ 0.11\\ [2pt]
818198 &  0.84$\pm$ 0.06 & -25.12$\pm$ 0.64 & 0.3 & $ 1.21_{-0.22}^{+0.09}$ &  1.30$\pm$ 0.11\\ [2pt]
818734 &  0.68$\pm$ 0.03 & -26.35$\pm$ 0.87 & 0.2 & $ 1.02_{-0.12}^{+0.10}$ &  0.93$\pm$ 0.11\\ [2pt]
818959 &  0.72$\pm$ 0.12 & -10.11$\pm$ 1.53 & 0.3 & $ 1.18_{-0.44}^{+0.11}$ &  0.73$\pm$ 0.13\\ [2pt]
819479 &  0.63$\pm$ 0.05 & -18.54$\pm$ 0.61 & 0.1 & $ 0.47_{-0.10}^{+0.23}$ &  0.08$\pm$ 0.11\\ [2pt]
819641 &  0.99$\pm$ 0.17 & -5.12$\pm$ 1.17 & 0.5 & $ 1.54_{-0.07}^{+0.07}$ &  1.26$\pm$ 0.15\\ [2pt]
819765 &  0.43$\pm$ 0.07 & -18.44$\pm$ 1.06 & 0.5 & $ 1.32_{-0.22}^{+0.09}$ &  1.57$\pm$ 0.11\\ [2pt]
823045 &  0.63$\pm$ 0.13 & -12.66$\pm$ 0.88 & 0.3 & $ 0.95_{-0.07}^{+0.07}$ &  0.95$\pm$ 0.11\\ [2pt]
823323 &  0.83$\pm$ 0.56 & -8.94$\pm$ 0.89 & 0.6 & $ 2.20_{-0.10}^{+0.09}$ &  1.84$\pm$ 0.12\\ [2pt]
823909 &  0.80$\pm$ 0.08 & -10.67$\pm$ 0.58 & 0.5 & $ 1.83_{-0.08}^{+0.08}$ &  1.49$\pm$ 0.11\\ [2pt]
824079 &  0.81$\pm$ 0.05 & -37.16$\pm$ 1.07 & 0.3 & $ 1.15_{-0.07}^{+0.07}$ &  1.54$\pm$ 0.11\\ [2pt]
824317 &  0.62$\pm$ 0.08 & -24.97$\pm$ 2.40 & 0.3 & $ 1.34_{-0.08}^{+0.08}$ &  1.36$\pm$ 0.12\\ [2pt]
824384 &  0.61$\pm$ 0.11 & -27.33$\pm$ 2.32 & 0.3 & $ 0.85_{-0.07}^{+0.07}$ &  1.33$\pm$ 0.12\\ [2pt]
824408 &  0.76$\pm$ 0.06 & -15.16$\pm$ 0.72 & 0.3 & $ 0.75_{-0.07}^{+0.07}$ &  0.73$\pm$ 0.11\\ [2pt]
824508 &  0.62$\pm$ 0.06 & -15.33$\pm$ 1.11 & 0.3 & $ 1.20_{-0.08}^{+0.09}$ &  1.29$\pm$ 0.11\\ [2pt]
824658 &  1.04$\pm$ 0.09 & -14.12$\pm$ 0.96 & 0.5 & $ 1.45_{-0.07}^{+0.07}$ &  1.66$\pm$ 0.11\\ [2pt]
824675 &  0.57$\pm$ 0.08 & -14.05$\pm$ 0.69 & 0.2 & $ 0.78_{-0.09}^{+0.10}$ &  0.72$\pm$ 0.11\\ [2pt]
824746 &  0.72$\pm$ 0.04 & -36.28$\pm$ 1.27 & 0.3 & $ 1.22_{-0.07}^{+0.08}$ &  1.27$\pm$ 0.11\\ [2pt]
824791 &  0.74$\pm$ 0.73 & -12.09$\pm$ 1.31 & 0.3 & $ 0.75_{-0.07}^{+0.08}$ &  0.70$\pm$ 0.12\\ [2pt]
824847 &  1.42$\pm$ 0.34 & -15.26$\pm$ 1.81 & 0.5 & $ 1.70_{-0.10}^{+0.29}$ &  1.58$\pm$ 0.12\\ [2pt]
825250 &  0.64$\pm$ 0.06 & -13.09$\pm$ 0.84 & 0.3 & $ 0.99_{-0.09}^{+0.08}$ &  1.00$\pm$ 0.11\\ [2pt]
825269 &  0.76$\pm$ 0.04 & -31.02$\pm$ 1.37 & 0.3 & $ 1.14_{-0.08}^{+0.07}$ &  1.40$\pm$ 0.11\\ [2pt]
825474 &  0.95$\pm$ 0.11 & -11.41$\pm$ 0.65 & 0.2 & $ 0.97_{-0.08}^{+0.08}$ &  0.82$\pm$ 0.11\\ [2pt]
826042 &  0.67$\pm$ 0.05 & -79.53$\pm$ 3.06 & 0.3 & $ 1.95_{-0.07}^{+0.07}$ &  2.12$\pm$ 0.11\\ [2pt]
826065 &  0.61$\pm$ 0.07 & -35.86$\pm$ 2.57 & 0.2 & $ 0.66_{-0.07}^{+0.07}$ &  0.93$\pm$ 0.11\\ [2pt]
826091 &  0.64$\pm$ 0.08 & -24.15$\pm$ 1.44 & 0.2 & $ 0.80_{-0.10}^{+0.27}$ &  0.84$\pm$ 0.11\\ [2pt]
826948 &  0.68$\pm$ 0.08 & -31.07$\pm$ 1.40 & 0.4 & $ 1.94_{-0.07}^{+0.07}$ &  1.94$\pm$ 0.11\\ [2pt]
827050 &  0.93$\pm$ 0.16 & -59.94$\pm$ 2.73 & 0.4 & $ 1.15_{-0.07}^{+0.07}$ &  2.21$\pm$ 0.11\\ [2pt]
827090 &  0.72$\pm$ 0.03 & -25.75$\pm$ 0.95 & 0.2 & $ 1.14_{-0.07}^{+0.08}$ &  1.10$\pm$ 0.11\\ [2pt]
827096 &  0.87$\pm$ 0.08 & -22.07$\pm$ 1.19 & 0.4 & $ 1.34_{-0.07}^{+0.08}$ &  1.58$\pm$ 0.11\\ [2pt]
829955 &  0.54$\pm$ 0.16 & -6.37$\pm$ 0.99 & 0.4 & $ 2.06_{-0.08}^{+0.07}$ &  1.27$\pm$ 0.13\\ [2pt]
830321 &  0.83$\pm$ 0.05 & -51.77$\pm$ 1.79 & 0.1 & $ 0.85_{-0.07}^{+0.07}$ &  0.75$\pm$ 0.11\\ [2pt]
830414 &  0.45$\pm$ 0.06 & -25.83$\pm$ 2.15 & 0.1 & $ 0.15_{-0.07}^{+0.07}$ &  0.50$\pm$ 0.11\\ [2pt]
831094 &  0.70$\pm$ 0.03 & -54.53$\pm$ 0.63 & 0.3 & $ 1.45_{-0.07}^{+0.07}$ &  1.96$\pm$ 0.11\\ [2pt]
831223 &  1.11$\pm$ 0.41 & -5.05$\pm$ 3.84 & 0.7 & $ 1.18_{-0.20}^{+0.10}$ &  1.11$\pm$ 0.35\\ [2pt]
831229 &  0.72$\pm$ 0.18 & -10.36$\pm$ 1.95 & 0.4 & $ 0.55_{-0.32}^{+0.14}$ &  0.63$\pm$ 0.14\\ [2pt]
831256 &  0.58$\pm$ 0.06 & -18.12$\pm$ 0.81 & 0.6 & $ 1.65_{-0.15}^{+0.24}$ &  1.76$\pm$ 0.11\\ [2pt]
831296 &  0.42$\pm$ 0.03 & -23.76$\pm$ 0.98 & 0.2 & $ 0.92_{-0.11}^{+0.44}$ &  0.87$\pm$ 0.11\\ [2pt]
831493 &  0.73$\pm$ 0.05 & -19.80$\pm$ 1.20 & 0.4 & $ 1.45_{-0.07}^{+0.07}$ &  1.72$\pm$ 0.11\\ [2pt]
831534 &  1.53$\pm$ 0.50 & -8.53$\pm$ 1.10 & 0.1 & $-0.15_{-0.07}^{+0.07}$ &  0.17$\pm$ 0.12\\ [2pt]
831675 &  1.51$\pm$ 0.26 & -6.62$\pm$ 0.60 & 0.5 & $ 1.64_{-0.07}^{+0.07}$ &  1.53$\pm$ 0.12\\ [2pt]
831848 &  0.99$\pm$ 0.20 & -25.87$\pm$ 1.78 & 0.4 & $ 1.75_{-0.07}^{+0.07}$ &  1.71$\pm$ 0.11\\ [2pt]
832184 &  1.30$\pm$ 0.36 & -16.48$\pm$ 0.65 & 0.4 & $ 1.66_{-0.52}^{+0.12}$ &  1.45$\pm$ 0.11\\ [2pt]
832277 &  0.69$\pm$ 0.07 & -24.77$\pm$ 1.27 & 0.4 & $ 1.43_{-0.08}^{+0.08}$ &  1.73$\pm$ 0.11\\ [2pt]
832385 &  0.32$\pm$ 0.08 & -30.10$\pm$ 2.76 & 0.5 & $ 1.58_{-0.14}^{+0.29}$ &  1.73$\pm$ 0.12\\ [2pt]
832708 &  0.70$\pm$ 0.05 & -26.13$\pm$ 1.03 & 0.2 & $ 1.15_{-0.07}^{+0.07}$ &  1.24$\pm$ 0.11\\ [2pt]
833167 &  0.80$\pm$ 0.04 & -36.15$\pm$ 1.27 & 0.2 & $ 1.07_{-0.16}^{+0.11}$ &  1.20$\pm$ 0.11\\ [2pt]
833209 &  1.20$\pm$ 0.08 & -31.71$\pm$ 1.64 & 0.2 & $ 1.43_{-0.07}^{+0.08}$ &  1.29$\pm$ 0.11\\ [2pt]
833707 &  0.89$\pm$ 0.10 & -10.61$\pm$ 1.04 & 0.4 & $ 1.14_{-0.08}^{+0.08}$ &  1.15$\pm$ 0.12\\ [2pt]
833862 &  0.52$\pm$ 0.11 & -23.83$\pm$ 1.32 & 0.6 & $ 1.42_{-0.07}^{+0.08}$ &  1.69$\pm$ 0.11\\ [2pt]
834100 &  1.13$\pm$ 0.09 & -44.41$\pm$ 3.31 & 0.2 & $ 0.96_{-0.07}^{+0.07}$ &  1.02$\pm$ 0.11\\ [2pt]
837355 &  0.71$\pm$ 0.04 & -47.27$\pm$ 3.80 & 0.1 & $ 0.55_{-0.07}^{+0.07}$ &  0.61$\pm$ 0.11\\ [2pt]
837433 &  0.71$\pm$ 0.13 & -9.18$\pm$ 0.74 & 0.3 & $ 0.95_{-0.07}^{+0.07}$ &  0.80$\pm$ 0.11\\ [2pt]
837491 &  0.80$\pm$ 0.04 & -30.12$\pm$ 1.08 & 0.2 & $ 1.08_{-0.09}^{+0.08}$ &  1.09$\pm$ 0.11\\ [2pt]
837613 &  0.50$\pm$ 0.08 & -20.38$\pm$ 1.45 & 0.4 & $ 1.15_{-0.08}^{+0.08}$ &  1.26$\pm$ 0.11\\ [2pt]
837931 &  0.63$\pm$ 0.05 & -19.21$\pm$ 0.76 & 0.0 & $ 0.05_{-0.07}^{+0.07}$ & -0.17$\pm$ 0.11\\ [2pt]
838455 &  0.22$\pm$ 0.06 & -6.93$\pm$ 1.64 & 0.4 & $ 1.47_{-0.08}^{+0.08}$ &  1.05$\pm$ 0.15\\ [2pt]
839193 &  1.09$\pm$ 0.30 & -15.99$\pm$ 1.38 & 0.3 & $ 1.25_{-0.07}^{+0.07}$ &  1.22$\pm$ 0.12\\ [2pt]
839379 &  0.61$\pm$ 0.05 & -12.90$\pm$ 0.54 & 0.2 & $ 0.65_{-0.07}^{+0.07}$ &  0.37$\pm$ 0.11\\ [2pt]
840112 &  0.26$\pm$ 0.11 & -4.55$\pm$ 0.64 & 0.2 & $ 0.62_{-0.11}^{+0.25}$ &  0.17$\pm$ 0.12\\ [2pt]
840266 &  0.91$\pm$ 0.06 & -28.39$\pm$ 1.12 & 0.3 & $ 1.31_{-0.11}^{+0.48}$ &  1.48$\pm$ 0.11\\ [2pt]
840390 &  0.47$\pm$ 0.05 & -15.25$\pm$ 0.79 & 0.4 & $ 1.25_{-0.07}^{+0.07}$ &  1.34$\pm$ 0.11\\ [2pt]
840437 &  0.62$\pm$ 0.04 & -30.57$\pm$ 1.08 & 0.2 & $ 1.05_{-0.07}^{+0.07}$ &  1.09$\pm$ 0.11\\ [2pt]
1254477 &  0.57$\pm$ 0.06 & -22.02$\pm$ 1.17 & 0.4 & $ 1.42_{-0.23}^{+0.09}$ &  1.63$\pm$ 0.11\\ [2pt]
\end{longtable}
\begin{flushleft} 
\justify
(1) Source HR-COSMOS identification number, (2) [O{\small II}]$\lambda3729/$[O{\small II}]$\lambda3726$ ratio, (3) [O{\small II}] rest-frame equivalent width, (4) color excess of the stellar continuum coming from the SED fitting process, (5) log(SFR) derived from the SED fitting process ($\S$ \ref{subsec:stellarmass}), (6)  log(SFR) derived from the [O{\small II}] EW corrected for internal extinction ($\S$ \ref{appendixC}).
\end{flushleft}
\end{longtab}
\end{document}